\documentclass{aa} 
\usepackage{float} 
\usepackage{graphicx}  
\usepackage{graphics}
\usepackage{txfonts}  
\usepackage{natbib}
\usepackage{adjustbox}
\bibpunct{(}{)}{;}{a}{}{,}
\usepackage[dvipsnames]{xcolor}
\usepackage[normalem]{ulem}
\usepackage{tabularx}
\usepackage{booktabs}
\usepackage{multirow}

\newcommand{\Msun}{M$_{\odot}$\xspace}

\newcommand{\vinf}{$v_\infty$\xspace}
\newcommand{\vsini}{$v_{\rm eq} \sin i$\xspace}

\usepackage{hyperref}
\hypersetup{
    colorlinks=true,
    linkcolor=blue,
    filecolor=magenta,      
    urlcolor=blue,
    citecolor=blue,
    pdftitle={},
    pdfpagemode=FullScreen,
    }
\usepackage[capitalise]{cleveref}
\usepackage{comment}

\begin{document}  

 
\title {Evolutionary models for the Very Massive Stars in the R136 cluster of 30~Doradus in the Large Magellanic Cloud}

\author{Z. Keszthelyi\inst{1,2} \and S.A. Brands\inst{1} \and A. de Koter\inst{1,3} \and N. Langer\inst{4,5} \and J. Puls\inst{6}}
\institute{
Anton Pannekoek Institute for Astronomy, University of Amsterdam, Science Park 904, 1098 XH, Amsterdam, The Netherlands 
           \and
Center for Computational Astrophysics, Division of Science, National Astronomical Observatory of Japan, 2-21-1, Osawa, Mitaka, Tokyo 181-8588, Japan\newline
          \email{zsolt.keszthelyi@nao.ac.jp} 
            \and 
Institute of Astronomy, KU Leuven, Celestijnenlaan 200D, 3001 Leuven, Belgium             
           \and 
         Argelander-Institut f\"ur Astronomie, 
         Universit\"at Bonn, Auf dem H\"ugel 71, 
         53121 Bonn, Germany
           \and
           Max-Planck-Institut f\"ur Radioastronomie, Auf dem H\"ugel 69, 53121 Bonn, Germany
           \and
LMU M\"unchen, Universit\"atssternwarte, Scheinerstr. 1, 81679 M\"unchen, Germany 
           }
\date{\today}

\abstract
{The cluster Radcliffe 136 in the Large Magellanic Cloud (LMC) contains a population of stars in excess of 100\,\Msun, including R136a1, the most massive star known. Very Massive Stars (VMSs) play an influential role in feedback processes and may potentially produce exotic supernova types and black holes of tens of solar masses. }
{The evolutionary history and final fate of the three most luminous stars, R136a1, R136a2, and R136a3, has been a puzzling issue. We aim to resolve this by dedicated stellar evolution models.}
{We compute rotating single-star \textsc{mesa} models and apply observationally constrained mass-loss rates during the early evolution and new theoretical Wolf-Rayet-type rates once the surface becomes enriched in helium. We consider various scenarios for internal angular momentum (AM) transport. We produce interpolated model grids and apply a Markov-Chain Monte Carlo analysis to compare our models with observations.}
{The nature of supernova progenitors strongly depends on mass loss and the AM coupling schemes. We predict no pair-instability and no gamma-ray burst progenitors from our fiducial model grid at LMC metallicity. 
The onset of Wolf-Rayet-type mass-loss rates on the main sequence leads to a rapid decrease in stellar mass and luminosity. The initially most massive model (800 \Msun) loses mass the most rapidly and becomes less massive than the initially least massive model (100 \Msun) in our grid. This mass turnover implies that the evolutionary history can only be inferred if additional constraints are available. We utilise the surface helium abundance, which poses a conundrum: R136a1, the most luminous star, is less enriched in helium than R136a2 and R136a3. We propose that this can be explained if both R136a2 and R136a3 were initially more massive than R136a1. From a rigorous confrontation of our models to spectroscopically-derived observables, we estimate an initial mass of 346$\pm41$~M$_\odot$ for R136a1, and $\gtrsim$500~M$_\odot$ for R136a2 and R136a3. 
}
{Even though VMSs are only present in the youngest clusters below 2 Myr of age, our study strengthens their role in local and galaxy evolution. At LMC metallicity, they will be observable as helium-enriched massive stars after their drastic mass loss, produced via single-star evolution. If the core collapse leads to a supernova, it will be of Type Ib/c. }

  
\titlerunning{Evolutionary models for VMSs in R136}
  
\authorrunning{Z. Keszthelyi et al.}  

\maketitle 

\section{Introduction}\label{intro}

Very Massive Stars (VMSs)\footnote{The precise definition of the mass limit for Very Massive Stars is not firmly established in the literature. Generally, VMSs are considered to have initial masses exceeding 60 or 100 \Msun, while Supermassive Stars are considered in excess of 1000 \Msun.} with $100$~M$_\odot <$~M$_{\rm ini}$~$< 1000$~M$_\odot$ \cite[e.g.,][]{vink2015,martins2015b,hirschi2017} dominate feedback processes and may contribute to chemical enrichment up to galactic scales through their strong stellar winds, intense ionising radiation fields and, ultimately, supernova explosions \citep[e.g.,][]{1977ApJ...218..148M,heger2002,doran2013,2021MNRAS.504.1039R,goswami2022,higgins2023,lahen2024}. VMSs that resulted from primordial star formation (Population III stars) are potential contributors to the re-ionisation of the universe \citep{bromm2001, 2014ApJ...781...60H,2015MNRAS.448..568H} and have likely played a crucial role in galaxy formation and evolution \citep[e.g.,][]{2009Natur.459...49B,kojima2021,eldridge2022}. 
%

The 30 Doradus region (a.k.a. the Tarantula Nebula) is a large and intrinsically bright star-forming region in the Large Magellanic Cloud \citep[LMC,][]{kennicutt1984,doran2013,nayak2023,kokusho2023} that hosts a large population of massive stars \citep{evans2011,schneider2018}. 30 Doradus has a metallicity approximately half that of the Milky way \citep[see][for a literature overview]{mokiem2007}, characteristic for giant starbursts in massive galaxies observed at high redshift \citep[e.g.][]{crowther2017}. 
The central cluster of 30 Dor, Radcliff 136 (R136) hosts the currently known most massive stars. Spectroscopic analysis of these stars yield current masses that reach up to 200--300\,M$_{\odot}$ \citep{crowther2010,crowther2016,bestenlehner2014,bestenlehner2020,brands2022,sabhahit2025}. Hence, in initial mass at least, they are representative of the very massive First Stars. 
As the spectral signatures of the most massive stars in the R136 cluster are dominated by very strong wind lines, they were originally thought to belong to evolved Wolf-Rayet stars \citep{parker1995}. However, Hubble Space Telescope (HST) observations of these stars were obtained by \cite{dekoter1997} who scrutinised their photospheric and atmospheric properties. This led to the understanding that these objects are not classical Wolf-Rayet stars, rather main sequence stars with particularly strong mass loss, displaying Wolf-Rayet-type spectral features. They belong to the specific subclass of Wolf-Rayet stars with the WNh designation, characterised by the presence of hydrogen and strong nitrogen emission in the spectrum.
Whether the winds of these objects are optically thin or thick is still not fully clear \citep{vink2012,bestenlehner2020,brands2022}.

In this study, we aim to contribute to our understanding of VMSs through tailored evolutionary modelling of the three most massive stars known, R136a1, R136a2, and R136a3, located in the R136 cluster. 
On the one hand, this will allow us to estimate their initial masses, that is, to place constraints on the stellar upper mass limit. Previous studies on the evolution of VMSs show that the strong dependence of mass loss on luminosity may potentially result in initially 250--1000\,\Msun\ stars to all end up with masses in the range of 200--300\,\Msun\ after about 1.5--2.5 Myrs of evolution \citep{2008A&A...477..223Y,kohler2015,higgins2022}. The most massive stars in R136 are about this massive and this old. We explore whether confronting our models to spectroscopically-determined observables can help break this degeneracy. 
On the other hand, we may address their future evolution, i.e., assess the possible supernova progenitor properties and supernova type \citep{smartt2009,smith2014}. At zero metallicity, VMS models of initially 140--260~M$_\odot$ are expected to produce pair-instability supernovae \citep{heger2002,chatzopoulos2012,takahashi2018,volpato2023}. This is likely not the case at solar metallicity \citep{2015ASSL..412..157H}, unless the progenitor could retain a substantial mass, for example, via magnetic mass-loss quenching \citep{petit2017,georgy2017}. 
Rotation induced internal mixing and angular momentum transport and mass and angular momentum loss by the stellar wind are considered as the most important physical mechanisms shaping the evolution of VMSs \citep[e.g.,][]{hirschi2004,Maeder2009a,brott2011,ekstroem2012,langer2012}. We seek to study how these processes impact the evolution and final fate of such stars. To this end, we explore different hypotheses describing internal angular momentum (AM) transport within a diffusive approximation.

Rotation was recognised as a main characteristic of stars a 100 years ago \citep{eddington1925}. It drives internal AM transport, which is generally assumed to be a consequence of \mbox{(magneto-)hydrodynamical} instabilities in stellar radiative zones and is inherently three-dimensional in nature. In one-dimensional models various simplifications are necessary to describe the effects of rotation \citep[e.g.,][]{zahn92,yoon2006,Maeder2009a}.
Internal AM prescriptions are much more difficult to directly compare with observational data than mass-loss schemes. This is because the main, `directly' measurable impact is the internal rotation profile of the star \citep{aerts2019}. 
Asteroseismology has been a powerful tool to make inferences about the internal structure of stars; however, due to significant turbulence in the photospheres of hot stars, the technique remains challenging for massive stars and, in particular, VMSs \citep[e.g.,][]{briquet2011,bowman2020a,bowman2020b}.
Several recent approaches aim at calibrating the efficiency of internal AM transport by requiring that by the end of the evolution the spin rates of compact objects are recovered \citep[e.g.,][]{suijs2008,kissin2018,fuller2019,ma2019,takahashi2021}. Complicating these latter approaches is that effects of internal AM transport properties (such as the necessary amount of core-envelope coupling) and AM loss by stellar winds are intricately connected. Constraints on internal AM transport in high-mass stars can only be established if the mass-loss rates are precisely determined. 

Pioneering works established that the winds of hot, massive stars are accelerated by radiation pressure on spectral lines \citep{lucy1970,cak1975,pauldrach1986}. Both massive stars and VMSs in the local Universe are subject to significant mass loss by radiative-line driving over the course of their evolution from the Zero Age Main Sequence (ZAMS) to core collapse \citep[e.g.,][]{puls2008,2022ARA&A..60..203V}. 
%
Mass-loss rates of hot stars have been uncertain. Observational studies of typical massive stars (i.e., up to $\sim$100~M$_\odot$) generally identify some deviations from theoretically determined rates, in part related to the treatment of inhomogeneities (clumps) in the trans-sonic outflows \citep[e.g.,][]{puls2015,hawcroft2021,backs2024}. 

Previously, large (order of magnitude) uncertainties on stellar mass-loss rates were a prohibitive issue to study the evolution of VMSs from the main-sequence to core collapse and several studies relied on the computation of pure helium cores to mitigate such issues \citep[e.g.,][]{heger2002,waldman2008}.
In recent years, new theoretical mass-loss prescriptions have been developed \citep{krticka2017,krticka2018,krticka2024,vink2018,sundqvist2019,gormaz2019,gormaz2021,gormaz2022,sander2020a,sander2020b,bjorklund2021} and improved atmospheric prescriptions and high-quality ultraviolet and optical spectroscopy have allowed to obtain wind-structure-corrected mass-loss rates, also for VMSs \citep{2012A&A...544A..67B,bestenlehner2020,hawcroft2021}.
In this study, we rely on two mass-loss prescriptions. During the early main-sequence evolution, when the surface helium abundance is close to its initial value, we adopt the empirical calibration of \citet{brands2022} to stars in R136. When the surface becomes helium rich, we apply the new \cite{sander2020b} rates, developed for Wolf-Rayet type stars.
In the former case, the typical uncertainty between current theoretical works and observations is a factor of 2--3. While this is remarkably good, even such small changes in the mass-loss rates can have a large impact on the evolution of high-mass stars \citep{keszthelyi2017b}. In the latter case, the uncertainty is of the order of a magnitude \citep[e.g.,][]{hainich2014,moriya2022}.
Therefore we also perform some additional tests with different rates.

We test our fiducial mass-loss rates and compare our models with different schemes of internal AM transport to observed stars in R136. This way we can mitigate, as much as possible, the complicating factor that both mass loss and internal AM transport affect the rotation of the star, which will determine the rotation of the remnant compact object. 
In all fairness, even when employing a tailored mass-loss prescription and exploring different strengths of internal AM transport, evolutionary modelling of VMSs remains challenging. This is largely due to these objects approaching the Eddington limit, with convection, magnetism, and envelope inflation also contributing to modelling uncertainties \citep[e.g.,][]{ishii1999,sanyal2015,sanyal2017}. 
This paper is structured as follows.  In \Cref{methods}, we present the \textsc{mesa} software instrument and outline the input parameters of our models, including various internal angular momentum transport prescriptions. In \Cref{res}, we detail the evolutionary model predictions from early to late stages, and in \Cref{obscomp} confront our models with observations of the three WNh stars in the R136 cluster. In \Cref{disc}, we put our findings in a broader context and discuss the implications of our results. Our conclusions are summarised in \Cref{conc}.

%
%
%
%
%
%
\section{Methods}\label{methods}

For our numerical model calculations we use Modules for Experiments in Stellar Astrophysics r22.11.01 \citep[\textsc{mesa};][]{paxton2011,paxton2013,paxton2015,paxton2018,paxton2019} and \textsc{mesa} Software Development Kit (\textsc{SDK}) version 22.10.1 \citep[][]{sdk}. We carry out test computations on the Dutch supercomputer Snellius\footnote{\url{https://servicedesk.surfsara.nl/wiki/display/WIKI/Snellius} } and final computations on the calculation server of the Japanese supercomputer XC50 \footnote{\url{https://https://cfca.nao.ac.jp/} }. 
\textsc{mesa} is a robust, open-source software instrument, which has been well-tested to model massive stars \citep{paxton2013,jones2015,martins2013b,keszthelyi2017b}. Several new investigations also use \textsc{mesa} to study models of VMSs \citep[e.g.,][]{vigna2019,graefener2021,sabhahit2022,higgins2022}. 
Our input files (including the \texttt{inlist} files and \texttt{run\_star\_extras}), as well as the computed models, can be downloaded from Zenodo.  
The \textsc{mesa} microphysics is summarised in Appendix \ref{sec:app_mesamicro}. Here we describe the main physical ingredients in our models. 

\subsection{Chemical composition}

The initial metallicity in the LMC, and more locally within the R136 cluster, is a topic of active investigations \citep[e.g.,][]{lah2024}. We follow \citet{dopita2019}, who combine the results of several different methods, including atmospheric modelling of hot stars, studies of supernova remnants and H~\textsc{ii} regions. For the initial elemental abundances of hydrogen, helium, and metals these are: $X_{\rm ini}= $ 0.73685, $Y_{\rm ini} = $ 0.25671, and $Z_{\rm ini}= $~0.00644 (see their table 5). 
The adopted initial CNO abundances in mass fraction are $X_{\rm C} = $9.25898 $\cdot$10$^{-4}$, $X_{\rm N} =$1.45717~$\cdot$10$^{-4}$, and $X_{\rm O} =$2.96143~$\cdot$10$^{-3}$, and all other metals follow their tabulated values. Thus, we do not simply scale the solar abundance pattern by the metallicity ratio $Z_{\rm LMC} / Z_{\odot} \sim 0.5$.
Isotopic ratios are adopted from \cite{lodders2003}. The same metallicity is used for the opacity calculations as for the chemical evolution of the models (unlike, e.g., \citealt{brott2011} and \citealt{kohler2015}).
%
%

\subsection{Convection, overshooting}

We assume a mixing length parameter $\alpha_{\rm MLT}\,=\,2.0$ (typically considered in the range of 1.5--2.0, e.g., \citealt{joyce2023}) and the ML1 scheme of convection in \textsc{mesa}, following \cite{bohm1958}. 
We use exponential overshooting, following the works of \cite{herwig2000,paxton2013}. We set the corresponding parameters such that the convective core is extended by approximately 20\% of the local pressure scale height. In convectively unstable (non-burning) shells, we apply over- and undershooting to avoid stiff boundaries. We do not use the semiconvective and thermohaline mixing options in \textsc{mesa}. Recently, \cite{blouin2024} found in detailed 3D simulations that semiconvective layers could be completely erased by the homogenising effect of overshooting. Therefore, while formally the Ledoux criterion is applied to calculate convective boundaries, in practice, the stability against convection is reduced to the Schwarzschild criterion. \cite{anders2022} found that the two criteria should be equivalent over evolutionary timescales, as multidimensional simulations provide new insights to the problem \citep[e.g.,][]{andrassy2024,leidi2024}.

\subsection{Rotational mixing}

Chemical element transport driven by rotational mixing is assumed to be a diffusive process, following the seminal works of \cite{pin1989} and \cite{zahn92}. In this approach, two scaling factors are commonly used. To account for the observed surface lithium depletion in solar-type stars and surface nitrogen enrichment in massive stars, the overall efficiency of chemical element transport compared to that of angular momentum transport is scaled by the free parameter $f_{\rm c}$. The mixing efficiency also strongly depends on composition gradients inside the star. To mitigate the effects of composition gradients, a factor $f_{\mu}$ is introduced to multiply $\nabla_{\mu}$ \citep[see also][]{maeder1998}. 
Although these parameters are uncalibrated and thus completely uncertain in the case of VMSs, we adopt $f_{\rm c} = 0.05$ and $f_\mu = 0.1$ \citep{yoon2006,brott2011,paxton2013}. The increase in $f_{\rm c}$ (compared to the value 0.033 used by \citealt{heger2000}) is adopted given the findings of \cite{markova2018}, who determined that the physical characteristics -- in particular, the surface abundances -- of massive O-type stars are better reproduced with a higher mixing efficiency (see also \citealt{martins2017}). The usual hydrodynamical mixing processes are assumed to operate: Eddington-Sweet circulation, shear mixing, and Goldreich-Schubert-Fricke (GSF) instability (\citealt{eddington1925,sweet1950,goldreich1967,fricke1968}, see also, e.g., \citealt{zahn92,talon1997,maeder1997}).

\subsection{Angular momentum transport\label{sec:angularscheme}}

Rotational core-envelope coupling and thus the internal angular momentum transport remains highly uncertain in VMSs. During the main-sequence evolution, the large core is convectively unstable. In these regions, AM transport is very efficient, leading to solid-body rotation. In the radiative stellar envelope, \mbox{(magneto-)rotational} instabilities could transport angular momentum; however, the exact physical scenario remains unclear.
We use \textsc{mesa}'s diffusive approximation of AM transport, which is based on the works of \cite{endal1978} and \cite{pin1989}, and is described via:
\begin{equation}\label{eq:mam2}
 \frac{ \partial \Omega  }{ \partial t}    =  \frac{\partial}{\partial m} \left[(4 \pi r^2 \rho)^2  \,  D_{\rm AM} \,  \frac{\partial \Omega}{\partial m}  \right] \, ,
\end{equation}
where $D_{\rm AM}$ is the total diffusion coefficient responsible for AM transport, and $r$, $\rho$, $m$, and $t$ are the radius, density, enclosed mass, and time, respectively. In convective regions, $D_{\rm AM}$ is calculated from the mixing length theory. In radiative stellar layers, $D_{\rm AM}$ is constructed as a sum of individual diffusion coefficients accounting for various rotationally-induced instabilities, though this approach is likely far too simplistic \citep[see discussion by][and references therein]{keszthelyi2022}. We consider three assumptions, describing moderate, strong, and perfect coupling. These are:

\begin{itemize}
    \item {\em Moderate coupling -- HY}: only hydrodynamical instabilities in stellar radiative zones (dominated by Eddington-Sweet currents) are considered. The transport is relatively efficient, but the angular velocity profile can decline more rapidly toward the stellar surface. 

    \item {\em Strong coupling -- TS}: angular momentum transport in radiative stellar zones conform the Tayler-Spruit dynamo \citep{tayler73,spruit2002}. This leads to near solid-body rotation on the main sequence.  

    \item {\em Perfect coupling -- SB}: Exact solid-body rotation ($\mathrm{d}\Omega/\mathrm{d}r = 0$) is enforced by a high diffusion coefficient ($D_{\rm AM} =$10$^{16}$~\,cm~\,s$^{-2}$) throughout the entire star.    

\end{itemize}
%

\subsection{Angular momentum loss}\label{sec:amloss}

In the diffusive scheme, angular momentum transport in the stellar interiors leads to replenishing the surface reservoir. The speed of this process depends on the diffusivity, which we assume with three different scenarios as explained above.  
The upper layers lose AM because stellar winds remove mass. In the stellar model, the mass removed is assumed to take away the associated specific AM in the near surface layers. This procedure leads to a more rapid decline of the surface rotational velocity compared to approaches where the wind induced AM loss is subtracted from the total AM reservoir of the star.

\subsection{Mass loss}\label{sec:mdot}

In accordance with most previous studies, we rely on the surface helium abundance to determine whether the stellar model would have optically thin or thick winds. This is a large over-simplification of the actual physical nature of winds; however, currently no model exists to describe and implement the transition in one-dimensional stellar evolution codes (though, for alternative approaches see \citealt{graefener2021} and \citealt{sabhahit2022}). Optically thin winds allow for directly inferring the photospheric conditions, since corresponding lines are formed below the sonic point. On the other hand, diagnostics of optically thick winds only provide direct constraints on the wind properties.
In the following, we apply an optically-thin wind prescription when $T_{\rm eff} > 10$\,kK and surface hydrogen mass fraction $X > 0.4$, and switch to an optically-thick wind prescription when $X < 0.4$, with linear interpolation between $X=$~0.44 and 0.36. We assume that these cases correspond to optically thin and thick winds, respectively. For stars with $T_{\rm eff} < 10$~kK the wind driving is still not fully understood. We adopt the empirical \cite{dej1988} rates if the models evolve into this domain. If the effective temperature drops below 4~kK, we switch to the empirical Red Supergiant rates determined by \cite{vanloon2005}. 

\subsubsection{Optically-thin wind}

A convenient way to represent the mass-loss rate of line-driven winds is to employ the wind momentum luminosity relation, WLR \citep{kudritzki1995,puls1996,kudritzki2000}. The key concept is that the wind momentum ($\dot{M}$~$v_{\infty}$, the product of mass-loss rate and terminal wind velocity) multiplied by the square root of the stellar radius is solely related to the stellar luminosity, i.e., 
\begin{equation}\label{eq:wlr1}
    D_{\rm mom} = \dot{M} v_\infty (R_\star/R_\odot)^{1/2} \propto L_\star^{1/\alpha'}~, 
\end{equation}
with $D_\mathrm{mom}$ the modified wind momentum, $R_\star$ and $L_\star$ the radius and luminosity of the star, and $\alpha'$ the difference of the so-called force multiplier parameters $\alpha$ and $\delta$ ($\alpha' = \alpha - \delta$, see, e.g., \citealt{pauldrach1986}). Underlying this relation is the assumption that for O-stars $\alpha' \approx 2/3$ (conform observations and theory; e.g., \citealt{puls2000}). In this case the mass-dependence of the `exact' expression drops out, and $D_{\rm mom}$ depends solely on luminosity. 
The scaling relation of \cref{eq:wlr1} is usually cast in logarithmic form: 
\begin{equation}\label{eq:wlr2}
    \log D_{\rm mom} = x \log L_\star + \log D_0~, 
\end{equation}
using the 10-based logarithm of \cref{eq:wlr1}, $x = 1/\alpha'$, and $\log D_0$ the offset to the linear fit.
We adopt $x = 2.00$ (implying $\alpha' = 0.5$, slightly different than the value 2/3 discussed above) and $D_{0} = 17.05$ from \citet{brands2022}, which have been derived for both massive stars and VMSs in the R136 cluster (see \citealt{keszthelyi2017b} for the first implementation of the WLR formalism in stellar evolution models and the impact of different values). Particularly for the most massive WNh stars (R136a1, R136a2, and R136a3), these parameters provide a reasonably good match to their observationally-inferred mass-loss rates with standard deviations of around a factor of two (see Figure 13 of \citealt{brands2022}). For this reason, our fiducial optically-thin mass-loss rates are given by Equation~\ref{eq:wlr2} with $x$ and $D_{0}$ as specified above.

\subsubsection{Optically-thick wind}

For the regime $T_{\rm eff} > 10$\,kK and $X < 0.4$, we apply the Wolf-Rayet (WR) type mass-loss rates recently determined by \cite{sander2020b}. 
Although these rates were derived for completely hydrogen-free atmospheres, our goal is simply to simulate a change in wind behaviour as the stellar surface becomes helium rich in our models.
In particular, the application of these rates will result in increased mass-loss rates compared to the ones obtained from the wind momentum luminosity relation specified above. The transition produces a relatively steep increase in mass-loss rates over time. We do not account for episodic mass loss, such as those that could characterise Luminous Blue Variables (LBVs), for example. 

\subsubsection{Rotational enhancement}

Rapid stellar rotation may increase the mass-loss rate \citep[recently, e.g.,][]{gagnier2019a,gagnier2019b,hastings2023}. We follow the implementation of \cite{keszthelyi2020} to include the rotational enhancement of theoretical mass-loss rates as described by \cite{maeder2000} (their Eq. 4.30): 
\begin{equation}
    \frac{\dot{M}}{\dot{M}(\Omega=0)} = \left( \frac{1 - \Gamma_e}{1 - \left( \frac{4}{9} \frac{v_{\rm rot}}{v_{\rm crit} } \right)^2 - \Gamma_e} \right)^{\frac{1}{\alpha'} -1 }
\end{equation}
where $\Gamma_e$ is the Eddington parameter for electron scattering and $\alpha'$ is the difference of force multiplier parameters introduced in our Eq.~\ref{eq:wlr1}. $v_{\rm rot}$ is the surface rotational velocity, and $v_{\rm crit}$ is its critical value.

\subsubsection{Other tests}

In addition to our "standard wind scheme", using the WLR for optically thin winds with the \cite{brands2022} values and the Wolf-Rayet-type rates of \cite{sander2020b} for optically thick winds, we explore i) scaling the mass-loss rates (both for optically thin and thick winds) throughout the entire evolution with a  factor of 0.5, and, ii) instead of switching to the \cite{sander2020b} rates, switching to the \cite{nugis2002} rates for optically thick winds. These additional tests are performed for one branch of models, namely, in the TS angular momentum transport scheme with MLT++ (see below) for an initial rotational velocity of 300~km\,s$^{-1}$.

\subsection{Terminal velocities\label{sec:vinf}}

The asymptotic maximum velocities with which hot stars eject gas range from $\sim$1500--3500\,km\,s$^{-1}$ \citep[e.g.,][]{kudritzki2000}. Typical uncertainties in the terminal velocities of the stars in R136 are 200--400\,km\,s$^{-1}$, with sizeable offsets of hundreds of km\,s$^{-1}$ between stars with otherwise similar properties \citep{brands2022}, a behaviour that is typical for comparable data sets \citep[e.g.,][]{kudritzki2000}. For this reason, we employ a prescription that captures the mean trend in \vinf rather than adopting direct measurements. To this end, we use established scaling relations from \cite{puls2000}, that is, the terminal wind velocity is self-consistently calculated in our evolutionary models as a function of the effective surface escape velocity: 
\begin{equation}
    v_\infty = f_{\infty} \cdot v_{\rm esc} = f_{\infty} \sqrt{\frac{2G M_\star}{R_\star} (1 - \Gamma_e)}
\end{equation}
with $G$ the gravitational constant, $M_\star$ the stellar mass, 
and $f_{\infty}$ a parameter depending on effective temperature. For typical O-type stars, its value is adopted in the range of 2.0--3.5 for $T_{\rm eff} > 20$~kK and 1.3 for $T_{\rm eff} < 20$~kK \citep{prinja1998,kudritzki2000,crowther2006,puls2008,markova2008,hawcroft2024}. Since a higher value provides a better match for the WNh stars, we adopt $f_{\infty}=3.5$ for the hot regime.

%
%
\subsection{Eddington limit}
\label{sec:EddingtonLimit}

For a star to be in hydrostatic equilibrium the inward-directed force due to gravity should be balanced by the outward-directed force by gas and radiation pressure\footnote{and in the case of rapid-rotation, by the centrifugal force}. If the force due to radiation pressure, $\mathbf{g_{\rm rad}}$, is so large that it exceeds the force of gravity, $\mathbf{g_{\rm grav}}$,
the star may inflate its envelope to recover equilibrium and/or to initiate an outflow. The stability criterion is formulated as:
\begin{equation}\label{eq:edd}
    \Gamma = \frac{\lvert \mathbf{g_{\rm rad}} \rvert }{\lvert \mathbf{g_{\rm grav}} \rvert} = \frac{L_{\rm rad}}{L_{\rm Edd}} = \frac{\kappa \cdot L_{\rm rad}}{4 \pi c \, G M_\star} < 1~, 
\end{equation}
where $L_{\rm rad}$ is the radiative luminosity of the layer being considered, $c$ is the speed of light, and $\kappa$ is the flux-weighted mean of the total opacity. 
At the photospheric radius, all energy is transported by radiation and $L_{\rm rad} = L_\star$.

In VMSs, the Eddington limit, i.e., $\Gamma = 1$, may be approached in sub-surface layers where the opacity peaks. This could lead to a dramatic inflation of the stellar envelope \citep{sanyal2015,sanyal2017}.
Reaching the Eddington limit may directly or indirectly lead to numerical problems in evolutionary model computations (e.g., increasingly small timesteps due to density inversion in the upper envelope). Two main methods are commonly considered to mitigate these problems:
{\em i)} When the star is in proximity of the Eddington limit, the mass-loss rate may be artificially increased. This reduces the stellar mass and consequently the luminosity, although for VMSs a nearly linear mass-luminosity relation implies that a significant amount of mass may need to be removed before stability is restored.
{\em ii)} When the star reaches the Eddington limit, it can become locally convectively unstable as radiation alone cannot transport the entire flux.
\textsc{mesa}'s MLT++ scheme artificially increases the convective luminosity at the expense of the radiative luminosity \citep{paxton2013}. This leads to reducing the local Eddington parameter. 
A similar restoring effect may be obtained if instead the efficiency of convection is artificially increased (for example, using $\alpha_{\rm MLT} \sim 5$). 
Since these are complex and unresolved questions, we compute models both with and without using MLT++.
Models that use MLT++ evolve with higher effective temperatures than models that do not adopt it. Since inflation is prevented in this case, the models remain more compact. In models with MLT++, we limit the maximum change in $\nabla T$ per time step with a scaling of 0.05 (\texttt{gradT\_excess\_max\_change} in \textsc{mesa}). Models without MLT++ can inflate their envelopes. We further elaborate on the model differences in Section~\ref{sec:ms}.

%
%
\subsection{Initialisation, final stages, and numerics}\label{sec:setup}

\textsc{mesa} contains pre-computed ZAMS models of a few representative metallicities up to 100\,\Msun. Since our models are outside of this mass range, we employed \textsc{mesa}'s option to create a pre-main sequence model without mass loss and rotation until the luminosity is primarily produced via core burning instead of gravitational contraction. Then, we load these pre-main sequence models and allow for the relaxation of the initial model by setting both mass loss and rotation (and other main control options). In our figures, the ZAMS corresponds to the time when the core hydrogen content has decreased by 0.3\% compared to its initial value\footnote{For example, \cite{graefener2021} uses 5\%, which can explain the offset in effective temperature seen between their and our models.}.
We compute models until the central temperature reaches $\log T_{\rm c} = 9.10$, which approximately corresponds to the end of carbon core burning and the onset of neon core ignition. We will refer to this as the carbon Terminal Age Main Sequence (C-TAMS) in the following. For VMSs, this stage is within a few months prior to core collapse, set by the timescale of the subsequent oxygen core burning.

We employ high time and mesh resolutions. Our stellar models typically have order of 1000 zones and timesteps smaller than $10^3$ years on the main sequence. 
It is well-known that rotating evolutionary models of high-mass stars are riddled with convergence problems. In particular, none of the models without MLT++ are able to reach convergence during the post-main sequence evolution. Some models without MLT++ show severe numerical issues on the main sequence. We limit the computations to reasonable model numbers and iterations. As such, in some cases the TAMS is not reached.
Some models with MLT++ and with initial rotational velocities below 300~km\,s$^{-1}$ also fail to converge in their helium core burning stage.
We choose to apply the same numerical setup consistently for all models in our grid and not to make arbitrary tweaks and adjustments on a model-by-model basis to resolve unconverged models. 
The only exception is the initially 630~M$_\odot$ model with MLT++, which is the only model that fails to converge from the pre-main sequence model on the ZAMS. Here, we explored several numerical choices to reach convergence; however, without success. 
We further comment on these issues when necessary but the detailed discussion of this is beyond the scope of the paper.

%
%
\renewcommand{\arraystretch}{1.3}
\begin{table}
    \caption{Summary of the computed models. \label{tab:gridsummary}}
    \begin{tabular}{p{1.9cm} p{6.3cm}}
\hline\hline 
Parameter & Input values/choices \\ \hline
$M_\mathrm{ini} \ [\mathrm{M}_\odot$] & 100, 110, 125, 140, 160, 180, 200, 225, 250, 280, 315, 355, 400, 445, 500, 560, 630, \quad 705, 795        \\
$v_\mathrm{ini}$ [km\,s$^{-1}$] & 100, 200, 300, 400, 500 \\ 
AM scheme & hydrodynamical (HY), Tayler-Spruit (TS), solid-body (SB) \\ 
MLT++ & with, without \\ 
Winds & $T_{\rm eff}>10$\,kK, $X > 0.4$: WLR with \cite{brands2022} fit,
 $T_{\rm eff}>10$\,kK, $X < 0.4$: WR-type winds from \cite{sander2020b},
 4\,kK\,$>T_{\rm eff}<$\,10\,kK: cool winds from \cite{dej1988}, $T_{\rm eff}<4$\,kK: RSG winds from \cite{vanloon2005} \\ 
Wind test & 0.5 overall reduction (for 1 branch) \\
 & $T_{\rm eff}>10$\,kK, $X < 0.4$: WR-type winds from \cite{nugis2002} (for 1 branch)\\ 
\hline
total & 19 x 5 x 3 x 2 + 2 x 19 = 608 models \\
\hline
    \end{tabular}
\end{table}
\subsection{Parameter space of the model grid}

Our model grid covers the initial mass range of \mbox{$M_\mathrm{ini} = 100-800$~M$_\odot$,} with a spacing of $0.05$ in logarithmic scale (from 2.0 to 2.9), and initial spin rates of $100-500$~km~s$^{-1}$, with a spacing of 100~km~s$^{-1}$. Three angular momentum transport schemes are considered (hydrodynamical, Tayler-Spruit, solid-body), as described above. We consider our fiducial mass-loss scheme as described in Section \ref{sec:mdot}. We consider models with and without MLT++. The parameter space of the grid is summarised in \Cref{tab:gridsummary}, with the initial masses rounded to the nearest value in 5~M$_\odot$.

%

%
%
%
%
\begin{figure*}
    \centering
    \includegraphics[width=0.47\textwidth]{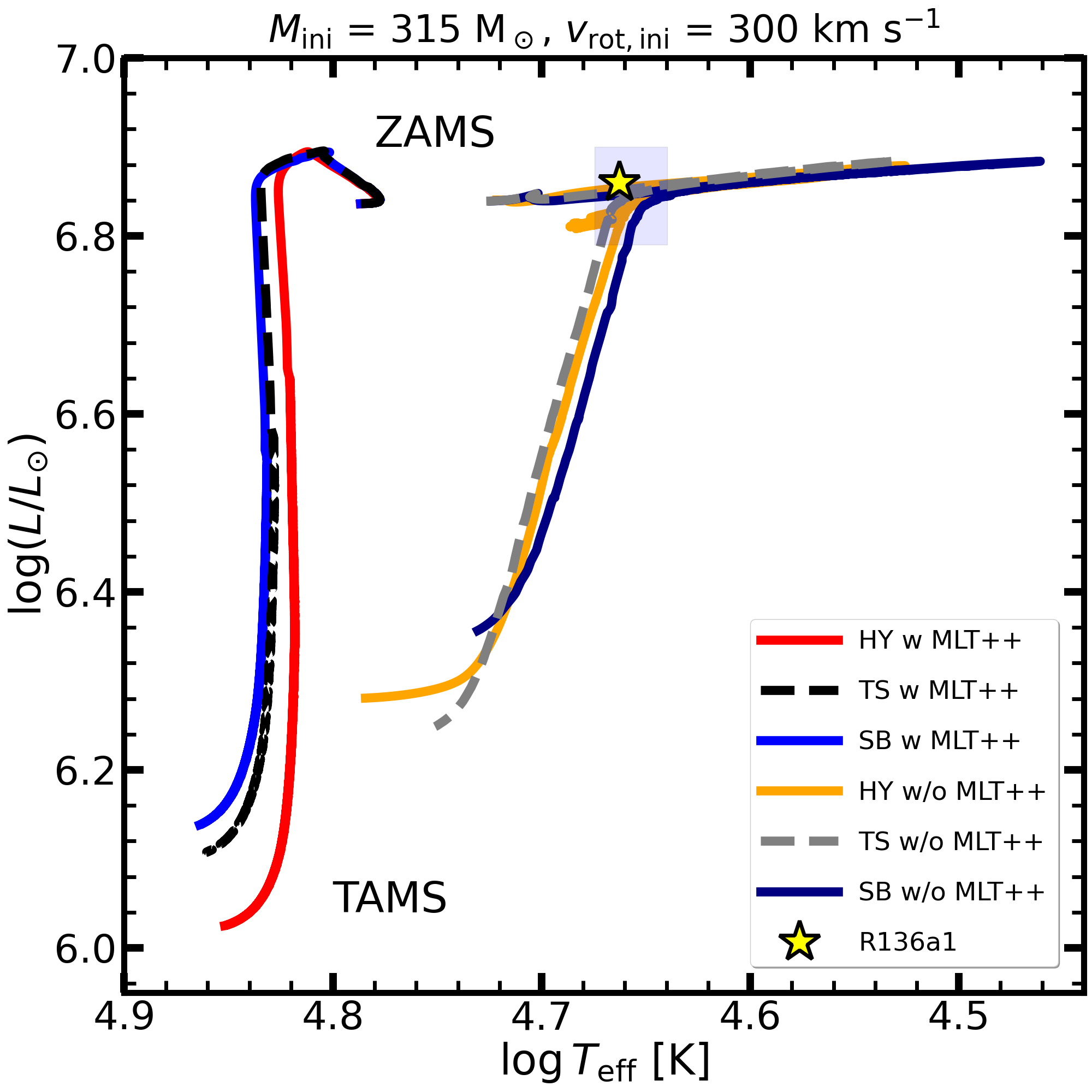} \includegraphics[width=0.47\textwidth]{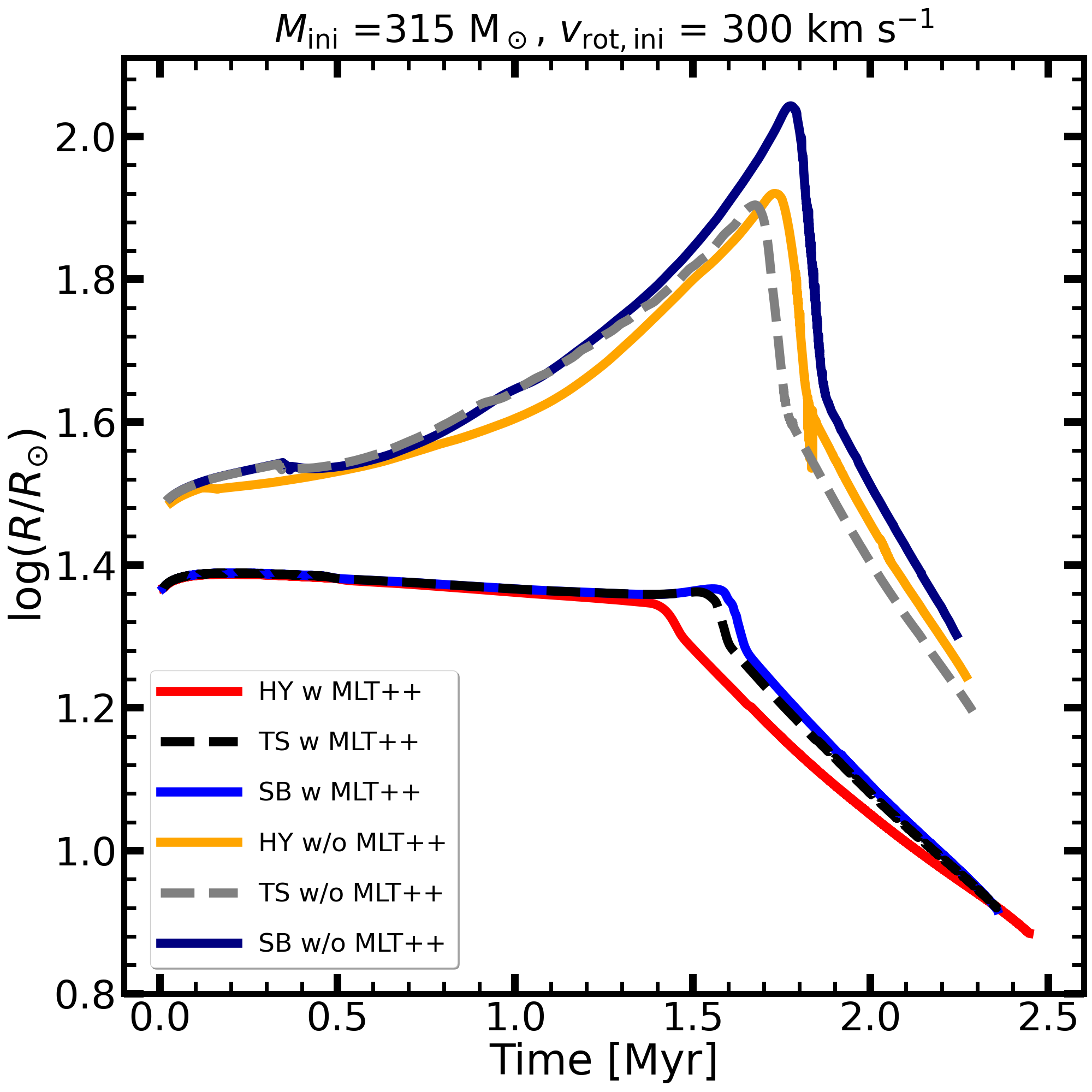}
    \caption{Comparison of evolutionary models with different internal angular momentum coupling schemes (HY: hydrodynamical, SB: solid body, TS: Tayler-Spruit) and with (w) or without (w/o) MLT++ in a typical VMS model with initial mass of 315~M$_\odot$ and initial rotational velocity of 300~km\,s$^{-1}$ during the main-sequence evolution from ZAMS to TAMS. {\em Left}: Hertzsprung-Russell diagram. For reference, the position of R136a1 according to \cite{brands2022} is indicated. The shaded region shows the formal uncertainty of the observations. {\em Right}: stellar radius vs. age. Numerical smoothing is included for visualisation purposes.}
    \label{fig:ms1}
\end{figure*}

%
%
%
%
\section{Results}\label{res}

In this section, we focus on the main properties of our new evolutionary models and present tracks both on the main sequence and beyond. We highlight differences that are caused by assumptions regarding mass loss, AM transport (core-envelope coupling), and the MLT++ schemes, in addition to the impact of initial mass and initial rotation. 


\subsection{Main-sequence characteristics of our models}\label{sec:ms}

\subsubsection{Impact of angular momentum schemes and MLT++ on the main sequence}\label{sec:hrd1}

In Figure \ref{fig:ms1}, we compare our evolutionary models for an initially 315\,M$_{\odot}$ star computed with different AM schemes and with/without MLT++ in the Hertzsprung-Russell Diagram (HRD).
The models with MLT++ evolve with significantly higher effective temperatures than models without MLT++.
%
Models with MLT++ remain more compact and contract throughout the entire main-sequence evolution. In contrast, models without MLT++ increase in stellar radius by inflating their envelopes. Contraction only begins at around 1.5 Myr, when the enhanced mass-loss rates due to optically thick winds drive the evolution toward much lower luminosities. In the HRD, this is reflected in an initially redward evolution, followed by a down-ward trajectory (which we discuss in the next section).
The various coupling schemes yield almost identical tracks, illustrating that, without asteroseismic measurements, the strength of core-envelope coupling is difficult to diagnose on the main sequence.
For reference, we indicated the position of R136a1 in the HRD, which makes it clear that main-sequence models with MLT++ are incompatible with it. Only models without MLT++ evolve to the effective temperature range that is consistent with the measured value of R136a1. 
However, our models without MLT++ do not successfully reach the C-TAMS, only models with MLT++ do. Since we need to rely on models with MLT++ to discuss the post-main sequence evolution, it is logical to ask to what extent the models without MLT++ would differ on the post-main sequence if they could be successfully computed. We find that the TAMS properties of models with and without MLT++ are broadly consistent, for example, in luminosity, mass, and rotation, with the notable exception of a $\sim$15~kK offset in effective temperature (for this specific model), which also implies different radii and surface gravities. For this reason, we argue that the findings regarding the post-main sequence evolution are applicable to the WNh stars in R136. That is, one might anticipate that models without MLT++ may have, at least qualitatively, a rather similar post-main sequence evolution as models with MLT++.

%
%
%
%
\begin{figure*}
    \centering    
    \includegraphics[width=0.9\textwidth]{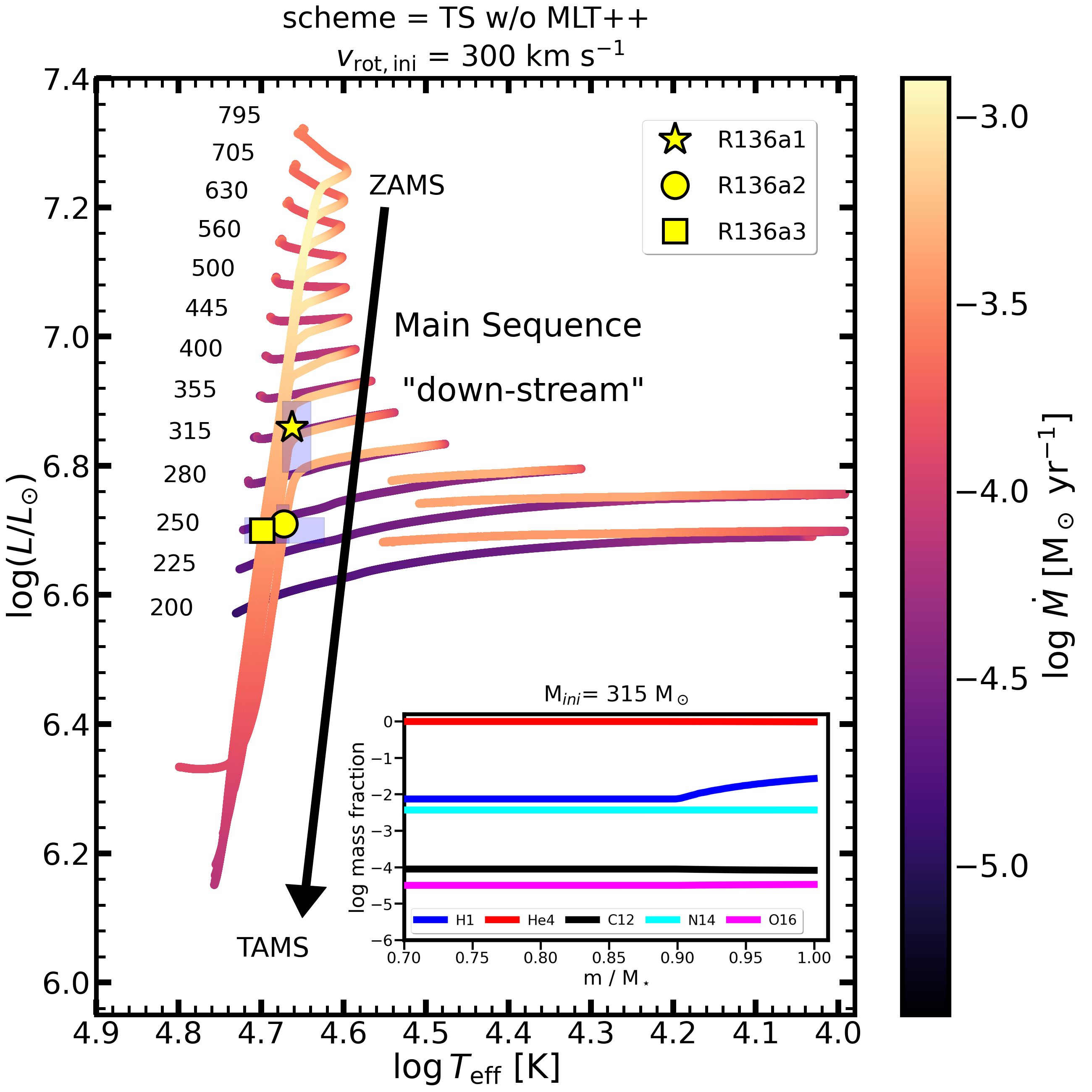}
    \caption{HRD of the main-sequence evolution of the most massive models in our grid within the TS scheme without MLT++. The initial rotational velocity is 300~km\,s$^{-1}$. Initial masses in solar units are indicated next to the starting point of the models. The colour-coding shows the logarithmic mass-loss rate. Stars with $M_\mathrm{ini} > 250$~\Msun follow a `down-stream' of rapidly decreasing luminosity evolution and become helium-rich objects by the TAMS. A mass fraction profile of the outer 30\% of the mass near the TAMS is shown in the inset, illustrating that they become helium-rich stars. The three most massive stars in R136 are indicated with symbols and the filled areas show uncertainties in their HRD positions according to \cite{brands2022}. Numerical smoothing is applied for visualisation purposes.}
    \label{fig:hrd_wrtr1}
\end{figure*}
\subsubsection{Producing helium-rich main-sequence stars from single-star evolution}\label{sec:wrtr}

We select the Tayler-Spruit coupling scheme and initial rotational velocity of 300~km\,s$^{-1}$ as our fiducial model set. In terms of AM scheme these models are bracketed by the less (HY) and more (SB) efficient schemes. A rotation speed of 300\,km\,s$^{-1}$ is the median value in our grid. Our fiducial models without MLT++ during their main-sequence evolution are presented in Figure~\ref{fig:hrd_wrtr1}.
Towards the end of main-sequence evolution, the surfaces of hot stars can become strongly enriched in helium, hereafter {\em helium-rich stars}\footnote{The term `helium stars' often refers to objects from low-mass progenitors or resulting from binary evolution; however high-mass progenitors are sometimes implied as well \citep[e.g.,][]{mcclelland2016,woosley2019,2020ApJ...904...56G}. }, if they undergo strong mass loss. 
Stars initially more massive than 250\,\Msun do not evolve to temperatures lower than 30\,kK. Instead, they converge to a similar solution in their $(L_\star,T_{\rm eff})$ behaviour, in a down-ward evolution in the HRD, before reaching the TAMS with high effective temperatures and strongly reduced luminosities compared to their ZAMS values. This behaviour can be directly linked to their mass-loss properties (colour-coded in the Figure~\ref{fig:hrd_wrtr1}). Once the surface hydrogen abundance drops below $X = 0.4$, mass loss is given by the WR-type prescription by \citet{sander2020b}. These rates are much higher than those in the prior phase of evolution and quickly cause a decrease in luminosity as $L \propto M^{\alpha}$, with $\alpha \sim 1$ in the mass range considered. As stars become much less luminous during this `down-stream' evolutionary trajectory, mass-loss decreases again because the predicted rates also scale with luminosity.

This evolutionary trajectory has been identified in previous works (referred to as `vertical evolution') and indeed been explicitly linked with an increased mass-loss rate, which triggers it (\citealt{higgins2022,higgins2023}, \citealt{sabhahit2022,sabhahit2023,sabhahit2025a}). In this work, we prefer to call this trajectory down-stream for two reasons. First, it is not completely vertical; there is a sizeable increase in effective temperature. Second, it also helps express that these models converge to a similar solution in the HRD, regardless of initial mass. 
The main differences in our work are the consideration of various initial rotation rates, the use of different mass-loss rates, and the different condition for the onset of optically thick winds. Namely, we use the surface helium abundance, whereas the above mentioned works use the \cite{vink2012} transition, which depends on the optical depth of the sonic point. Despite these quantitive differences, we find a generally good agreement regarding the predicted evolutionary paths. 

Given these considerations, our prediction is that all stars at LMC metallicity in excess of initially 250~M$_{\odot}$ should evolve towards helium-rich stars already on the main sequence. This result is robust against the adopted initial rotation rate and rotational coupling scheme. 

As we discuss later, two out of three WNh stars, R136a2, and R136a3, might be in a down-stream evolution, even though their mass-loss rates are lower than expected from the Wolf-Rayet type wind prescription of \cite{sander2020b}. In our scenario, the Wolf-Rayet type rates are needed to initiate the steep drop in mass and luminosity.
For initial rotational velocities higher than 300\,km\,s$^{-1}$, the stars might experience quasi-chemically homogeneous evolution. This too may result in main-sequence stars with a strongly enhanced surface helium abundance \citep[e.g.,][]{martins2013}. As such, from single-star evolution scenarios two channels may be identified to potentially lead to producing compact, helium-rich stars \citep{yoon2005,woosley2006,meynet2007,georgy2009}. According to \citet{shenar2020} and \citet{pauli2022}, in high-metallicity environments, the majority of such objects may be produced via binary evolution, though low-metallicity environments might favour the single-star scenario. In Section \ref{sec:bin}, we briefly address the question of binarity.

%
%
%
\begin{figure}
    \centering
    \includegraphics[width=0.38\textwidth]{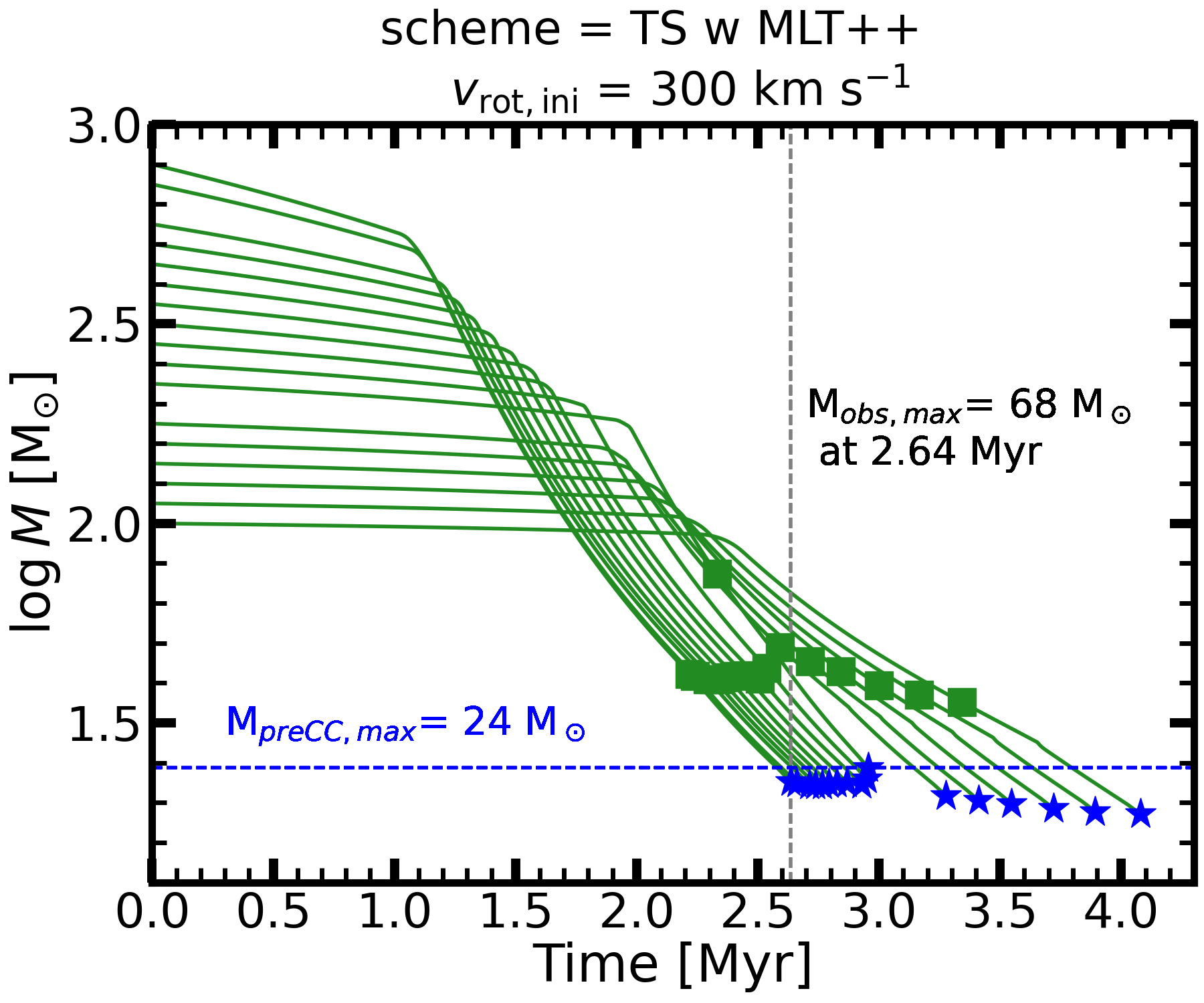}    \includegraphics[width=0.38\textwidth]{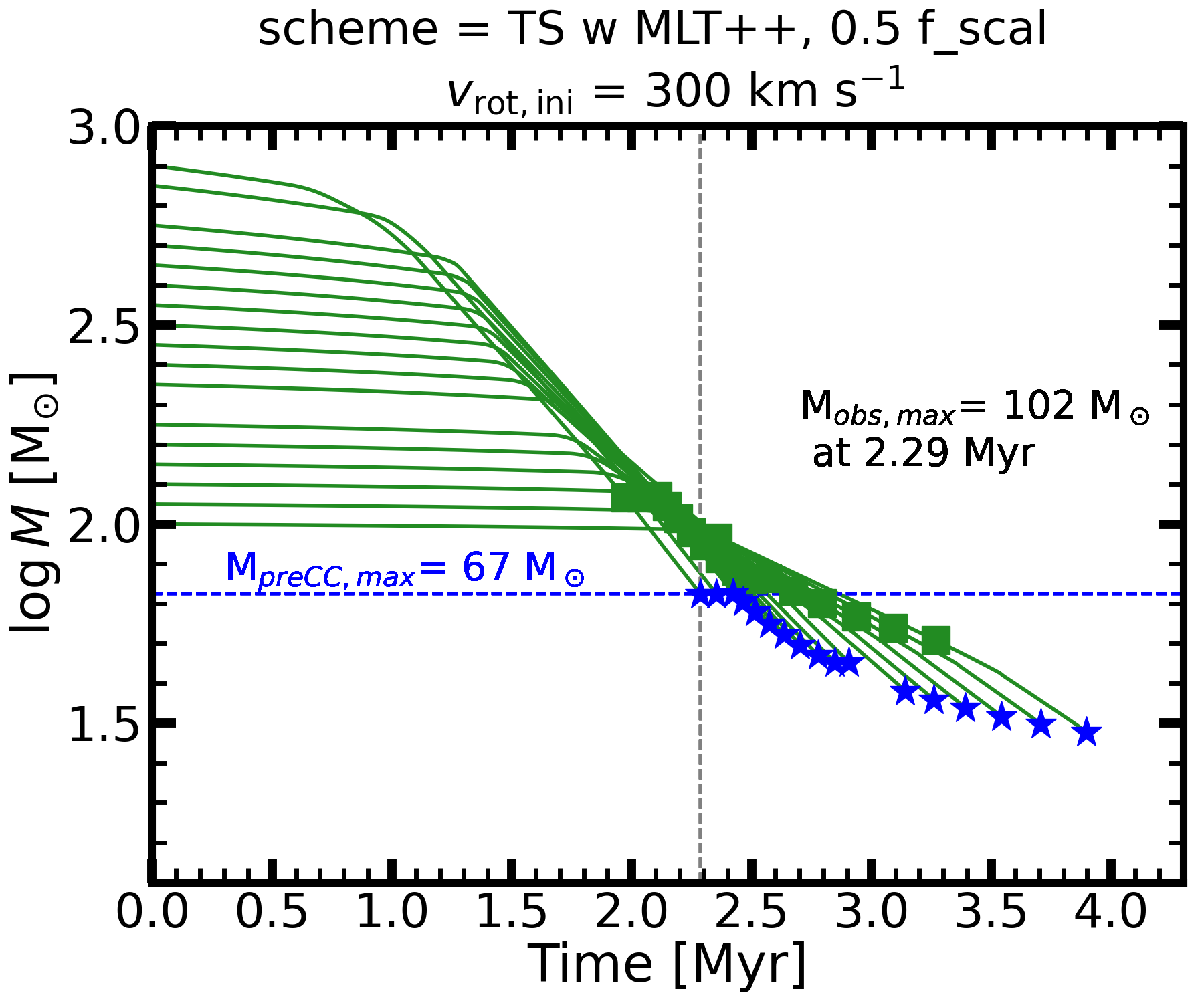}
    \includegraphics[width=0.38\textwidth]{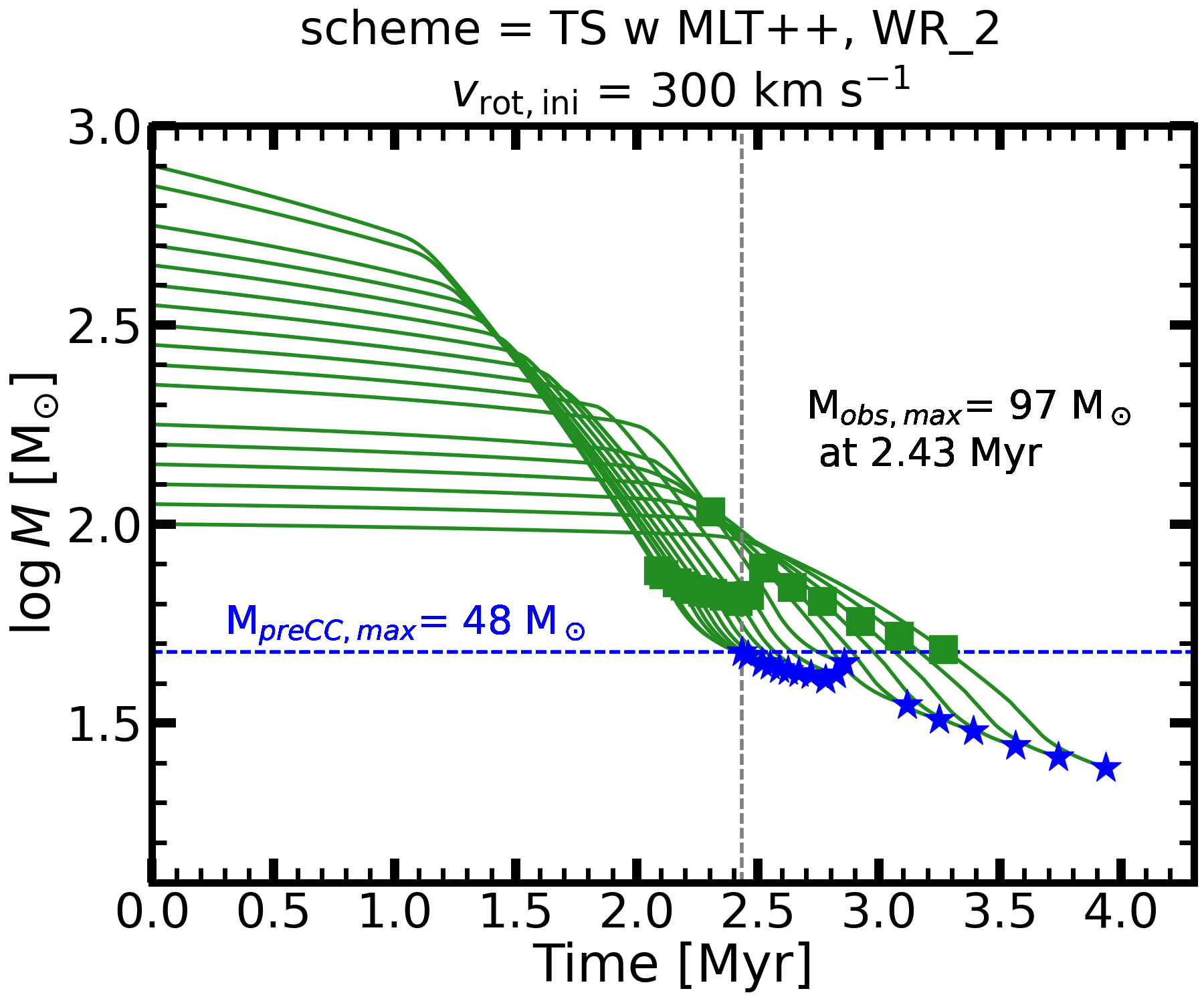}
    \caption{Evolution of stellar mass as a function of time. Models within the Tayler-Spruit coupling scheme with MLT++ are shown for an initial rotational velocity of 300~km\,s$^{-1}$. The grey vertical line indicates the time when the first supernova takes place. $M_{\rm obs, max}$ indicates the maximum stellar mass observable at that time. $M_{\rm preCC, max}$ is the highest pre-core collapse mass by the end of the evolution of all models. Green squares mark the TAMS and blue stars denote successfully reaching the C-TAMS phase. \textit{Top panel}: our fiducial models. \textit{Middle panel}: Reduced mass-loss rates over the entire evolution by a factor of two. Note that this allows for producing supernova progenitors with more than double the mass. \textit{Lower panel}: switching to the \cite{nugis2002} rates instead of the \cite{sander2020b} rates. 
    \label{fig:massevol}}
\end{figure}
\subsubsection{Mass evolution on the main sequence}\label{sec:masstime}

The evolution of stellar mass over time in the fiducial models\footnote{Here, the $M_{\rm ini}=630$~M$_\odot$ model did not converge from the ZAMS model, thus it is not included. Additionally, we omit the $M_{\rm ini}=200$~M$_\odot$ model from the figure because it produces a qualitatively different evolutionary trajectory than the other models in the grid.} is presented in Figure~\ref{fig:massevol}. We show models with MLT++ to best connect to a discussion of post-main sequence mass evolution (in \cref{sec:postms}), where the assumption of MLT++ in \textsc{mesa} is unavoidable for computing models up to the supernova progenitor stage. As demonstrated in Section~\ref{sec:hrd1}, the effective temperature is affected by this choice, however, the mass evolution is only weakly impacted. In particular, for models with M$_{\rm ini}$~$> 250$~M$_\odot$ the TAMS masses are practically the same with and without MLT++. Models without MLT++ in the range of M$_{\rm ini} =$ 100 to 200\,M$_\odot$ tend to lose somewhat less mass in their main-sequence phase than corresponding ones with MLT++, because they experience a more extended redward evolution during which the mass-loss rates are lower. The switch in mass-loss treatment once the surface hydrogen mass fraction becomes low ($X < 0.4$) can be directly identified as a kink or discontinuity in $\dot{M}$, with a much shallower decrease in mass until around 1.2--2.4\,Myr, depending on initial mass. 
For example, in our fiducial scheme (upper panel of Figure~\ref{fig:massevol}), the model with  $M_{\rm ini} \sim 795$\,M$_{\odot}$ loses about $300\,M_{\odot}$ in the first 1.2 Myr. Then, this models loses another $300\,M_{\odot}$ within only 0.2 Myr once the optically-thick WR-type winds kick in. The TAMS is reached at an age of about 2.2 Myr, with a mass of only $\sim 40\,M_{\odot}$.

As initially more massive models have stronger early main-sequence mass loss ($\dot{M} \propto L_\star^{2}$ in our WLR approach) and shorter nuclear timescales to produce surface helium enrichment via rotational mixing, the switch to higher, WR-type mass-loss rates occurs sooner than for stars with lower initial mass. For example, a steeper decline in stellar mass occurs at 1.2\,Myr for $M_{\rm ini} = 795$\,M$_{\odot}$, but only at $\sim$2.5\,Myr for $M_{\rm ini} = 100$\,M$_{\odot}$. 
As a consequence, at some moment during the main-sequence evolution an initially higher mass star will have a lower current mass ($M_{\rm curr}$) than an initially lower mass star (see also, e.g., \citealt{kohler2015,higgins2022}). 
For the tracks shown in \Cref{fig:massevol} such a `$M_{\rm ini}$--$M_{\rm curr}$ turnover' first happens at about 1.4\,Myr. At that time, several of the models have a current mass of $\sim$200--300\,M$_{\odot}$ from initially 200--800\,M$_{\odot}$ progenitors. Though these models have about the same current mass, other properties (e.g., core mass, rotation, and surface abundances) may differ more appreciably, thus they are not physically indistinguishable. 
The occurrence of $M_{\rm ini}$-$M_{\rm curr}$ turnover is robust against varying the initial rotation or coupling schemes, though the quantitative details do depend on such properties (see Appendix \ref{fig:massevol_app}).

To gauge the impact of adopted mass-loss rates, we also show results with initially the same setup but, first, decreasing the overall mass-loss rates by a factor two (middle panel), and second, keeping our fiducial mass-loss rates for the optically-thin winds but adopting the \cite{nugis2002} rates instead of the \cite{sander2020b} rates when switching to optically-thick WR-type winds (lower panel). In these cases, the C-TAMS masses are significantly higher than in the fiducial case.

%
%
%
%
\begin{figure*}
    \centering
    \includegraphics[width=0.33\textwidth]{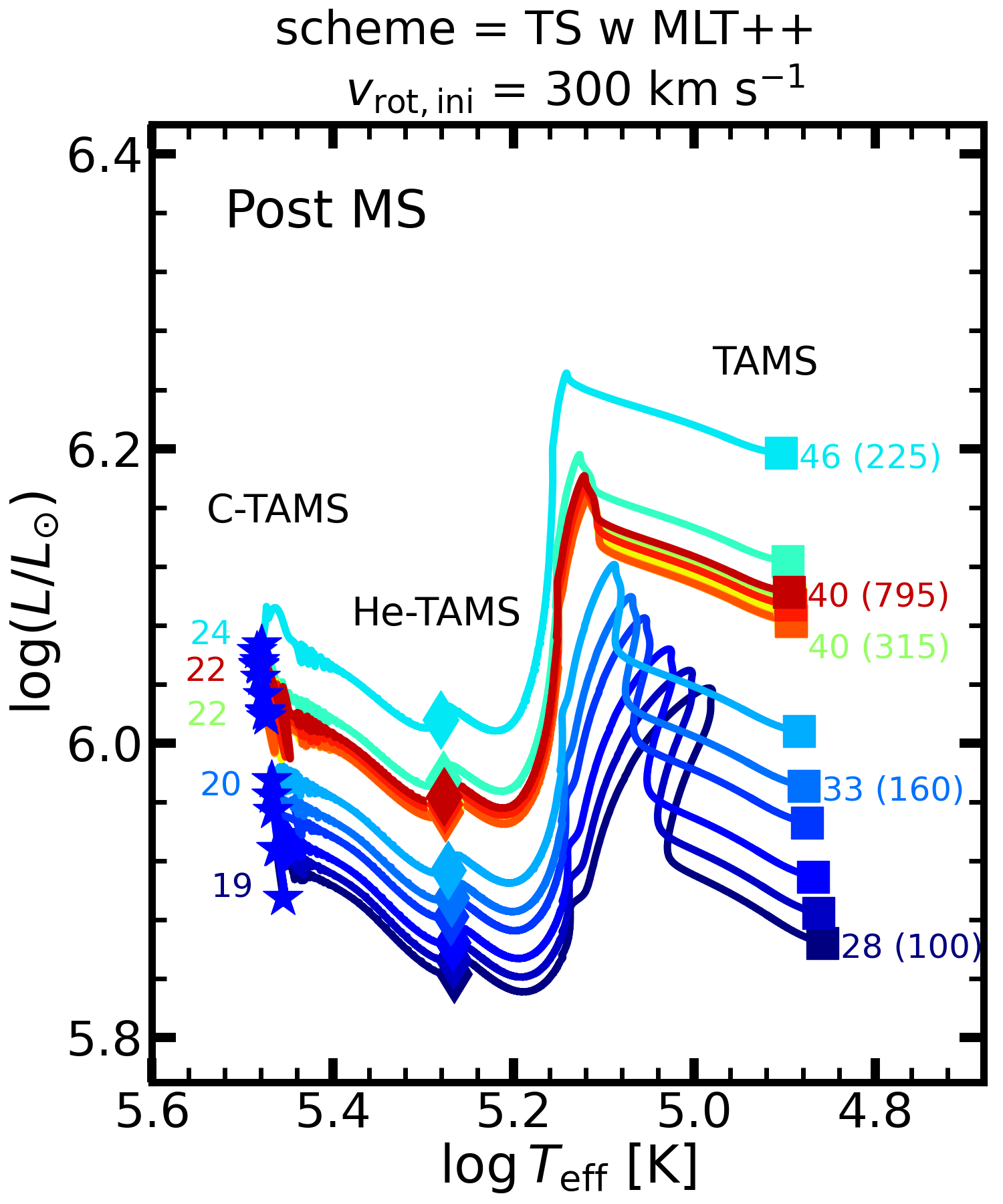}\includegraphics[width=0.33\textwidth]{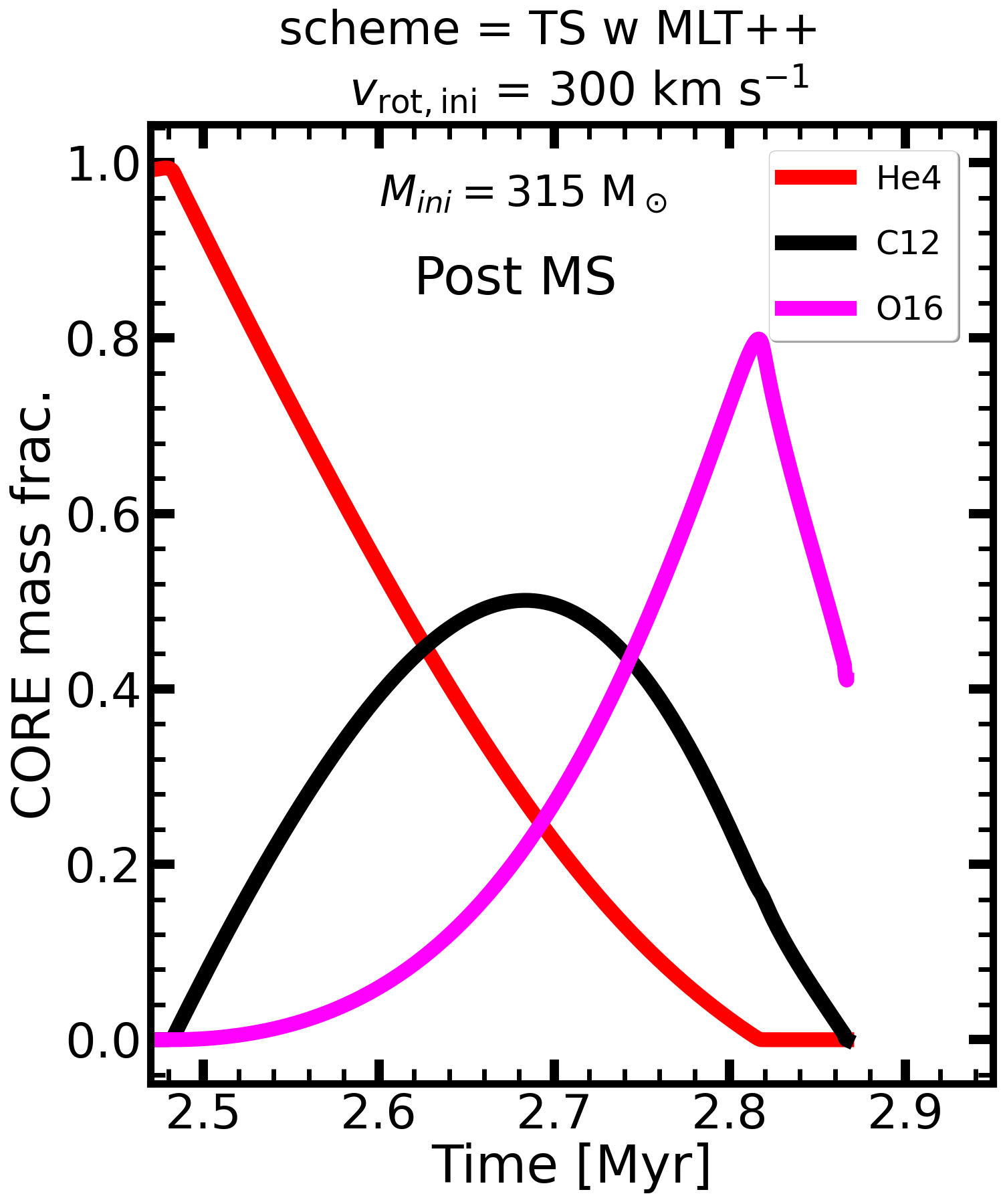}\includegraphics[width=0.33\textwidth]{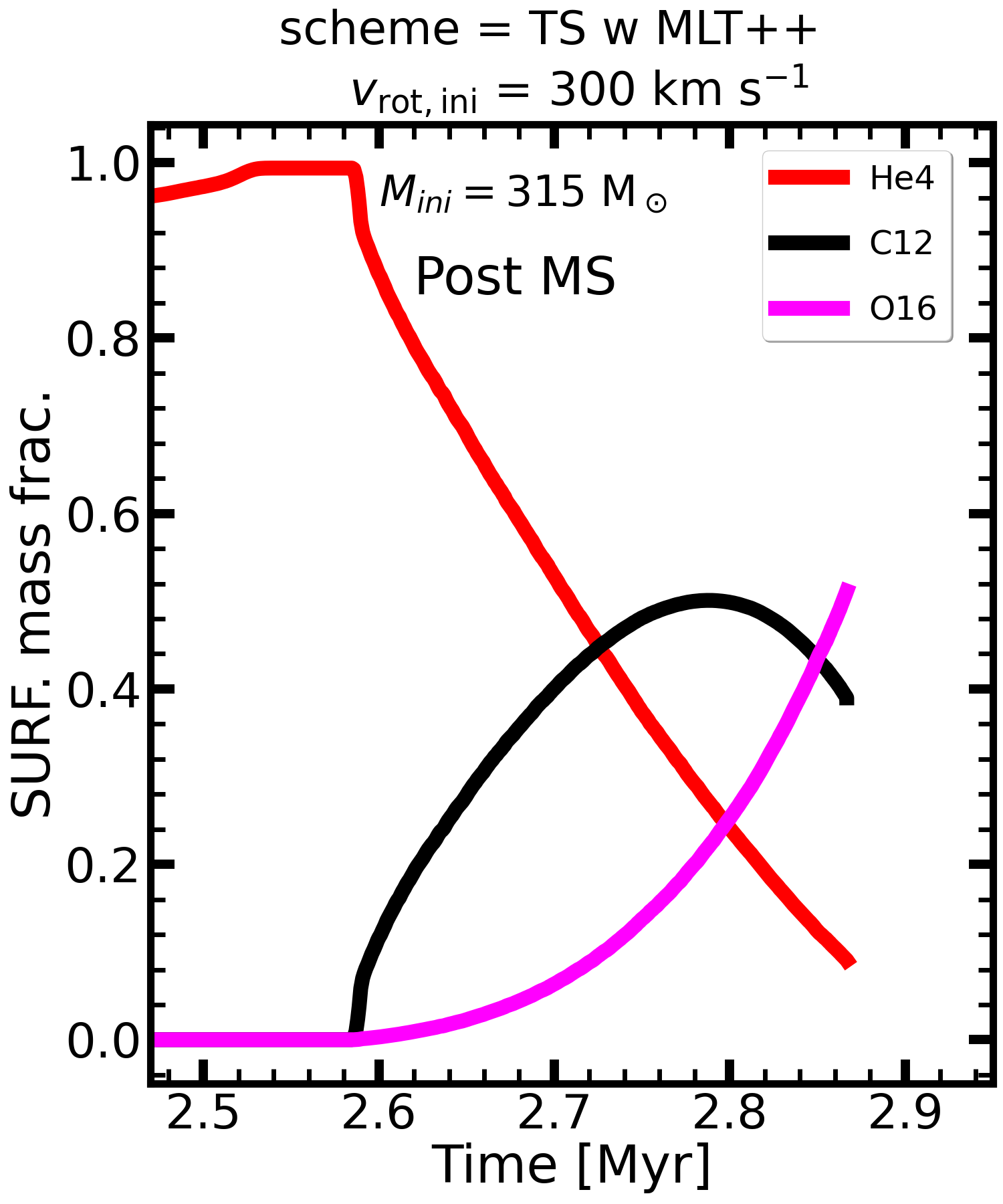}
    \caption{Post-main sequence evolutionary tracks in the Taylor-Spruit scheme with $v_{\rm ini}=$~300~km\,s$^{-1}$. \emph{Left panel:} HRD from the Terminal Age Main Sequence (TAMS with squares), through the helium-TAMS (diamonds), until carbon-TAMS (blue stars). The colours denote different initial masses. On the right-hand side of this panel, the TAMS masses are shown in solar units and the ZAMS masses are indicated in brackets. Corresponding to these cases, the masses at the C-TAMS are also shown in solar units on the left-hand side. \emph{Middle and right panels:} post-main sequence evolution of core and surface helium, carbon, and oxygen mass fractions for the initially 315~M$_\odot$ model.
    \label{fig:postMS}}
\end{figure*}
%
%
%

\subsection{Post-main-sequence evolution } 
\label{sec:postms}

%
%
Using \Cref{fig:massevol}, we begin discussing the post-main sequence mass evolution and predicted final masses from our models.
The most massive model has the most rapid evolution, reaching the C-TAMS in 2.64 Myr in our fiducial scheme, assuming a 300~km\,s$^{-1}$ initial rotational velocity and the TS scheme with MLT++. When the mass-loss rates are lowered, the stellar mass remains higher, and the fusion in the core is more rapid. 
The time to reach the C-TAMS is 2.29 Myr and 2.43 Myr in the case of models with a factor of 2 lower overall rates and different WR rates, respectively. 

The further time evolution to core collapse is rapid. 
From the models in our fiducial grid, the most massive stellar model at 2.64 Myr, when the first supernova explodes, has a maximum observable mass $M_{\rm obs,max}=$~68~M$_\odot$. We predict that in 2--3\,Myr co-evally formed  clusters, VMSs would only appear as helium-rich stars in their post-main sequence evolution with significantly lower masses -- potentially, by hundreds of solar masses -- than their initial mass. This highlights that observations of VMSs is challenging, and R136 represents a prime opportunity to study VMSs at an age under 2 Myr. 
Considering the HY, TS, and SB angular momentum schemes and the various initial rotational velocities in our model grid, none of the maximum observable stellar masses would exceed 130~M$_\odot$ by the time the first supernova takes place (see also \Cref{sec:app1}). 
Changing the treatment of mass loss would allow for higher masses (102 and 97 M$_\odot$ in our two tests shown in the middle and lower panel of \Cref{fig:massevol}, respectively). 
We should highlight that these quantitive results depend on numerics as some models fail to converge (\Cref{fig:massevol_app}). However, the qualitative trends are robust and indicative of the changes in rotational and mass-loss properties.
We further elaborate on the implications of these results, particularly in terms of constraining initial mass functions from present-day mass functions in \Cref{disc}.

Using \Cref{fig:massevol}, we also investigate $M_{\rm preCC,max}$, the highest stellar masses at the C-TAMS in our model grid, to get an indication for the expected most massive core-collapse supernova progenitors at LMC metallicity assuming our model setup.
This results in a maximum pre-supernova mass of 24~M$_\odot$ in our fiducial scheme. Interestingly, it does not originate from the initially most massive progenitor (see below). For various AM schemes and initial rotational velocities, the maximum C-TAMS masses typically range between about 20 and 50~M$_\odot$ (\Cref{fig:massevol_app}). This highlights the crucial role of mass loss in shaping the post-mains sequence evolution and final fates of these stars.
Decreasing the overall mass-loss rates by a factor of two leads to 67~M$_\odot$, whereas using different WR-type rates results in 48~M$_\odot$ for the maximum pre-supernova mass. Thus even small changes in the mass-loss rate, that is, within currently best measured uncertainties, can drastically impact the model predictions in the pre-supernova stage.

We report additional findings for the post-main sequence evolution of our fiducial models in Figure \ref{fig:postMS}.
The HRD (left panel) shows that all these models produce hot and luminous post-main sequence objects. In the full range of explored initial masses, the initially 225\,M$_{\odot}$ stellar model leads to the highest TAMS mass of 46~M$_\odot$ and produces the most massive supernova progenitor with 24~M$_\odot$.
The initially most massive model (795~M$_\odot$) becomes a 40~M$_\odot$ helium-rich star at the TAMS and evolves to a 22~M$_\odot$ supernova progenitor. 
The initially least massive 100\,M$_{\odot}$ star is also the least massive supernova progenitor, with a C-TAMS mass of 19\,M$_{\odot}$. Clearly, there is no linear relation between initial and final mass. 

For the adopted mass-loss properties, the fairly low final masses imply that VMSs at LMC metallicity are not expected to form exceptionally massive black holes, regardless of whether or not episodic or eruptive mass loss in the very final stages leading up to core collapse may remove larger amounts of mass \citep[e.g.,][]{hillier2019} and/or the core-collapse event itself leads to the ejection of the outer envelope. 

We briefly discuss the final core and surface chemical composition for the initially 315\,M$_{\odot}$ stellar model. We pick this initial mass as it might be a possible progenitor for the WNh stars in the 30\,Doradus central cluster. This model reaches the SN progenitor stage in less than 3\,Myr, of which core helium and carbon burning take up less than 0.5\,Myr. The star becomes very compact and hot during its post-main sequence evolution. Though the core becomes oxygen rich (middle panel of Figure \ref{fig:postMS}), the surface abundances do not immediately reflect on the core composition (right panel). At the C-TAMS, the surface abundances of carbon and oxygen are about equal, with approximately a 10 percent mass fraction of helium still being present.
The predicted surface C/He ratio is characteristic for WC and WO stars; the predicted O/He ratio, however, is a factor of a few higher \citep[e.g.,][]{sander2012,tramper2015}. If this picture is at least qualitatively correct, we may expect that the most massive stars in R136 become extremely hot and luminous, hydrogen-poor objects in their post main sequence evolution. In the next section, we elaborate on the final fates that we infer from our model calculations.

\subsection{Expected final fate} 
\label{sec:final_fate}

Very Massive Stars are considered potential progenitors of super-luminous supernovae, pair-instability supernovae, gamma-ray bursts, and massive black holes of tens of solar masses recently identified in gravitational wave detections \citep[e.g.,][]{kuroda2018,uchida2019,moriya2020,nicholl2020,abbott2023}. Their physical nature, evolution, and future are therefore of considerable interest. We discuss some aspects of this problem.

%
%

%
%
\begin{figure*}
    \includegraphics[width=0.47\textwidth]{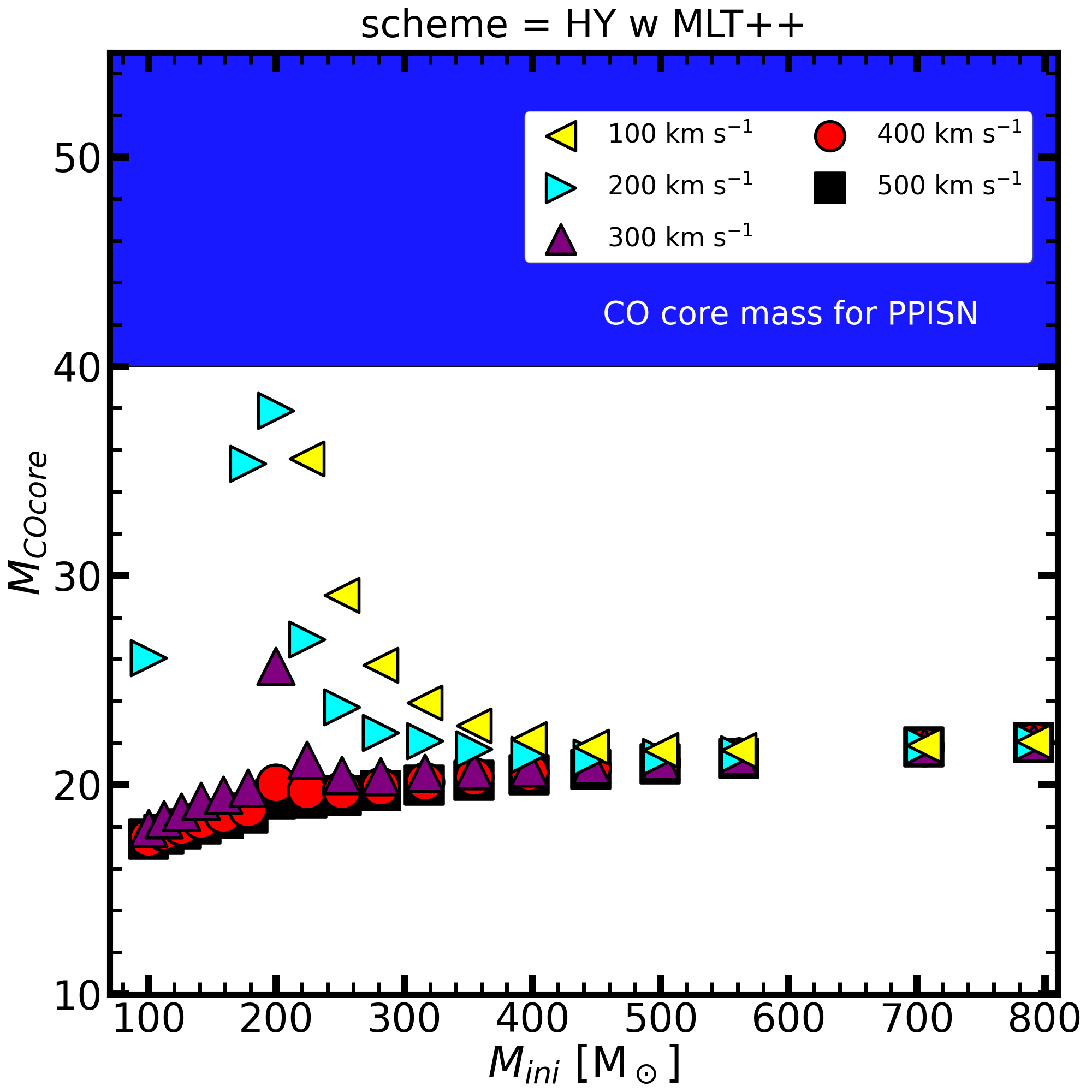}\includegraphics[width=0.47\textwidth]{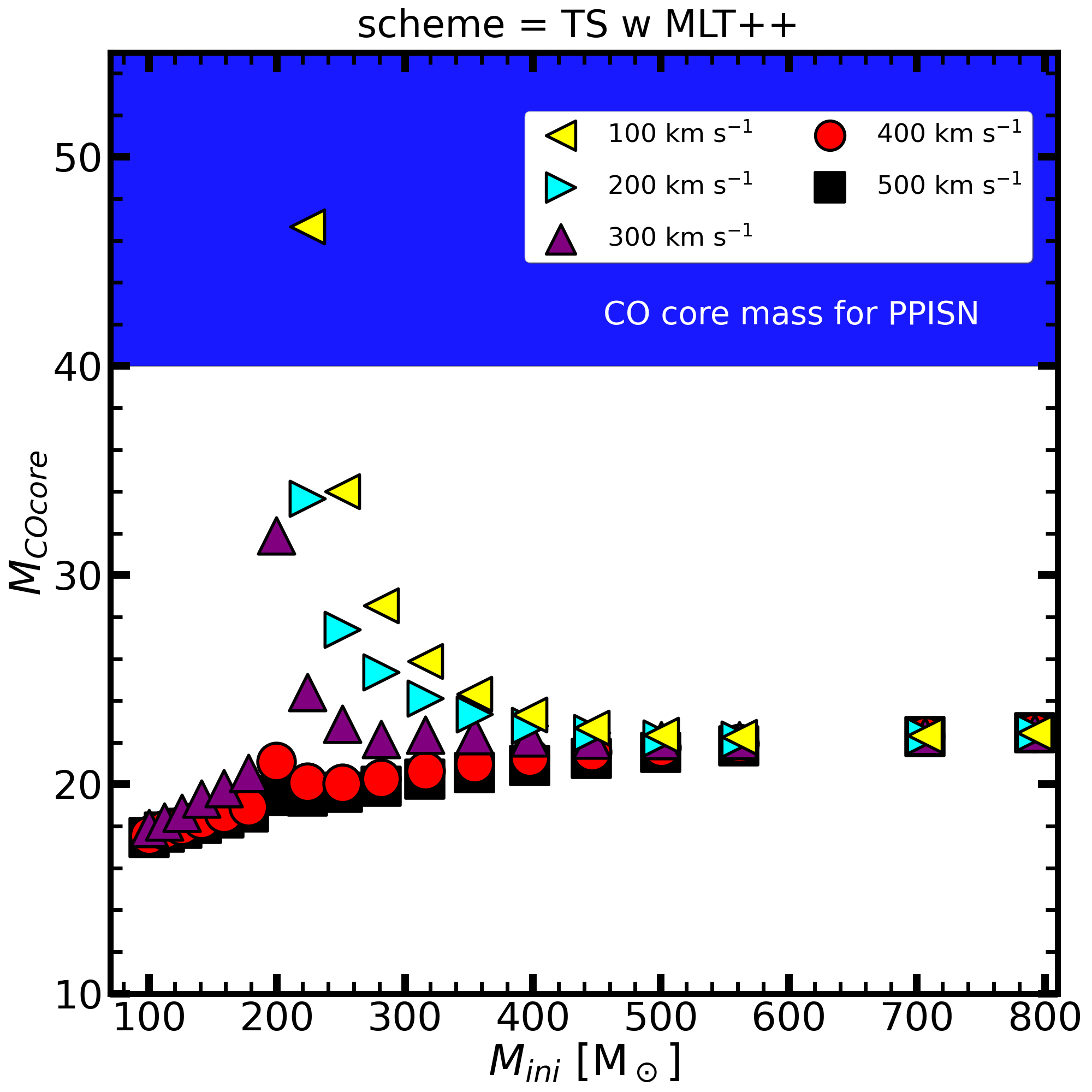}
    \includegraphics[width=0.47\textwidth]{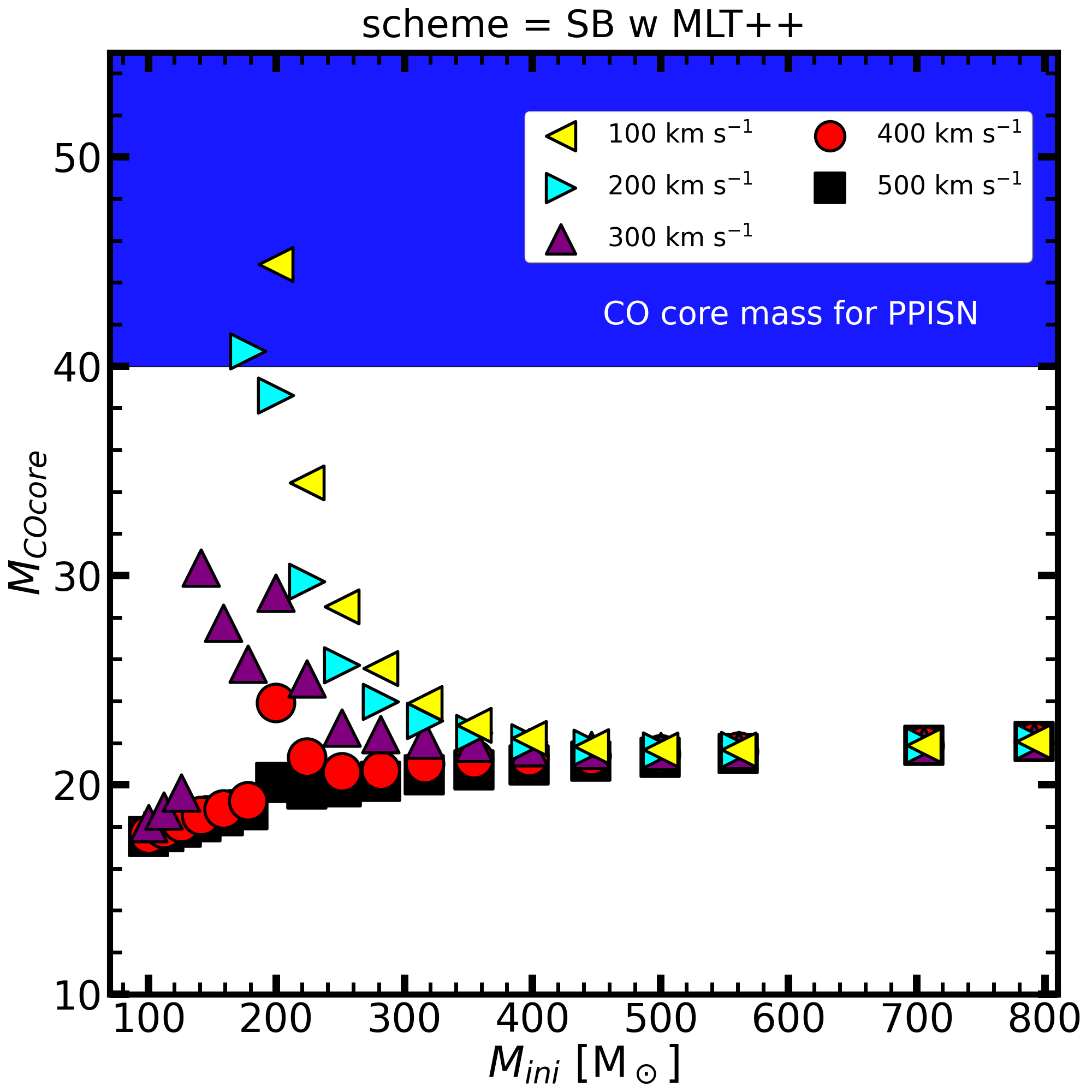}\includegraphics[width=0.47\textwidth]{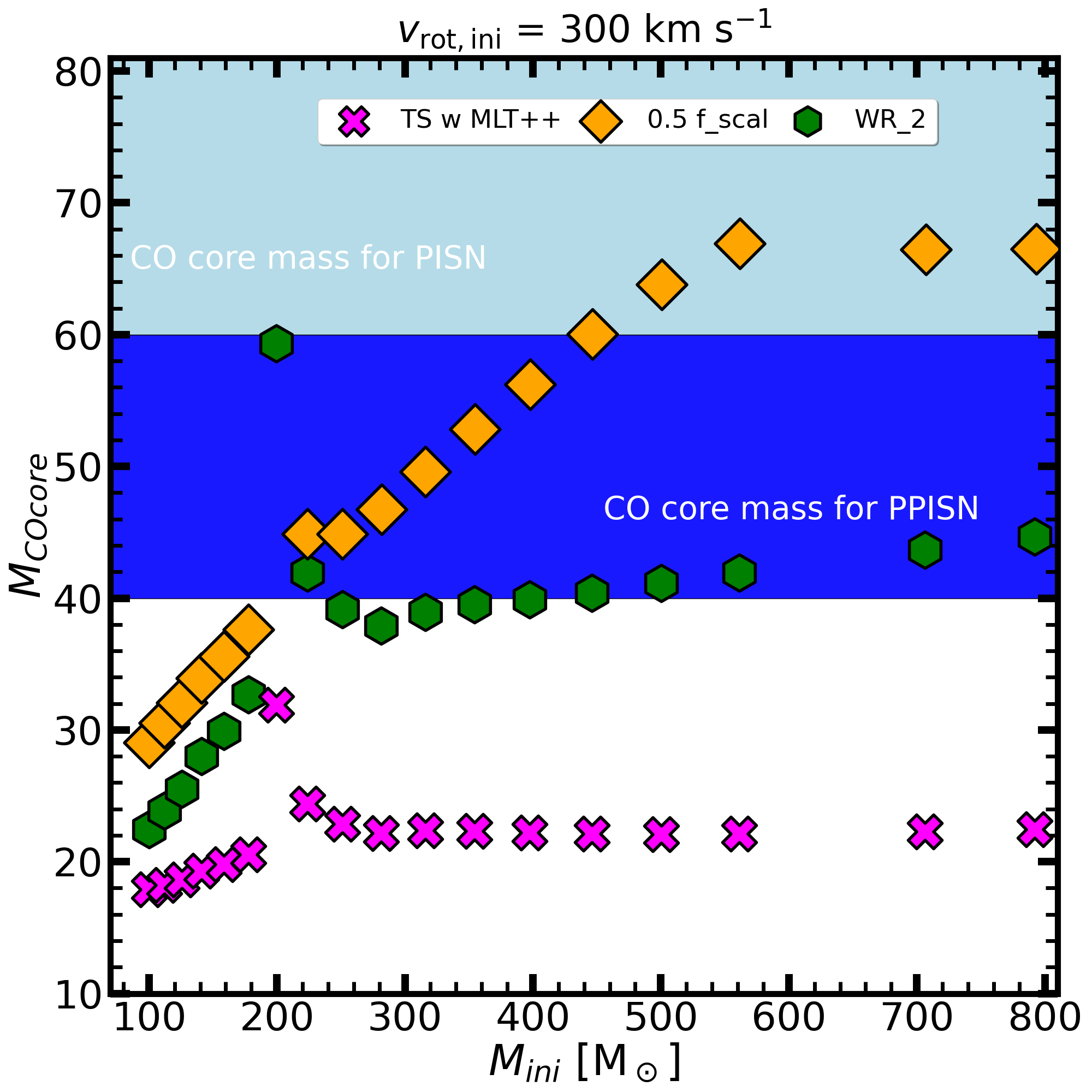}
    \caption{Maximum carbon-oxygen core mass as a function of initial stellar mass. The HY, TS, and SB coupling schemes are shown on the top left, top right, and lower left, respectively. These panels include all initial rotational velocities calculated in our grid (shown with different symbols and colours). The lower right panel shows models with a 300 km\,s$^{-1}$ initial rotational velocity in the TS coupling scheme, varying only the mass-loss schemes (magenta cross: nominal scheme; orange diamond: reduction of overall mass-loss rates by a factor two; green hexagon: switching to different Wolf-Rayet type rates).
    The shaded regions denote the approximate limits for pulsational pair instability (blue) and pair instability (light blue) supernovae. Note the different vertical axis scale in the lower right panel.}
    \label{newfig:cocoremass}
\end{figure*}

%
%
\subsubsection{Escaping pair-instability} 
\label{sec:pisn}

If encountered, the pair-instability mechanism is expected to completely disrupt a massive star without black-hole formation \citep{bond1984}. The explosion mechanism leading up to a pair-instability supernovae is thought to be well understood \citep[e.g.,][]{langer2012}. For such an event to occur, the core mass at the He-TAMS (consisting of mainly carbon and oxygen) needs to be in the range of 60--130\,\Msun or at least 40\Msun for pulsational pair instability \citep{eleid1986,heger2002}, though these estimates may depend on core rotation of the star \citep{chatzopoulos2012,chatzopoulos2013}. 
In the (P)PISN scenario, the star can combine a relatively low ($\log \rho_{c} \approx 5\, [\mathrm{g\,cm^{-1}}]$) density with a relatively high temperature ($\log T_{c} \approx 9\,  [\mathrm{K}]$), allowing for electron-positron pair production.
%
%
In Figure \ref{newfig:cocoremass}, we show the carbon-oxygen core mass at the final step of our computations in our models (c.f. Section~\ref{sec:setup}) as a function of initial mass and initial rotation. When using the hydrodynamical or Tayler-Spruit rotational coupling schemes, we find that the CO core masses are systematically below the limit to produce (pulsational) pair-instability supernova. The solid-body rotating models allow for PPISN progenitors in a few cases, notably in the initial mass range around 200~M$_\odot$. In this case, the initial rotation plays an important role in achieving higher or lower CO core masses. In fact, for initial rotation rates less than 500~km\,s$^{-1}$, the CO core mass at LMC metallicity does not increase linearly with the initial mass. This is in contrast with lower-metallicity models
\citep[e.g.,][]{yoshida2011,takahashi2018}. For initial masses in excess of 400~M$_\odot$, the dependence on initial rotation or core-envelope coupling scheme becomes negligible to determine the maximum CO core mass because the strong mass loss produces slowly rotating stars early in the evolution. 
Our results are drastically impacted by the choice of mass-loss schemes (lower right panel of Figure \ref{newfig:cocoremass}). When reducing the mass-loss rates by a factor of two, the stellar masses remain higher (c.f. Figure~\ref{fig:massevol}) and consequently much higher CO core masses can be achieved. In this case, several models lead to PPISN and, for the highest initial masses, PISN. This parameter test also reproduces the results in near zero metallicity VMS models, namely, the CO core mass increases as a function of initial mass. When switching to the \cite{nugis2002} Wolf-Rayet type rates, we find that the trends remain as in our fiducial model; however, the predicted CO core masses are systematically larger, typically by about a factor of 2. 
Overall, our results fit the general picture that PISN may not be expected for a metallicity representative of the LMC \citep[e.g.,][]{langer2007}. Strong winds remove sufficient mass that most models in the 100 to 800 M$_\odot$ initial mass range will have CO cores below 40 M$_\odot$ (see also \citealt{yusof2021,higgins2021}). 
Nonetheless, mass loss does not only depend on rotation and metallicity but also on magnetic fields. If VMSs at LMC metallicity harboured strong magnetic fields that could quench their mass loss, then they might remain more massive, as shown for solar metallicity models \citep{petit2017,georgy2017}.

%
%

\begin{figure}
    \centering
    \includegraphics[width=0.45\textwidth]{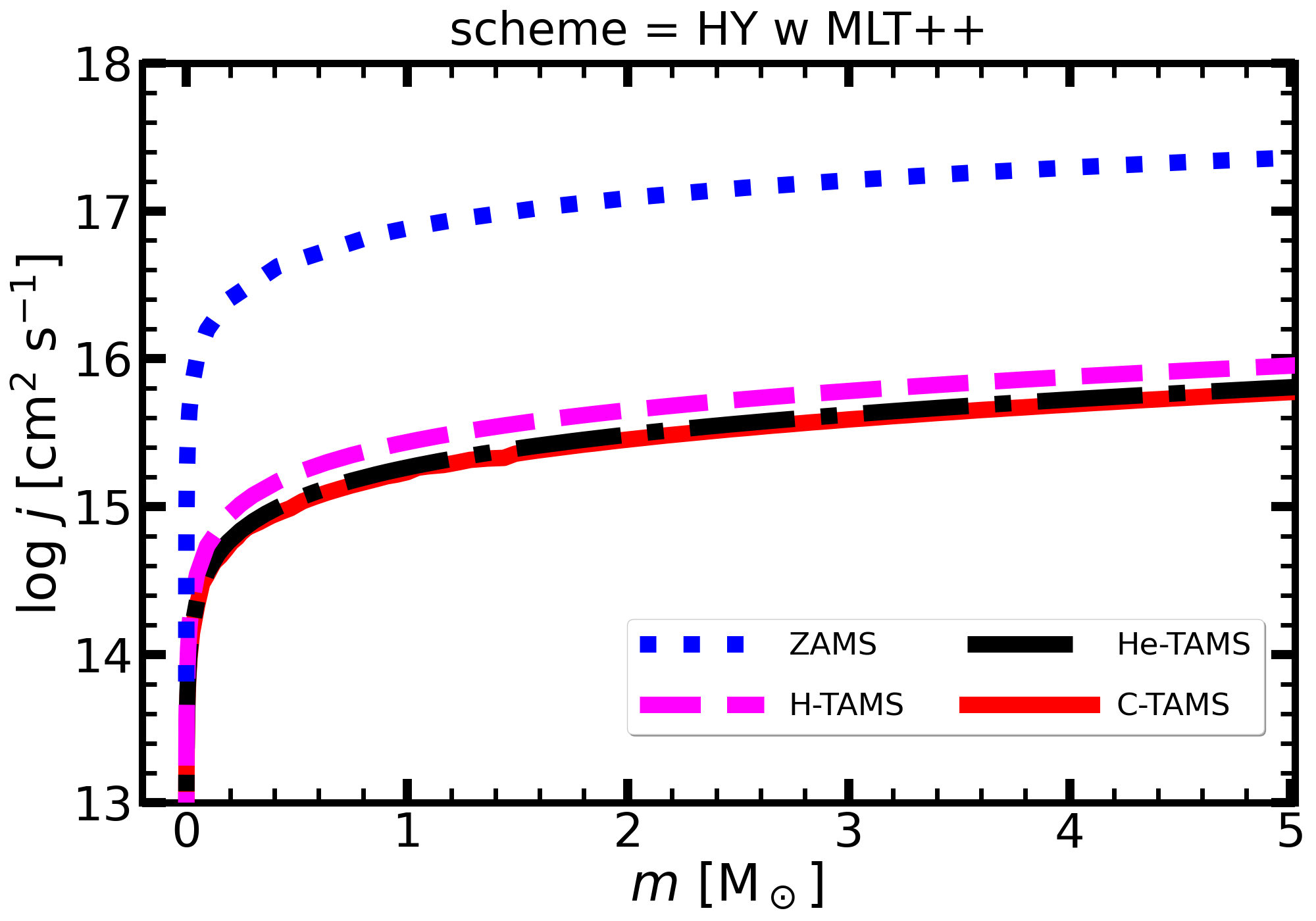}
    \includegraphics[width=0.45\textwidth]{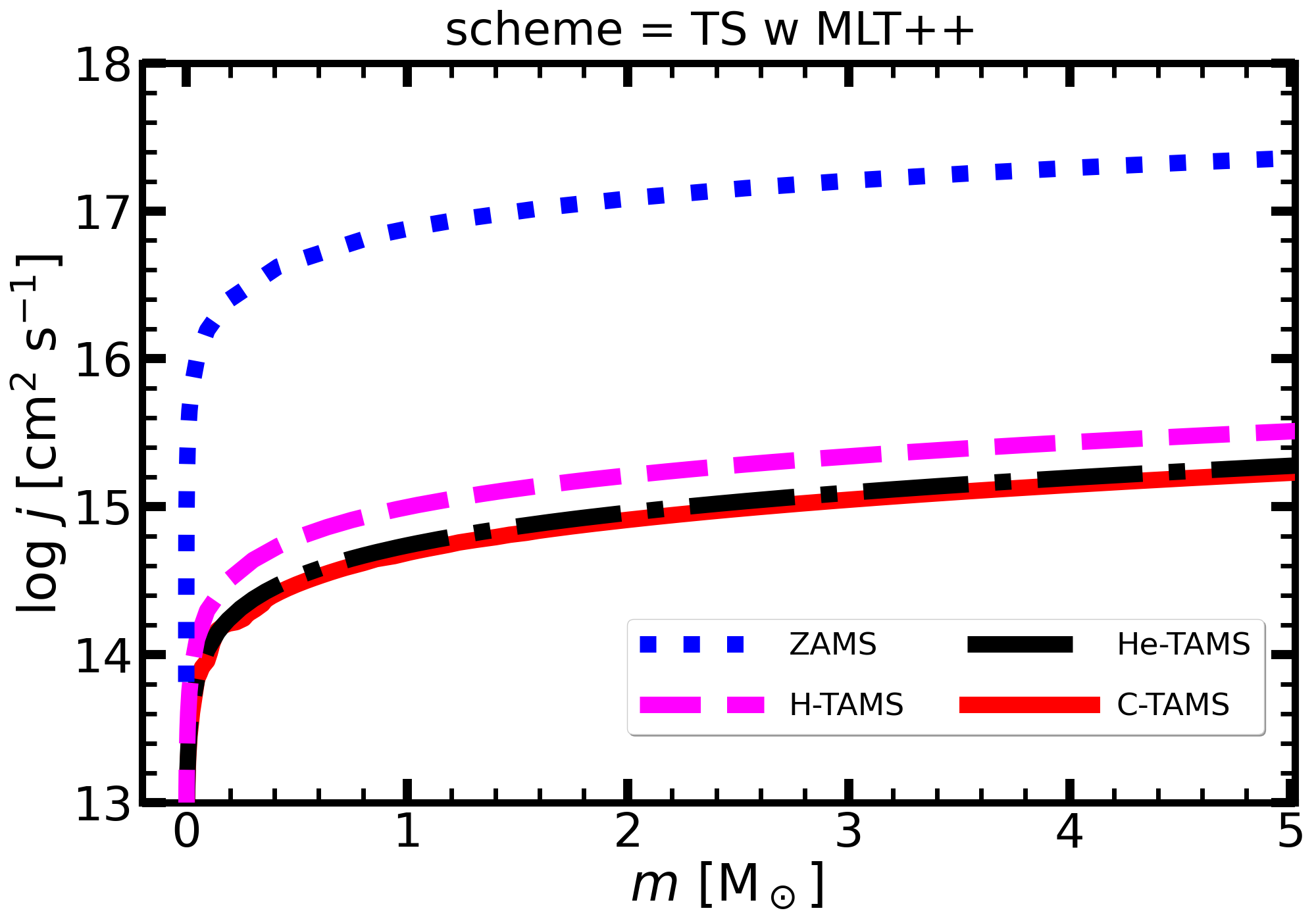}
    \includegraphics[width=0.45\textwidth]{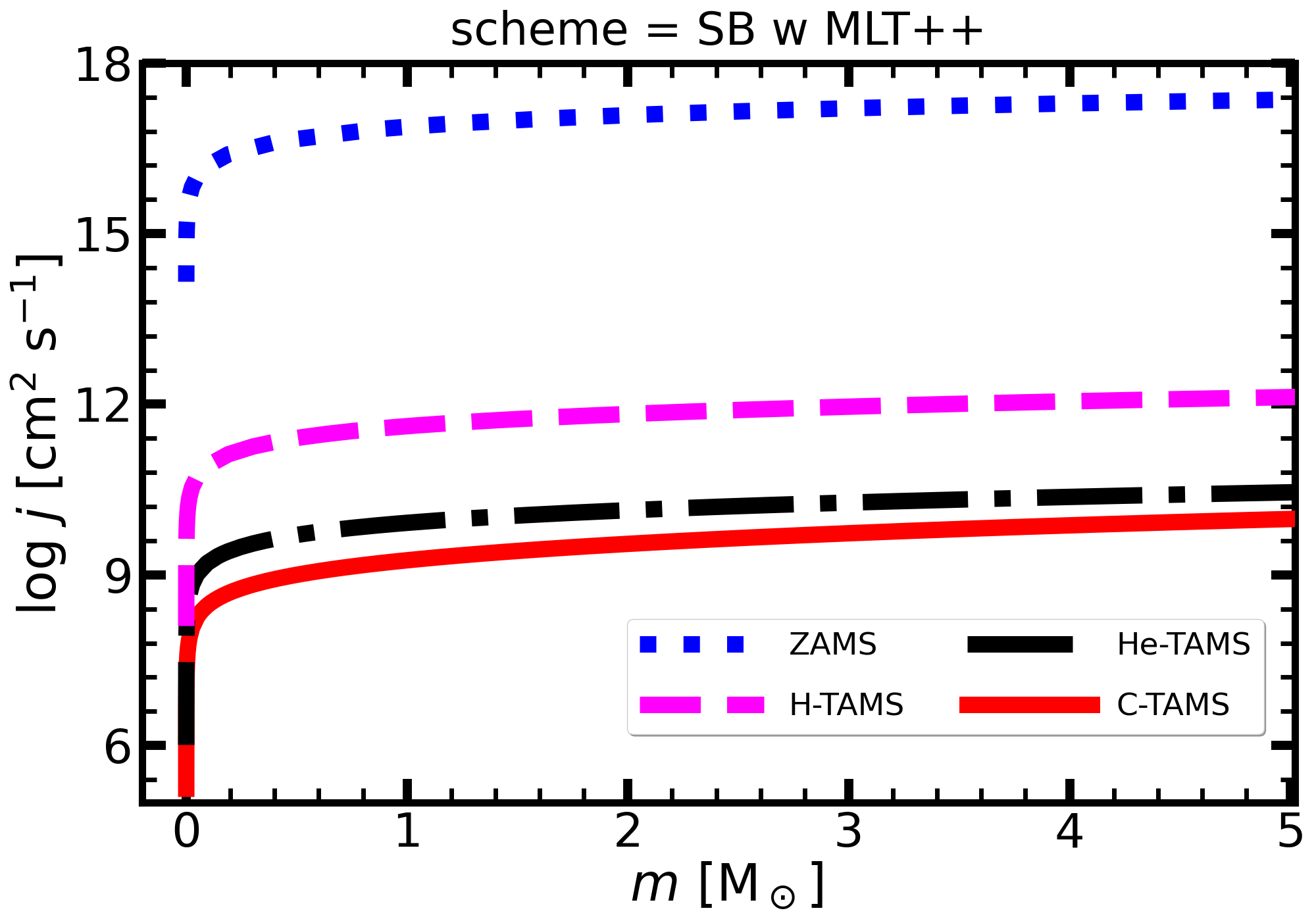}
    \caption{Specific angular momentum in the central 5 solar masses of the stellar model at representative evolutionary stages. 
    Shown are the initially 315 M$_\odot$ models with an initial rotational velocity of 300~km\,s$^{-1}$, using the hydrodynamical (top panel), Tayler-Spruit (middle panel), and solid-body (lower panel) coupling schemes. MLT++ is used in these models. Note the different vertical axis scale in the lower panel.
    \label{fig:jrot}}
\end{figure}
%
%

\begin{figure*}
    \centering
     \includegraphics[width=0.49\textwidth]{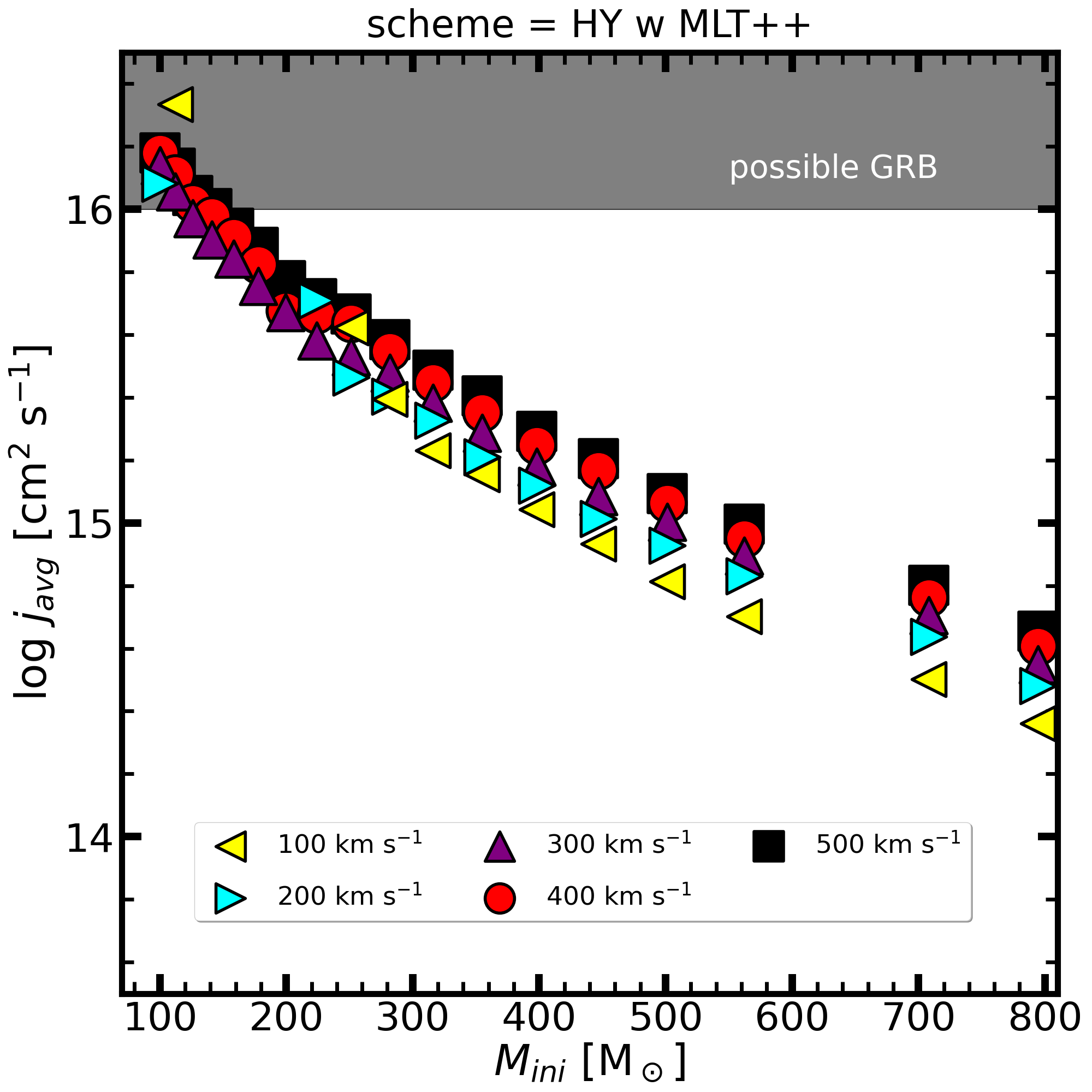}\includegraphics[width=0.49\textwidth]{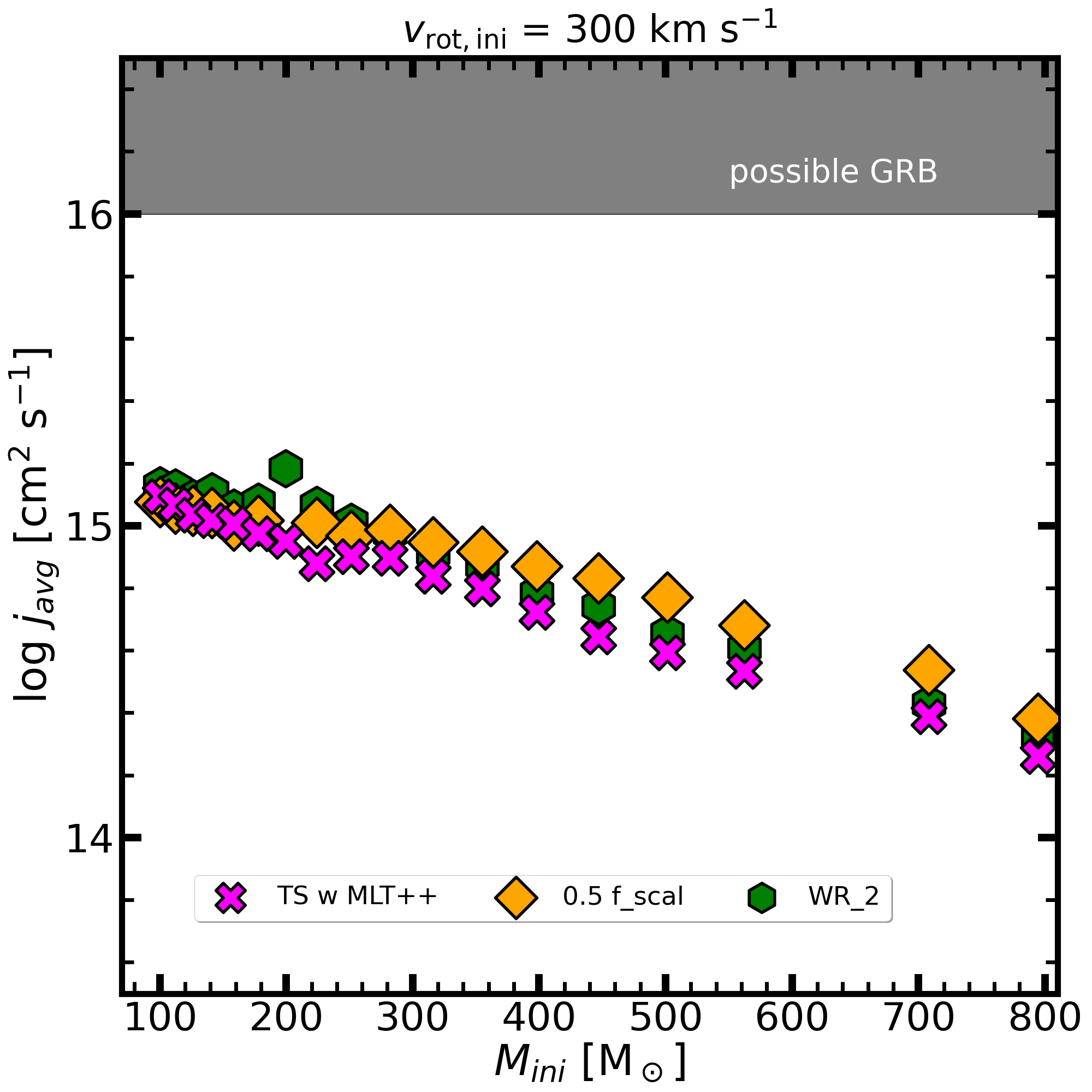}
    \caption{Average specific angular momentum within the central 5 solar masses of the core at the end of carbon core burning versus the initial mass. Left: hydrodynamical coupling scheme. Right: Our experiment with different mass-loss rates. The symbols denote different initial rotational velocities (left) or different mass-loss schemes (right). The approximate limit to produce GRBs is shown with the gray region. 
    \label{fig:jrotsum}}
\end{figure*}
%
%

%
%
\subsubsection{Progenitors of long-duration gamma-ray bursts} 
\label{sec:grb}

Angular momentum transport, while uncertain in current stellar evolution models, crucially impacts the final outcome of VMS evolution. If sufficient AM can be retained, the star may produce an enigmatic long-duration gamma-ray burst (GRB). 
Current estimates are still rudimentary; however, an average specific angular momentum of $10^{16}$~cm$^{2}$~s$^{-1}$ in the central region of the star shortly before collapse might be considered to lead to long-duration~GRBs \citep{yoon2006,yoon2012,aguilera2018}.
%
%
In Figure \ref{fig:jrot}, we demonstrate how the specific angular momentum reservoir within the inner 5 solar masses is depleted from ZAMS until the C-TAMS in case of a representative model. 
The HY model with moderate core-envelope coupling can retain more of its central angular momentum by approximately an order of magnitude compared to the TS model. Most of the angular momentum from the core is lost during the main sequence. 
The model with Tayler-Spruit coupling scheme also loses most of its central angular momentum on the main sequence. Subsequently, the specific angular momentum profile remains approximately constant. This result is in good qualitative agreement with previous findings obtained for models up to an initial mass of 100~M$_\odot$ at Z~$\approx 1/50 $~Z$_\odot$ \citep{aguilera2018}.
The solid-body rotating model loses a drastic amount of angular momentum over the course of its evolution, leading to a decrease in its central specific angular momentum profile by 8 orders of magnitude from ZAMS to C-TAMS.

%
%
In Figure \ref{fig:jrotsum}, we summarise these findings for the hydrodynamical coupling scheme, using the entire parameter space in initial rotational velocities and initial mass. As expected, models in the Taylor-Spruit coupling scheme and with perfect solid-body rotation are systematically below the threshold to produce GRBs as they have little angular momentum remaining in their cores. 
We find that models with initial masses of approximately 100--200~M$_\odot$ in the HY scheme could become GRB progenitors. However, the TS and SB schemes remove a larger amount of angular momentum from the central regions. Therefore, we do not expect GRBs from these models. In general, the initial rotation has a relatively modest impact on these findings. In our numerical experiments with a given scheme and initial rotation (TS, 300~km\,s$^{-1}$), the different wind physics have a modest impact on these results. Thus, no GRB is expected when the mass-loss rates are lowered by an overall factor of 2 or when switching to different WR-type winds in the TS scheme. We conclude that the main constraint is posed by the use of various AM transport schemes to predict the production of long-GRBs.


%
%
\begin{figure}
    \centering
    \includegraphics[width=0.45\textwidth]{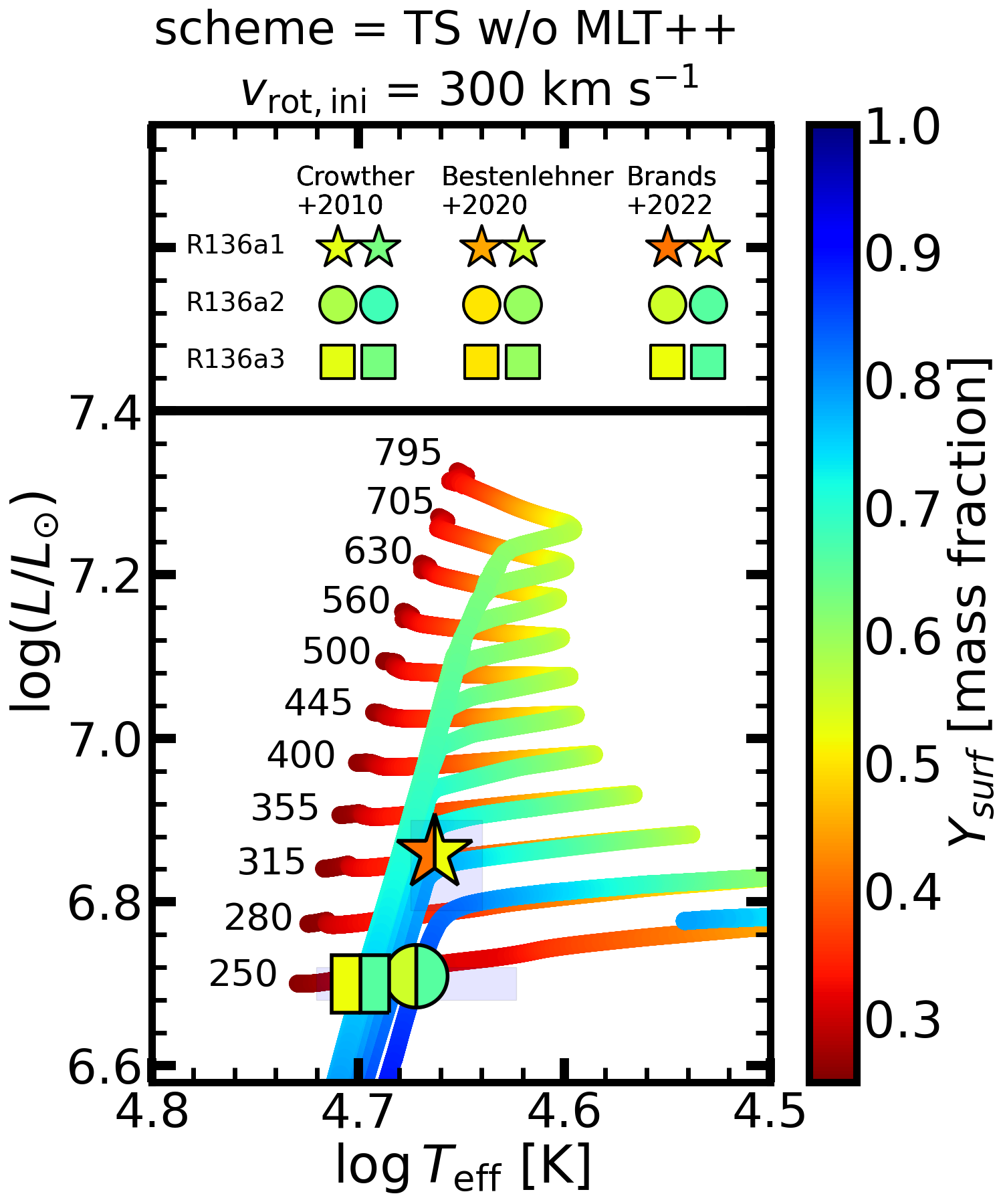}
    \caption{HRD colour-coded with the surface helium mass fraction in our models (using the Taylor-Spruit scheme without MLT++ with an initial rotational velocity of 300~km\,s$^{-1}$) and in observations of the WNh stars R136a1 (star), R136a2 (circle), and R136a3 (square). The initial mass of the evolutionary tracks is indicated in solar units. We apply smoothing to avoid numerical noise. Uncertainty ranges (lower and upper limits) in the measured helium abundances from three authors are indicated in the upper part (see also Table~\ref{tab:Obs}). For the nominal values, we adopt the \cite{brands2022} results and show the lower and upper limits as the left and right side of the filled markers. The shaded regions show the formal uncertainty in $\log L$ and $T_{\rm eff}$ from \cite{brands2022}.
    }\label{fig:hrd_he}
\end{figure}
%

%
%
\begin{figure*}
    \centering
    \includegraphics[width=0.99\textwidth]{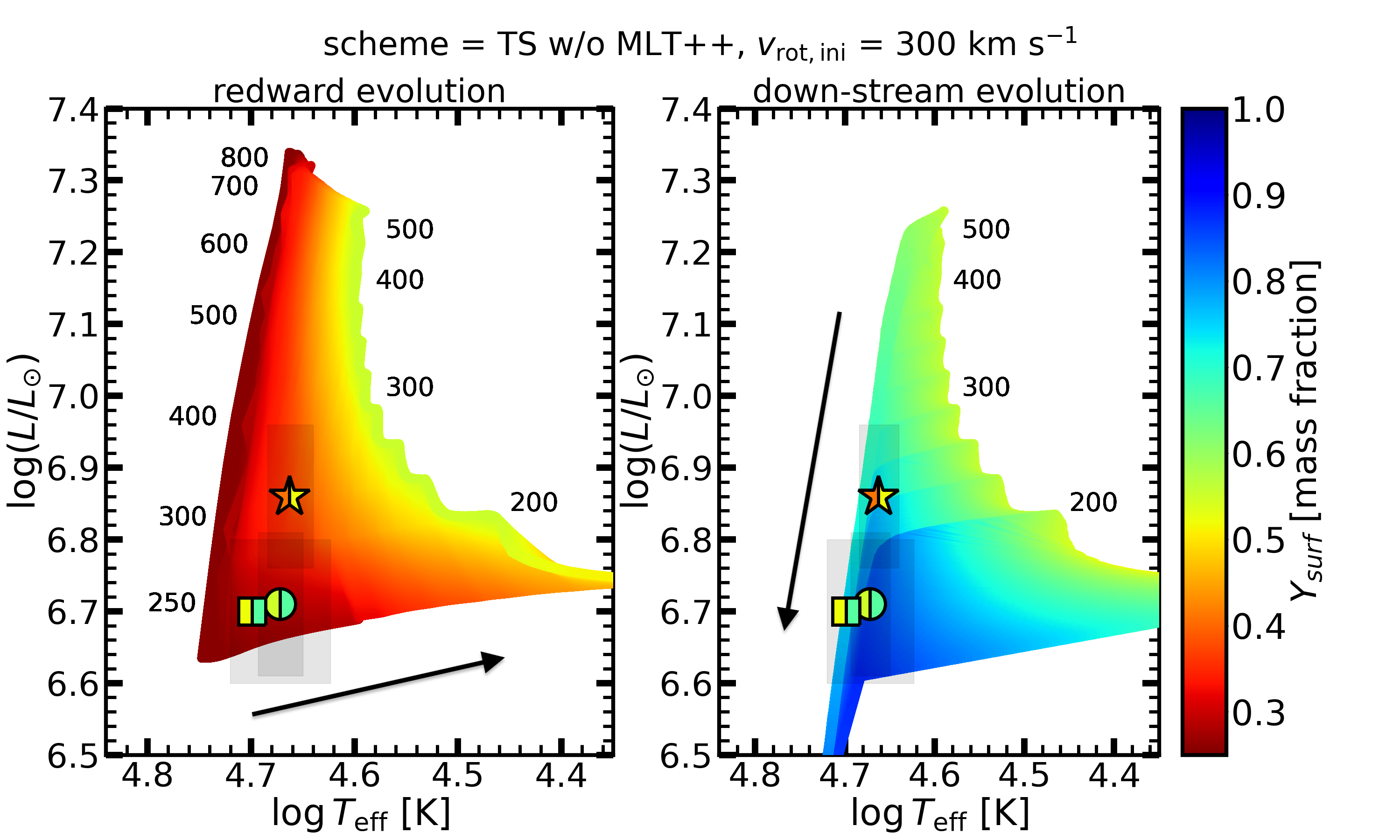}
    \caption{Interpolated models on the HRD with colour-coding showing the surface helium mass fraction. The \citealt{brands2022} observations are adopted as in Figure~\ref{fig:hrd_he}, but the uncertainty in $\log L$ and $T_{\rm eff}$ are updated for the MCMC analysis (see text). The left panel shows models on a redward trajectory, the right panel shows models in a down-stream evolution. The values to the left indicate the initial stellar mass in solar units. The values to the right show the mass in solar units before the down-ward turn in evolutionary path. The arrows indicate the direction of the evolution.}
    \label{fig:interp1}
\end{figure*}
%

%
%
%
\setlength{\tabcolsep}{1.3pt}
\begin{table*}[]
    \centering
    \caption{Empirically-determined properties of the WNh stars in R136. \label{tab:Obs}}
    \begin{tabular}{l@{\hspace{12pt}} c@{\hspace{10pt}} c@{\hspace{10pt}} c c}
        \hline\hline
         & R136a1 & R136a2 & R136a3 & Reference \\ \hline\hline
         %
         \multirow{3}{*}{$\log (L/L_\odot)$} & 6.94 $\pm$ 0.09 & 6.78 $\pm$ 0.09 & 6.58 $\pm$ 0.09  & [1] \\
         & 6.79 $\pm$ 0.10 & 6.75 $\pm$ 0.10 & 6.63 $\pm$ 0.10  & [2] \\
         & 6.86 $\pm^{0.04}_{0.07}$ & 6.71  $\pm^{0.03}_{0.03}$ & 6.70  $\pm^{0.02}_{0.02}$  & [3] \\ \hline
         %
        \multirow{3}{*}{$T_{\rm eff}$ [kK]} & 53 $\pm$ 3.0 & 53 $\pm$ 3.0 & 53 $\pm$ 3.0  & [1] \\
         & 46 $\pm$ 2.5 & 50 $\pm$ 2.5 & 50 $\pm$ 2.5  & [2] \\
         & 46 $\pm^{1.3}_{2.4}$ & 47 $\pm^{1.0}_{0.5}$ & 50 $\pm^{2.5}_{8.0}$ & [3] \\ \hline
         %
         \multirow{3}{*}{$Y_{\rm surf}$ [mass fraction]} & 0.58 $\pm$ 0.05 & 0.63 $\pm$ 0.05 & 0.58 $\pm$ 0.05  & [1] \\
         & 0.50 $\pm$ 0.05 & 0.55 $\pm$ 0.05 & 0.55 $\pm$ 0.05  & [2] \\
         & 0.45 $\pm^{0.05}_{0.06}$ & 0.60 $\pm^{0.06}_{0.05}$ & 0.60 $\pm^{0.06}_{0.08}$ & [3] \\ \hline
         %
         \multirow{3}{*}{$\log \dot{M}$ [M$_\odot$yr$^{-1}$]} & $-4.29 \pm$ 0.08 & $-4.34 \pm$ 0.08 & $-4.43 \pm$ 0.08  & [1] \\
         & $< -3.80 \pm$ 0.20 & $< -3.84$ $\pm$ 0.20 & $< -3.83 \pm$ 0.20 & [2] \\
         & $-4.57 \pm$ 0.13 & $-4.48 \pm$ 0.12 & $-4.64 \pm$ 0.08 & [3] \\ \hline
         %
         %
         \hline
    \end{tabular}
        \begin{minipage}{0.9\textwidth}
        \vspace{5pt}
        \small \textbf{Notes.} [1] \citet{crowther2010}, [2] \citet{bestenlehner2020}, [3] \citet{brands2022}. Mass-loss rates from \citet{bestenlehner2020} are upper limits.
        \end{minipage}
\end{table*}

%
%
\begin{figure}
    \centering
    \includegraphics[width=0.49\textwidth]{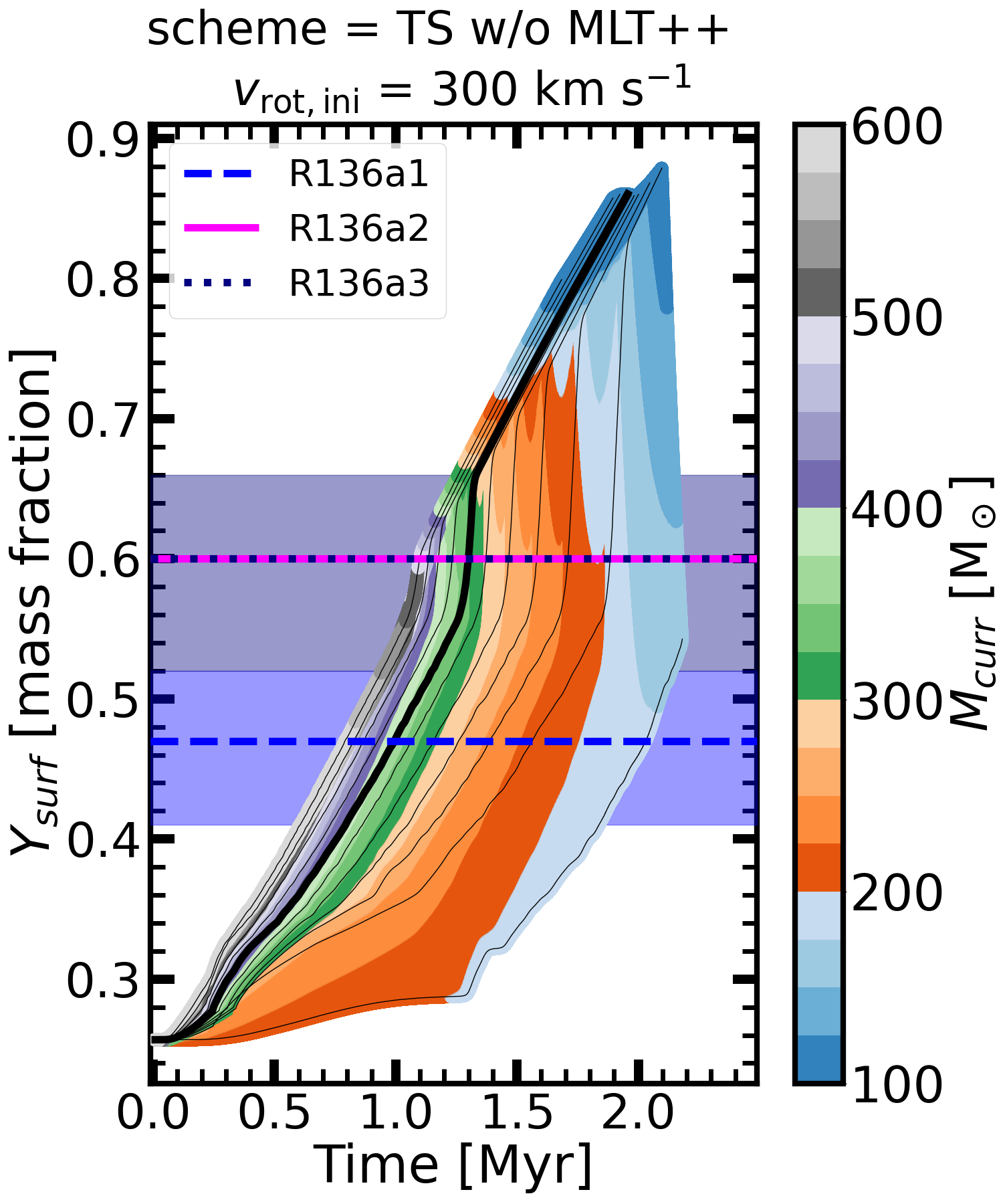}
    \caption{Surface helium mass fraction as a function of time. The evolutionary models are shown with black lines with decreasing initial mass from left to right. The model with $M_{\rm ini}=~500$~M$_\odot$ is highlighted with thicker line. The colour-coding shows the interpolated models as a function of their current mass on a logarithmic scale. The observed surface helium mass fractions by \citealt{brands2022} are indicated for the three WNh stars (R136a1 -- gray, R136a2 -- magenta, R136a3 -- navy, see also Table~\ref{tab:Obs}). The shaded region shows the associated uncertainty.}
    \label{fig:interp2}
\end{figure}
%
%

%
%
\section{Comparison to observations}\label{obscomp}

In this section, we compare our models without MLT++ (see Figures \ref{fig:ms1} and \ref{fig:hrd_wrtr1} and related discussion) by means of interpolation and Markov-Chain Monte Carlo analysis with the observed properties of the most massive stars presently known: R136a1, R136a2, and R136a3. Our primary focus is on simultaneously reconciling the observed luminosity, effective temperature, and surface helium abundance with theoretical predictions. We also discuss constraints on mass-loss rates and surface rotation.

\subsection{Empirical characterisation}

Using HST optical spectroscopy, \cite{crowther2010} found that the current masses of several stars in R136 exceed 150~M$_\odot$. 
\cite{bestenlehner2014,bestenlehner2020} also studied the most luminous stars in 30 Doradus, finding current masses for the WNh stars of the order of 200--300~M$_\odot$. These investigations used the {\sc cmfgen} model atmosphere code \citep{hillier1990,hillier2019}.
\cite{pauldrach2012} studied R136a3 and the evolutionary history of VMSs. While they require a lower effective temperature (44 kK) to match the spectrum, the Helium abundance is in agreement with determinations by other authors.
Most recently, \citet{brands2022} analysed new HST optical and UV spectroscopy of luminous stars in the core of the R136 cluster. 
The spectral analysis was performed with the model atmosphere code {\sc fastwind} \citep{1997A&A...323..488S,2005A&A...435..669P,2018A&A...619A..59S} and the genetic fitting algorithm Kiwi-GA \citep{brands2022}. 
The spectral resolution used by \citet{brands2022} was relatively low for quantitative spectroscopy ($\lambda/\Delta\lambda = 5840-7700$ for the optical observations and $\lambda/\Delta\lambda = 1250$ for the UV spectra). For many of the stars, the signal-to-noise ratio is modest (S/N $\sim$ 20--30), however, for the very bright WNh stars that we focus on in this paper the S/N was in the range of 50--70. This is sufficient to obtain constraints on stellar parameters including abundances and wind properties. 
Recently, \cite{sabhahit2025} used the \textsc{PoWR} stellar atmosphere code and constrained the clumping corrected mass-loss rate ($\log \dot{M}= -4.69 $~M$_\odot$~yr$^{-1}$) and spectroscopic mass ($233$~M$_\odot$) of R136a1. They use the luminosity from \cite{bestenlehner2020}, which is 0.07 dex lower than the one found by \cite{brands2022}. 
Indeed, the luminosity determination of these stars is still under some debate. For example, \cite{rd2017} argue that the commonly adopted high luminosities could be overestimated, which would lead to lowering the current mass estimates (see also \citealt{kalari2022}).

In our comparison, we adopt the luminosity, effective temperature, and surface helium abundance of three WNh stars, R136a1, R136a2, and R136a3 from the study of \citet{brands2022}, and summarise the key parameters in Table~\ref{tab:Obs}. Since the listed uncertainties are the formal measurement uncertainties, we opt to increase these values to account for more general, systematic uncertainties when comparing with our models. Therefore, the uncertainty on the luminosity is considered to be $\pm 0.1$~dex. For the uncertainty on the effective temperature, we adopt $5 \%$ of the nominal measurement value, unless the formal uncertainty is larger.

In Figure~\ref{fig:hrd_he}, we compare the three parameters (luminosity, effective temperature, and surface helium mass fraction) between our models and the most massive WNh stars. 
This shows that R136a1 is compatible with a redward evolution of an $\approx$300~M$_\odot$ ZAMS progenitor, considering the \cite{brands2022} measurements. On the other hand, the surface helium abundance obtained by \cite{crowther2010} is notably higher, and would imply a down-stream evolution. Thus the uncertainty in surface helium abundance strongly impacts the evolutionary interpretation of R136a1. 
R136a2 and R136a3 have high surface helium abundances, which disfavours the interpretation that they could lie on the initially 250~M$_\odot$ track and evolve redward. The observed surface helium abundances are robust: \citet{crowther2010}, \citet{bestenlehner2020}, and \cite{brands2022} find no value lower than 0.50 for the surface helium mass fraction of R136a2 and R136a3 (Table~\ref{tab:Obs}). 
%

\subsection{The conundrum}

From the empirical characterisation, a conundrum emerges:
from the three WNh stars, R136a1 is the most luminous but the least helium enriched (according to the analyses of \citealt{bestenlehner2020} and \citealt{brands2022}, c.f. Table~\ref{tab:Obs}). This is in tension with standard stellar evolution theory because the initially most massive star should evolve the most rapidly, and thus be the most helium-enriched. 
Previous inferences yielded the highest initial mass for R136a1 and lower initial masses for R136a2 and R136a3 \citep[e.g.,][]{crowther2010,brands2022}. This is incompatible with the surface helium abundances. 

There are multiple ways to resolve this conundrum, including rapidly rotating stars (Section \ref{sec:5.1}) and stellar mergers (Section \ref{sec:bin}).
Based on our model calculations, we outline another possible resolution. Namely, R136a2 and R136a3 could have been initially more massive than R136a1, undergone a mass-turnover (Section~\ref{sec:masstime}), and currently be observable as less massive, less luminous, but more helium-enriched stars.

\subsection{Interpolated models}\label{sec:interp}

Using the evolutionary models without MLT++ in the TS and SB schemes, we generate interpolated model grids\footnote{The HY scheme is omitted because of the larger numerical noise and convergence problems affecting several models towards the TAMS.}. First we normalise the main sequence time, and map the physical parameters onto the normalised time axis. As demonstrated in e.g. Figures \ref{fig:hrd_wrtr1} and \ref{fig:hrd_he}, only models in their main-sequence phase cover the parameter space compatible with observations. 
Then a dense, high-resolution grid of 20 million data points is obtained by linearly interpolating in time (20,000) and initial mass (1,000). The process is described in more detail in Appendix \ref{sec:app_interp}.

Figure~\ref{fig:interp1} shows one branch of the interpolated models, constructed from the evolutionary tracks in the TS scheme without MLT++, with an initial rotational velocity of 300~km\,s$^{-1}$. 
For clarity, we display the interpolated models in two panels, showing only the models in a redward evolutionary trajectory (left panel) and in a down-stream evolution (right panel). The boundary is determined by the surface hydrogen mass fraction (0.44), corresponding to a surface helium mass fraction of approximately (0.56). This is where the adopted the mass-loss rates begin to follow the Wolf-Rayet type rates, which leads to a drastic change in evolutionary pathways. 
The interpolated models demonstrate a smooth behaviour between the evolutionary tracks and show a continuous coverage of the parameter space.

\subsection{Minimum ages}\label{sec:min_age}
An absolute lower limit on the stellar age can be established by assuming that the measured helium abundance at the surface is the same as the core abundance. This is because in a single star, the surface helium enrichment is a result of the core-produced nucleosynthesis. Figure~\ref{fig:interp2} shows the surface helium mass fraction for the evolutionary models, interpolated models, and observed WNh stars.
Our stellar evolution models with initial masses of 500--800~M$_\odot$ have surface helium abundances that evolve almost in parallel with the core helium abundance. Thus, in these models mass loss is the dominant process since it removes the upper layers of the star, revealing its (near) core composition.
Models with initial masses of 200--500~M$_\odot$ only show surface helium excess with a delay compared to the core composition. This means that rotational mixing still plays a relevant role. However, the enrichment in this case depends on the uncertain efficiency of mixing in the stellar interior (see also, Figure \ref{fig:postMS} for an example). 
Based on these constraints, and considering the lower limits on the observed surface helium abundances, we obtain that the minimum age of R136a1 has to be 0.5--0.8 Myr (from 800 to 500~M$_\odot$ progenitors, respectively), whereas the ages of R136a2 and R136a3 have to be at least 0.9--1.2 Myr (also from 800 to 500~M$_\odot$ progenitors, respectively). For progenitors with initial masses below 500~M$_\odot$, the minimum ages are higher since they require more time to produce the helium excess in their cores, and subsequently reveal it at the stellar surface.

%
%
%
\setlength{\tabcolsep}{1.3pt}
\begin{table*}[]
    \centering
    \caption{Estimated parameters of the WNh stars from the MCMC analysis. \label{tab:Res}}
    \begin{tabular}{l@{\hspace{12pt}} c@{\hspace{10pt}} c@{\hspace{10pt}} c c}
        \hline\hline
         & R136a1 & R136a2 & R136a3 \\ \hline\hline
         Age [Myr] & 1.02 ($\pm 0.09$, $\pm 0.16$)  & 1.45 ($\pm 0.01$, $\pm 0.11$) & 1.48 ($\pm 0.02$, $\pm 0.14$)\\ 
         \hline        
         $M_{\rm curr}$~[M$_\odot$] & 291 ($\pm 34$, $\pm 46$ )  & 195 ($\pm 4$, $\pm 35$) & 184 ($\pm 6$, $\pm 40$)\\ 
         \hline
         $M_{\rm ini}$~[M$_\odot$] & 346 ($\pm 42$)  & $\gtrsim$500 ($\pm 46$) & $\gtrsim$500 ($\pm 53$)\\          
         \hline
    \end{tabular}
            \begin{minipage}{0.9\textwidth}
        \vspace{5pt}
        \small \textbf{Notes.} The first value in the brackets denotes the standard deviation between the solutions (both the $\chi^2$ and Maximum A Posteriori) from the 10 interpolated grids (see text). The second value for the age and current mass denotes the mean of the standard deviations from the individual posterior probability density distributions.
        \end{minipage}
\end{table*}

\subsection{MCMC analysis}\label{sec:mcmc}

We use the Python package \texttt{emcee} to set up a Markov-Chain Monte Carlo (MCMC) framework with uniform priors over the interpolated grid. A Gaussian likelihood is defined based on the combined $\chi^2$ distribution of the three main parameters ($\log L$, $\log T_\mathrm{eff}$, and $\log Y_{\rm surf}$). We examine both the combined $\chi^2$ minima and the Maximum A Posteriori (MAP) models, which are based on the joint posterior probability (jpp) distributions. The procedure is further detailed in Appendix \ref{sec:app_mcmc}. 
Here, we assess how this Bayesian framework, comparing observations with 10 interpolated model grids of two schemes and five initial rotational velocities, allows for estimating the ages and the initial and current masses of the three WNh stars. 
Since the solutions are consistent across the 10 interpolated models grids, i.e., the AM scheme and initial rotation does not significantly impact this quantitive comparison (Appendix Figure \ref{fig:sum} and Tables \ref{tab:mcmc_grid_table_TS} and \ref{tab:mcmc_grid_table_SB}), we report best solutions as the mean values resulting from the individual comparisons to each star. The uncertainty of the age and current mass is considered as both the one resulting from the standard deviation when calculating the mean from the 20 individual solutions ($\chi^2$ minima and highest jpp for 10 interpolated model grids) and the mean of the standard deviations resulting from the individual posterior probability distribution. 
The former one can be understood as a scatter between results obtained from the different grids and statistical interpretations. The latter one is a combined estimate from all posterior probability distributions. These vary slightly between the interpolated model grids (Tables \ref{tab:mcmc_grid_table_TS} and \ref{tab:mcmc_grid_table_SB}).
The initial mass is a derived estimate from the models with the lowest combined $\chi^2$ and highest jpp, and as such, it does not have a posterior probability distribution and a formal uncertainty (see Appendix \ref{sec:app_mcmc} for further discussion). Here, we only estimate the uncertainty as the scatter between the solutions produced by the different grids. Table \ref{tab:Res} summarises our findings. 

%
%
\begin{itemize}
   \item\textbf{{R136a1:}} We consistently find that this star is in a redward evolution, evolving towards lower effective temperatures at almost constant luminosity. The $\chi^2$ distributions have well-pronounced minima and the MAP models provide solutions close to these minima. 

   \item\textbf{{R136a2 and R136a3:}} The down-stream evolution is compatible with the observations of these two stars. The age and current mass are well-constrained. However, in the down-stream channel, models with a large range of initially high masses converge to similar solutions. Although the MCMC analysis selects a formal $\chi^2$ minimum and highest jpp, which are in the range of 600--800~M$_\odot$, the specific values are statistically less robust compared to the case of R136a1. This is because the $\chi^2$ distribution against the initial mass shows a very broad minimum in excess of approx. 500~M$_\odot$ for both R136a2 and R136a3. Therefore, we only estimate the initial mass with a lower limit.
\end{itemize}

%
%
%
%
%
\subsection{Posterior probability predictions for winds and rotation}

\subsubsection{Stellar winds}

For R136a1, the posterior probability distribution of the mass-loss rates (around $\log\dot{M}\sim-4.2$~M$_\odot$yr$^{-1}$) approximately recovers the empirically inferred values (Table \ref{tab:Obs}). In this case, the models employ the WLR to calculate the optically-thin rates.
For R136a2 and R136a3, the mass-loss rates inferred from the posterior probability distributions are systematically, about an order of magnitude, higher than the spectroscopically derived rates by \cite{brands2022}. In these cases, the models apply the Wolf-Rayet type rates. While further investigating and calibrating the model details (e.g., lowering the WR rates, changing the switch condition between optically thin and thick winds) might provide a solution to match the empirically determined mass-loss rates, it is also necessary that the models reproduce the high effective temperatures. This will need to be explored in future works.

\subsubsection{Rotation}

In all cases, our models predict that the strong winds take away sufficient angular momentum that results in low surface rotational velocities after the ZAMS. In a down-stream evolution, the change in stellar radius is modest. The solid-body rotating models spin down the least rapidly since in this case the entire stellar reservoir loses angular momentum. As demonstrated in Section~\ref{sec:grb}, the winds are powerful to significantly deplete this reservoir. In all AM transport schemes and for all initial rotational velocities without MLT++, we find models that are slowly rotating ($v_{\rm rot} \leq 50$~km\,s$^{-1}$) after $\sim$1~Myr. 

The observationally inferred \vsini\ values are non-negligible (though within uncertainty, they are compatible with slow-rotation). However, the role of macroturbulence has not been well-determined. Therefore we propose that future efforts should focus on disentangling the components of line broadening to help distinguish between evolutionary scenarios.

\subsection{Summary}

Our model comparison with the three WNh stars shows that self-consistent solutions are possible with single-star models, when relying on three spectroscopically inferred quantities.
In particular, we use the surface helium abundances, which have a strong constraining power. This is because in the single-star channel, the helium observed at the surface, has to be produced in the core of the star. 
For R136a2 and R136a3 the surface helium abundances are higher than the amount produced in the core of an initially $\approx$250~M$_\odot$ model during the redward evolution. This rules out such progenitors within $\approx$~1.5 Myr of evolution, when using our modelling assumptions.
Based on our detailed comparison we find approximately the same ages for R136a2 and R136a3 (1.45, 1.48 Myr) and argue that both of these stars are undergoing a down-stream evolution, rapidly decreasing their mass and luminosity. In this process they reveal their helium-enriched, near core composition. On the other hand, R136a1 is the youngest star (1.02 Myr) out of the three WNh stars and we infer that it is still in a redward evolution.
It is important though that this result strongly depends on the measured helium abundance, which is the least consistent among various authors for R136a1. 
Based on our model calculations and MCMC analysis, we find that while currently R136a1 is the most massive star, initially both R136a2 and R136a3 could have been more massive. We estimate that R136a2 and R136a3 could originate from ZAMS progenitors of $\gtrsim$500~M$_\odot$.


%
%
%
%
\section{Discussion}\label{disc}

%
%
\subsection{Previous works on VMS evolutionary modelling at LMC metallicity and main uncertainties}\label{sec:5.1}

We briefly outline previous works that computed VMS evolutionary models and address some of the physical and technical complexities of modelling VMSs. While VMS models are much studied, especially at galactic and (near-)zero metallicity, we only focus on models computed for an LMC metallicity so that we can make some direct comparisons with our study. 

%
%
\cite{yusof2013}, \cite{eggenberger2021}, and \cite{martinet2023} used the Geneva stellar evolution code to compute VMS models at LMC metallicity. These models account for an advecto-diffusive scheme in the angular momentum transport and do not allow for density inversions in the stellar envelope. 
\cite{martinet2023} demonstrate that at LMC metallicity, VMS models do not maintain rapid rotation on the main sequence and (P)PISN progenitors may only be expected from initially slowly-rotating models. Our study is in agreement with these results (Figure~\ref{newfig:cocoremass}).
\citeauthor{martinet2023}
find that the abundances of R136a1, R136a2, and R136a3 are challenging to explain with models up to $M_{\rm ini} = 300$~M$_\odot$. Our best-matching solutions indeed imply higher initial masses.

%
%
\cite{martins2017,martins2022} used the \textsc{starevol} code \citep{decressin2009} to study VMS evolution. They considered non-rotating models and found evolutionary trajectories that proceed redward on the HRD, resulting in inflating envelopes. The mass-loss implementation follows the work of \cite{graefener2021}, utilising the \cite{vink2001} rates for optically thin winds, and an empirical fit by \cite{bestenlehner2014} for optically thick winds; the switch between the two determined by the sonic point optical depth.
Due to strong mass loss, their models become helium enriched regardless of rotation. However, an important difference is that the redward evolution in their study results in effective temperatures that are much lower than inferred for the WNh stars in R136. Therefore, our study is in agreement in terms of utilising higher, optically-thick winds to explain chemical enrichment, but in disagreement in terms of the general evolutionary trajectories in the HRD. 

%
%
The PARSEC code \citep{bressan2012} has been utilised to compute grids that include VMSs at LMC metallicity \citep{costa2021,costa2025}. Their mass-loss implementation follows the \cite{vink2001} rates for optically-thin winds, and a luminosity-dependent scaling from \cite{sander2019} for optically-thick winds. The switch between the two is considered as a function of the surface hydrogen abundance (\citealt{costa2025} use $X<$0.3, whereas we use $X<$0.4.). Figure 12 of \cite{costa2025} shows the observed stars in R136 with their evolutionary tracks. This reveals almost horizontal evolutionary trajectories, with VMSs evolving redward to the Red Supergiant phase, and then undergoing a blueward evolution. This highlights that both the early mass-loss rates and the details to switch to higher rates strongly impact the model predictions.

%
%
\cite{kohler2015} used the Bonn evolutionary code and extended the \cite{brott2011} grid to VMSs. Their evolutionary models assume diffusive angular momentum transport (similar to our study) and extrapolate the \cite{vink2001} mass-loss rates for optically-thin winds in higher-mass VMSs. Insofar, these tracks have been most widely used to derive physical parameters of the WNh stars in R136 \citep[e.g.,][]{bestenlehner2020,brands2022}. For the best-matching solutions, their models maintain a high rotational velocity for $\sim$~1 Myr. Thus the chemical enrichment and high effective temperatures are mostly associated with quasi-chemically homogeneous evolution due to rapid rotation. However, we find that rotation always becomes slow, thus physically mass loss is the driving mechanism that leads to these effects. 
This difference in rotational velocities, to a large extent, originates from the numerical treatment of angular momentum removal. While \cite{kohler2015} subtract the specific angular momentum loss from the total reservoir, we only remove specific angular momentum from a rich, near-surface reservoir (Section \ref{sec:amloss}). 
Therefore the initial masses we require for R136a2 and R136a3 are systematically and significantly higher than those inferred from the \cite{kohler2015} models.  

%
%
Several authors used the \textsc{mesa} code to study VMS models. \cite{choi2016} computed a large, general grid (i.e., not specifically tailored for the physics of Very Massive Stars), and utilised the \cite{vink2001} and \cite{nugis2002} rates for optically thin and thick winds, respectively. Since they adopt the MLT++ routine, their models are generally not consistent with the effective temperatures measured for the WNh stars in R136 (as we demonstrate in our Section~\ref{sec:ms}). However, the feature of a down-stream evolution is also seen in their tracks. 

\cite{graefener2021} also used \textsc{mesa} models with similar modelling assumptions as \cite{abel2019}. The mass-loss rates are given by the \cite{vink2001} rates for optically-thin winds, whereas an empirical scaling is adopted for optically-thick winds from \cite{bestenlehner2014}. This combination results in a redward evolution, similar to the study of \cite{martins2022}. Higher optically-thick mass-loss rates are required to produce a down-stream evolution from VMS models (their Figure 11). \cite{graefener2021} also tested models with and without MLT++. Our study is consistent in terms of models with MLT++ evolving at higher effective temperatures. However, the main evolutionary trajectories are rather different, likely due to various choices in numerical treatments and mass-loss implementation.

\cite{higgins2022} studied non-rotating \textsc{mesa} models, finding that strong mass loss leads to a down-stream evolution in luminosity, whereas lower rates allow for a redward trajectory in the HRD (their Figure 3). \cite{higgins2022} identify the mass turnover to be at $\sim$~1.7 Myr. Our results agree qualitatively; however, we find that the way initially more massive stars become less massive than initially less massive ones is dependent on initial rotation and AM transport scheme (Figures \ref{fig:massevol} and \ref{fig:massevol_app}). In our fiducial scheme, we predict the turnover to begin at $\sim$~1.4 Myr for models in excess of $M_{\rm ini} \sim 400$~M$_\odot$. 

Recently, \cite{sabhahit2022,sabhahit2023} scrutinised the transition from optically thin to optically thick winds in VMS models, using \textsc{mesa}. Similar to \cite{graefener2021}, the transition is considered to depend on the optical depth of the sonic point. In agreement with \cite{higgins2022}, \cite{sabhahit2022} also find that utilising the canonically-considered \cite{vink2001} rates for O-type stars leads to a redward evolution, whereas stronger mass-loss rates result in a down-stream channel. In their study, they utilise observational results from \cite{bestenlehner2014}. Considering the newer studies of \cite{bestenlehner2020} and \cite{brands2022}, we believe that their down-stream evolution models would also provide quantitatively similar conclusions to our study. Namely, that the initial mass of the WNh stars can potentially be much higher than previously considered.

%
%
Physical processes, such as convection, magnetism, rotation, and mass loss are key to properly investigate the nature of VMSs. Their inferences rely on various diagnostics that often yield only indirect constraints. A necessary simplification in stellar evolution codes is the adaptation of appropriate parametric prescriptions to account for inherently three dimensional phenomena. However, we emphasise that stellar evolution codes often differ in detailed numerical treatments. Therefore, even when similar prescriptions are adopted (e.g., same mass loss and AM coupling scheme), the exact numerical treatment (e.g., how many layers are ascribed to lose mass and angular momentum) may differ considerably. This can ultimately result in significant differences between models computed with various codes that may be difficult to trace and quantify. Finally, physical processes considered in stellar evolution models often act on different timescales. The numerical hydrodynamic solvers, in principle, can reach convergence with time steps appropriate for the given problem. Nonetheless, choosing these time steps can be complex and indeed, it has been demonstrated that resolution can affect modelling results \citep{lau2014}.

In summary, while all previous studies agree in the importance of stellar winds for VMSs and several scenarios have been explored with different numerical choices, the high effective temperatures in combination with strong surface helium enrichment has received less scrutiny. This is nonetheless essential to explain the nature of the WNh stars in the R136 cluster.

%
%
\subsection{A single or binary nature of R136a1, R136a2, R136a3 }\label{sec:bin}

Confronting observations of R136a1, R136a2, and R136a3 to our models establishes that the full realm of properties of the WNh stars can be reconciled with single-star evolution within observational uncertainties.
While dedicated surveys show that binarity is common amongst typical O and B-type stars \citep[e.g.,][]{almaida2017,villasenor2021,shenar2022}, 
recently, both \cite{kalari2022} and \cite{shenar2023} found a lack of evidence for close companion stars in the case of R136a1, R136a2, and R136a3. Along with earlier inferences of \cite{crowther2010,crowther2016}, this supports the hypothesis that these stars are single.
If the WNh stars in R136 were merger products, we may expect that their inferred ages would be younger than those of other stars in the cluster, the characteristics of rejuvenation being a function of time of merging \citep{schneider2014}. However, R136 is a very young cluster and invoking, at least, main-sequence stellar mergers does not seem required to explain the observed characteristics. In fact, main-sequence stellar mergers would further aggravate the helium problem. For example, if a star with current mass of $\approx$200~M$_\odot$ was formed via the equal-mass merger of two initially $\approx$100~M$_\odot$ stars, then the timescale to produce the excess helium would take much longer because the fusion in the core of a 100~M$_\odot$ star is much slower. In our fiducial setup without MLT++, the initially 100~M$_\odot$ model takes 1.59 (2.07) Myr to produce a core helium mass fraction of 0.45 (0.6). 
If the merger was unequal mass, then the lower-mass companion would certainly not have a high helium abundance to enrich the surface of the primary star.
Nonetheless, the formation of stars in excess of initially $\sim 300$~M$_\odot$ in the local Universe is an unresolved problem.


%
%
\subsection{The formation of stars of $M_{\rm ini} > 300\,M_{\odot}$}

Whether stars like R136a1, R136a2, and R136a3 form from a single molecular cloud filament or are the result of a series of proto-stellar mergers is still not fully understood \citep[e.g.,][]{bonnell2002,zinnecker2002,zinnecker2007,vergara2025}. \citet{kuiper2018} performed hydrodynamical simulations and showed that if a sufficient amount of gas can be concentrated in a small enough volume, stars of several hundred solar masses may form. This suggests that nature merely needs to produce the right initial conditions for VMS formation to occur in the process of massive cluster formation.
\cite{fukui2017} argue that, in analogy to molecular cloud-cloud collisions, the tidal interaction between the LMC and SMC created a large-scale atomic gas flow that led to a strongly compressed region, which evolved to become R136. Formation of stars in excess of 100\,M$_{\odot}$ may then have taken place in very high-density hubs of filaments in such a compressed region \citep[e.g.,][]{fukui2015,kumar2020}. It seems vital that magnetic pressure supports these structures against gravitational collapse on the cloud scale \citep[e.g.,][]{hwang2022,badmaev2022} and also against fragmentation on the core scale (\citealt{ commercon2011,commercon2022, machida2008,mignonrisse2021,oliva2022}, although see, e.g., \citealt{harada2021}). Remaining puzzles about the exact formation of VMSs involve the uncertainty of the mass accretion rate \citep[e.g.,][]{haemmerle2016,haemmerle2021}. 

The assembly of R136 is only a recent event in the star formation history of the larger 30 Doradus Nebula (which, projected in the sky, measures about $150 \times 200$\,pc). 30 Doradus is likely undergoing multiple bursts of star formation \citep[e.g.][]{schneider2018b,khorrami2021}, extending over several tens of Myr as witnessed by the 20--25\,Myr old clusters Hodge\,301, SL\,639, and NGC\,2100 \citep{britavskiy2019}. The burst that formed R136 possibly resulted in the formation of two sub-clusters, with a second more diffuse clump, currently at some 5\,pc to the north-east, forming about a Myr later \citep{sabbi2012,stoop2024}. This may be in line with star clusters being the final products of the hierarchical merger of smaller sub-structures \citep{bonnell2003,bate2009,federrath2010}. The somewhat older population surrounding R136 \citep[of ages 3--7\,Myr;][]{cignoni2015} may too be related to R136, in that self-regulation from massive stars in this older part triggered the formation of R136 \citep{dominguez2022,fahrion2024}. 

Whether this complex history of star formation in 30 Doradus somehow contributed to shaping the conditions for currently up to 300\,M$_{\odot}$ stars to be formed in the central cluster remains to be investigated \citep{castro2018}. This may have been essential, as theoretical works indicate that stellar masses are not stochastically sampled from the IMF \citep{yan2022}. That is, many small clusters do not yield the same IMF as a single massive cluster of the same combined number of stars. This may explain the finding that nearby dwarf galaxies show top-light IMFs, i.e., the most massive stars in their star-forming regions are substantially lower in mass \citep[at $\sim$ 40--60 \Msun;][]{pflamm-altenburg2007,pflamm-altenburg2009} compared to those in R136. 

An important implication from this work and other studies addressing the evolution of VMSs in R136 is that if the clusters that are studied are older than 2.5\,Myr, stars of hundreds of solar masses will have evolved off from the main sequence. Our models predict that they should be identified as very hot and luminous helium-rich stars with present-day masses below 50~\Msun if their initial mass was in excess of 200~\Msun. However, this phase lasts for only about 0.5 Myr until core-collapse. This underpins the significant uncertainty of inferring IMFs from present-day mass functions in young clusters, with potentially large implications for galaxy evolution \citep[e.g.,][]{goswami2022,martins2023,upadhyaya2024}, and highlights the importance of large-scale programs aiming to address the upper-mass-limit problem from a star-formation perspective \citep[e.g.,][]{motte2022}.


%
%
\subsection{Stellar feedback}

Our scenario implies that both R136a2 and R136a3 could have lost hundreds of solar masses into the surrounding ISM. Ejecting 300 M$_\odot$ with a wind terminal velocity of 3500 km\,s$^{-1}$ yields a kinetic energy of $\approx 10^{52}$ erg, which is an order magnitude larger than the kinetic energy of a typical core collapse supernova. This opens the questions whether the ejected mass or the energy output from the WNh stars could be observed in 30 Dor.
Unlike supernovae, which release energy and mass in an instant, the stellar feedback from the WNh stars would be much more gradual over $\approx$~1.5 Myr. During this time, the energy injected to the ISM can thermalise and the mass can disperse over the scales of up to a 100 pc \citep[e.g.,][]{garciasegura1996,mackey2015,vanmarle2015,haid2018, meyer2020,geen2023,lancaster2025}. In this time frame, feedback from other stars, which are less massive but more numerous, will also affect the same region. 
Observational studies show evidence of large-scale stellar feedback in 30 Dor \citep[e.g.,][]{wong2022, grishunin2024}. However, disentangling the contribution from the WNh stars may be challenging. To test our evolutionary scenario based on the ISM properties, further observations and dedicated stellar feedback studies are necessary.

%
%
%
%
%
\section{Conclusions}\label{conc}

In this work, we compute and analyse a new grid of stellar evolution models for an LMC environment, adopting empirical mass-loss rates from state-of-the-art spectroscopic analysis of \citet{brands2022} and theoretical Wolf-Rayet type rates from \cite{sander2020b}. We confront our models with the three most massive stars in the R136 cluster in 30 Doradus in the LMC; stars with present-day masses between 150--300\,\Msun. This provides the most direct and stringent constraints and predictions on the evolution of VMSs. 
Our main findings are:

\begin{enumerate}
    \item During the main-sequence evolution of VMSs, assumptions on internal angular momentum transport which may be characterised by moderate, strong, and very strong coupling between the core and the envelope, have little effect on observable properties. This hampers investigations into this transport. However, the subsequent, post-main sequence evolution strongly depends on it.
    \\
    \item Assumptions on angular momentum transport impact the \emph{internal rotational} properties (i.e., CO core mass and core rotation), with consequences for supernova progenitors. For models with initial masses up to 400~M$_\odot$, the initial rotational velocity from (100 to 500~km\,s$^{-1}$ as explored here) plays a significant role in determining the pre-supernova evolution. Our models using the Tayler-Spruit and solid-body coupling schemes avert initially very rapidly spinning stars away from a possible gamma ray burst ending, whereas GRB progenitors are possible from models with initial masses between 100 and 200~M$_\odot$ using the hydrodynamical coupling scheme. The \emph{surface rotation} brakes on a short timescale, therefore we argue that chemical enrichment in VMSs is primarily the result of strong mass loss revealing the near core composition. 
    \\ 
    \item We discuss that strong mass loss from VMSs can produce helium-enriched stars if $M_{\rm ini} \ga 200$\,\Msun. These objects shed so much mass already on the main sequence that they follow a down-stream evolution in luminosity ($\log (L/{\rm L}_{\odot}) \sim$6.1--6.3) to become hot (50--60\,kK), helium-rich stars with masses of 25--50~M$_\odot$. We expect that they will avoid the pair instability regime. However, changing the mass-loss rates by only a factor of two (which is the uncertainty of the currently best determinations for optically thin winds), the quantitive results would drastically change regarding the pre-supernova model predictions. For example, in our fiducial model, it would lead to doubling the stellar mass at core-collapse. We cannot make firm predictions whether the models would produce a supernova; however, if they did, it would be of Type Ib/c.
    \\ 
    \item We detail the impact of the mass turnover: initially more massive stars will lose mass more rapidly and become less massive than initially less massive ones. For example, in our fiducial scheme at an age of 2 Myr, an initially 600~M$_\odot$ star will have a current mass of 50~M$_\odot$, whereas an initially 200~M$_\odot$ star will have a current mass of 150~M$_\odot$. In fact, by the time the first supernova takes place in a cluster, it seems very difficult to find stars in excess of 150~M$_\odot$ -- unless, of course, star formation is not coeval. This significantly impacts studies that aim to constrain the IMF from present day mass and luminosity functions that, even in the best cases, are typically considered in a few Myr old clusters. 
    \\ 
    \item Previous observations lead to a conundrum about the evolutionary history of the three WNh stars. R136a1, is the currently most luminous, but least helium enriched. We propose that the resolution to this is that both R136a2 and R136a3 were initially more massive, shed a large amount of mass, and became less massive than R136a1. 
    \\
    \item We produce interpolated models from our evolutionary model grid, and by means of MCMC analysis, we confront our model predictions with spectroscopically inferred quantities (luminosity, effective temperature, and surface helium abundance) of the three most massive stars known, R136a1, R136a2, and R136a3. We find matching solutions in the single-star channel. We argue that the surface helium abundances, measured by three authors independently, provide robust constraints 
    on the evolutionary history of the WNh stars. For R136a2 and R136a3, we find that the best-matching models are those in a down-stream evolution, implying initial masses of $\gtrsim$500~M$\odot$. 
    \\
    \item In the down-stream evolutionary scenario, the rate of the decrease in luminosity is approximately 1 dex / Myr. This would roughly correspond to a 25 ppm increase in the visual magnitude over 10 years. At the same time, we expect no significant changes in the effective temperature, contrary to a redward evolutionary channel, in which T$_{\rm eff}$ would continue to decrease. Therefore, our study strongly supports the long-term photometric and spectroscopic monitoring of these stars and consistent parameter determination by means of stellar atmosphere codes. 
    \\
    \item For R136a1, we find compelling solutions in a typical, redward evolutionary trajectory, implying an initial mass between 300 and 400~M$_\odot$. We strongly recommend revisiting the helium abundance measurement since it is the least consistent for this star and it plays a large role in this interpretation. We infer that R136a1 is somewhat younger than R136a2 and R136a3, with estimated ages of \mbox{1.02$\pm0.16$,} 1.45$\pm0.11$, and 1.48$\pm0.14$ Myr, respectively. These age estimates are broadly consistent with previous studies \citep{lennon2018,brands2022}, and would also suggest that R136a2 and R136a3 had more time to enrich their surfaces with helium.
    \\ 
\end{enumerate}

These findings underpin the extraordinary role of mass loss and rotation in shaping the lives of Very Massive Stars, their fates as supernova progenitors, and their contributions to the chemical and dynamical evolution of their host environments. By confronting our new theoretical model predictions with state-of-the-art spectroscopic analyses, we find that single-star models can explain the most massive stars known in the Universe. This work not only advances our understanding of VMSs but also opens new avenues to explore their puzzling formation mechanism and their role of chemical enrichment on galactic scales.

%
%
%
%

%
%
%
%
%
\section*{Data availability}\label{data}

A full Reproduction Package is shared on Zenodo at \mbox{\url{https://zenodo.org/records/15702085}}. This record contains the computed stellar structure and evolution models, the template files, as well as, the shell and python scripts.

\begin{acknowledgements}

We thank the anonymous referee for the constructive report, which helped us to further improve the manuscript.
We thank the \textsc{mesa} developers for making their code publicly available. 
We appreciate helpful discussions with Fabian Schneider, Kaila Nathaniel, Masato Kobayashi-san, Takashi Ito-san, Takashi Moriya-san, Tomoya Takiwaki-san, Koh Takahashi-san, and Ylva G\"otberg.
This work made use of the Dutch national e-infrastructure with the support of the SURF Cooperative using grant no. EINF-1050 and EINF-2818. 
Numerical computations were furthermore carried out on the general-purpose PC cluster at the Center for Computational Astrophysics, National Astronomical Observatory of Japan.
This publication is part of the project `Massive stars in low-metallicity environments: the progenitors of massive black holes' with project number OND1362707 of the research TOP-programme, which is (partly) financed by the Dutch Research Council (NWO). 

\end{acknowledgements}  

\bibliographystyle{aa} 
\bibliography{biblio.bib} 

\begin{appendix}

%
%
\section{Mass evolution}\label{sec:app1}

\begin{figure*}
    \centering
    \includegraphics[width=0.3\textwidth]{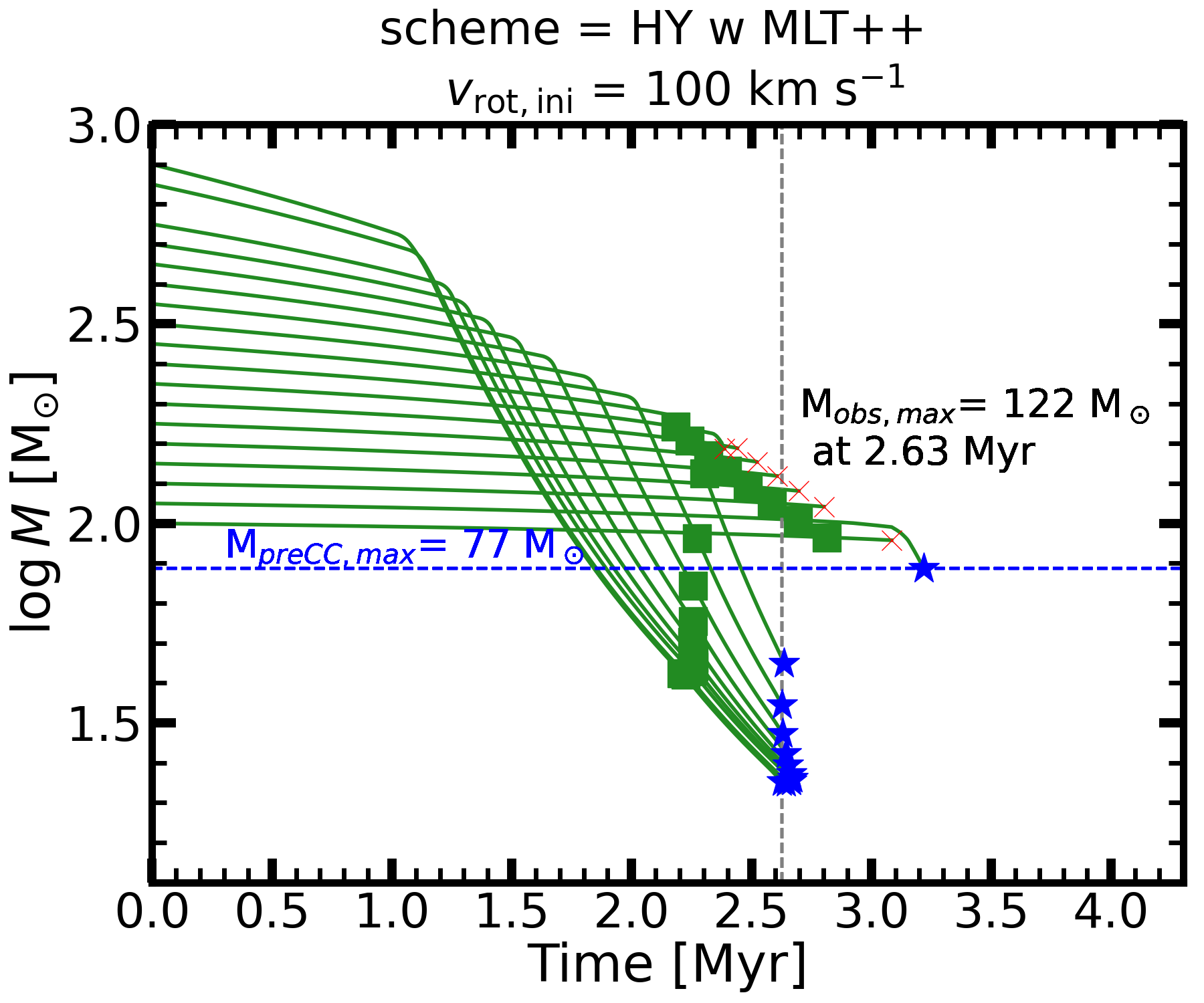}\includegraphics[width=0.3\textwidth]{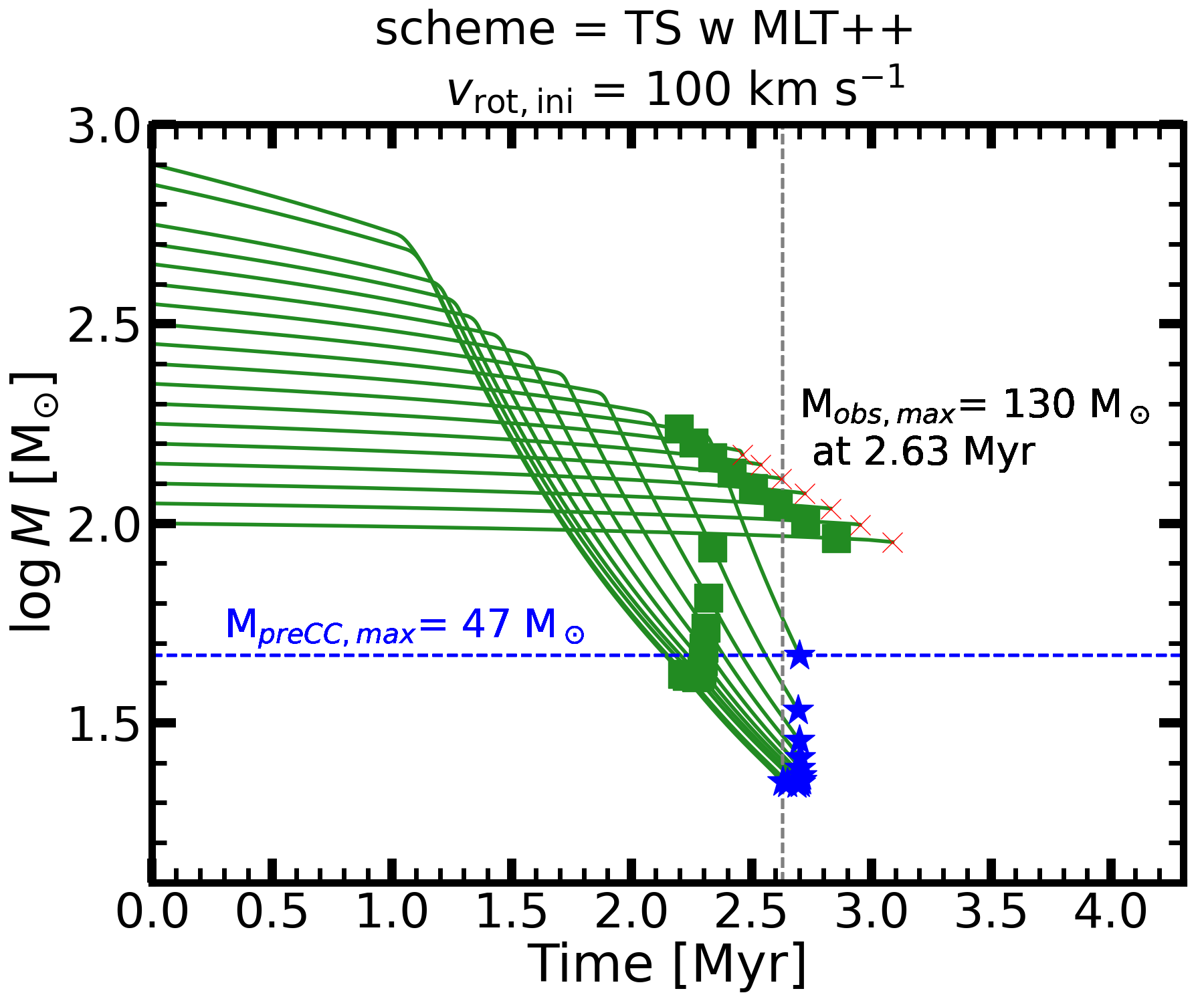}\includegraphics[width=0.3\textwidth]{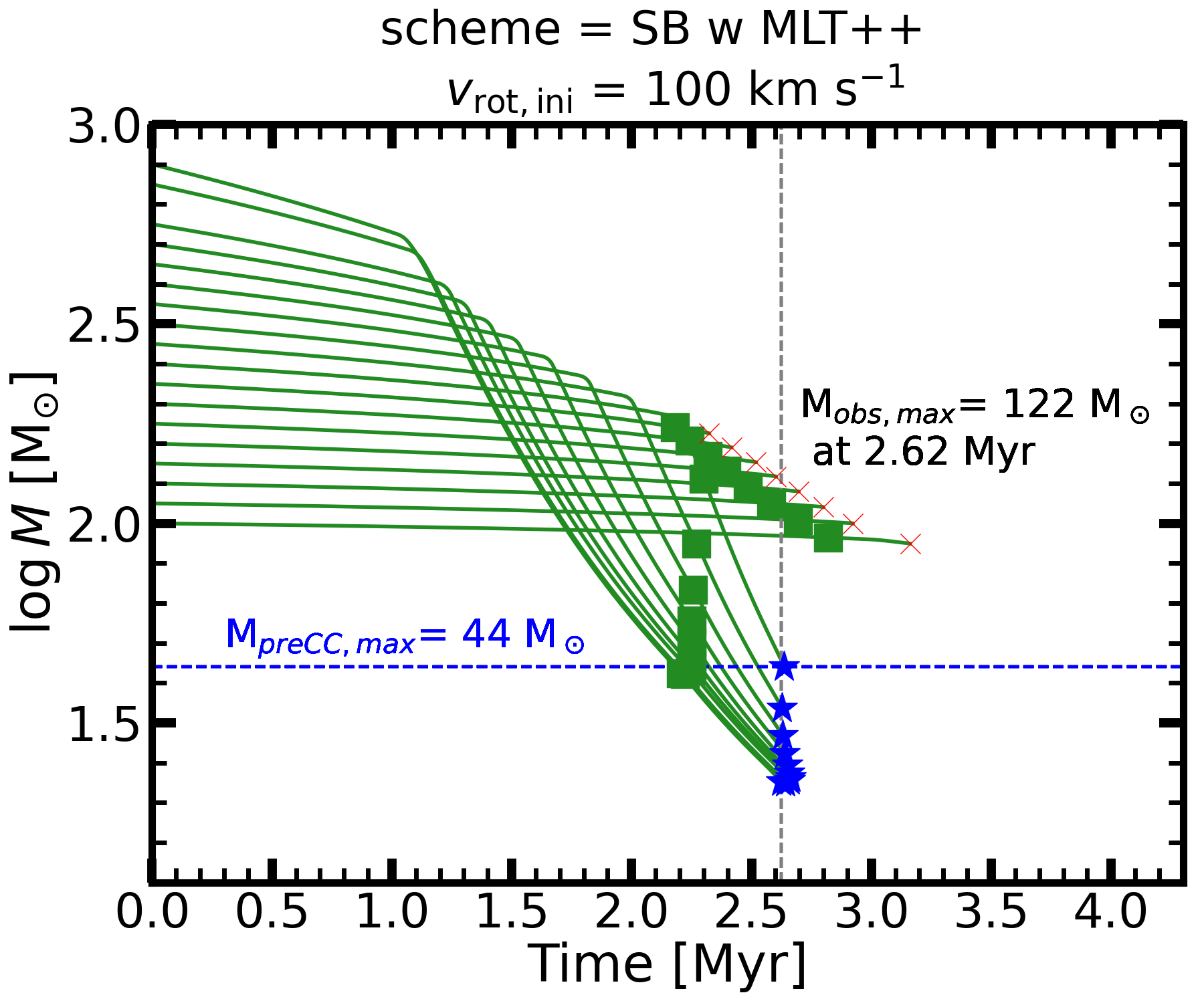}
    \includegraphics[width=0.3\textwidth]{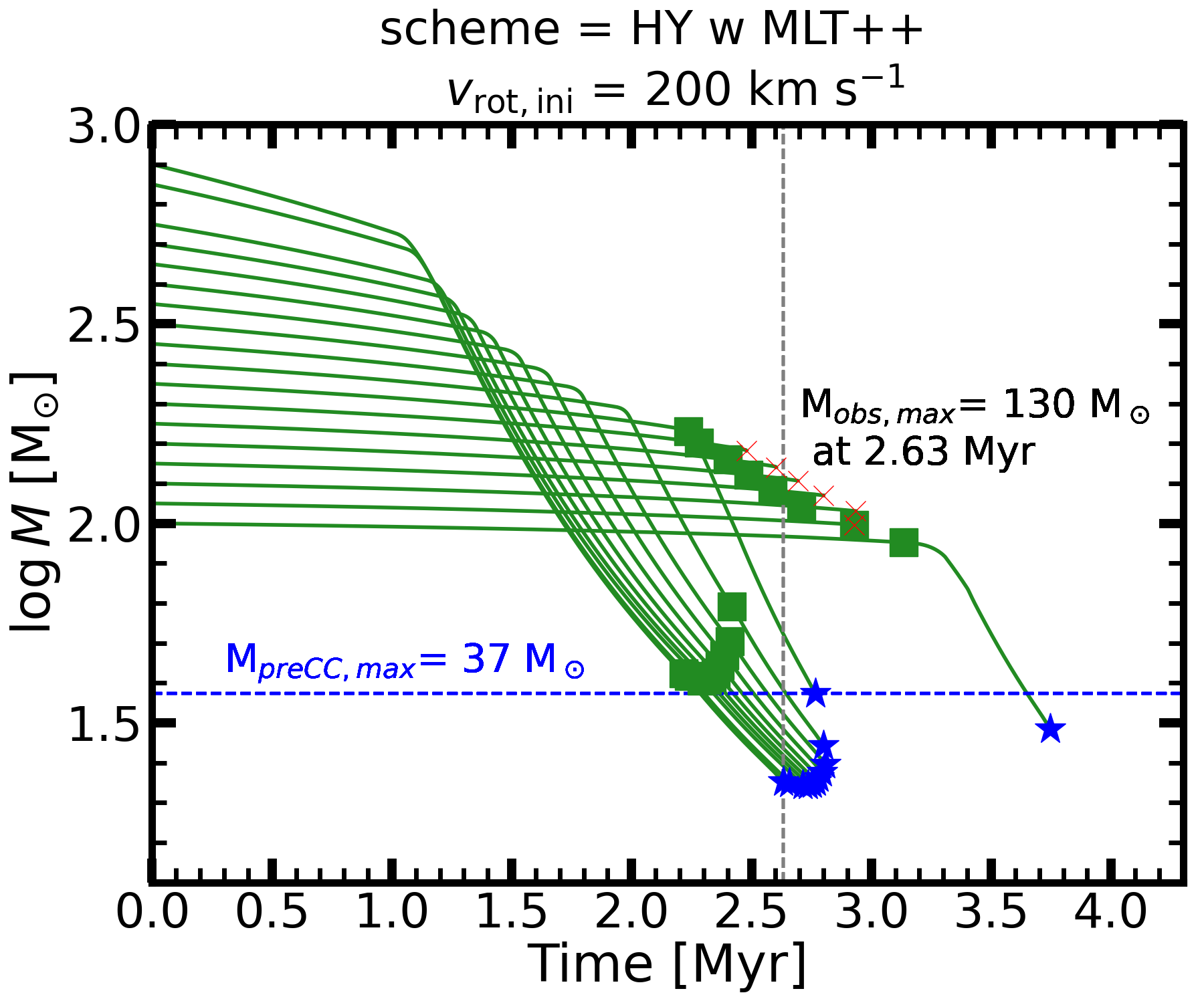}\includegraphics[width=0.3\textwidth]{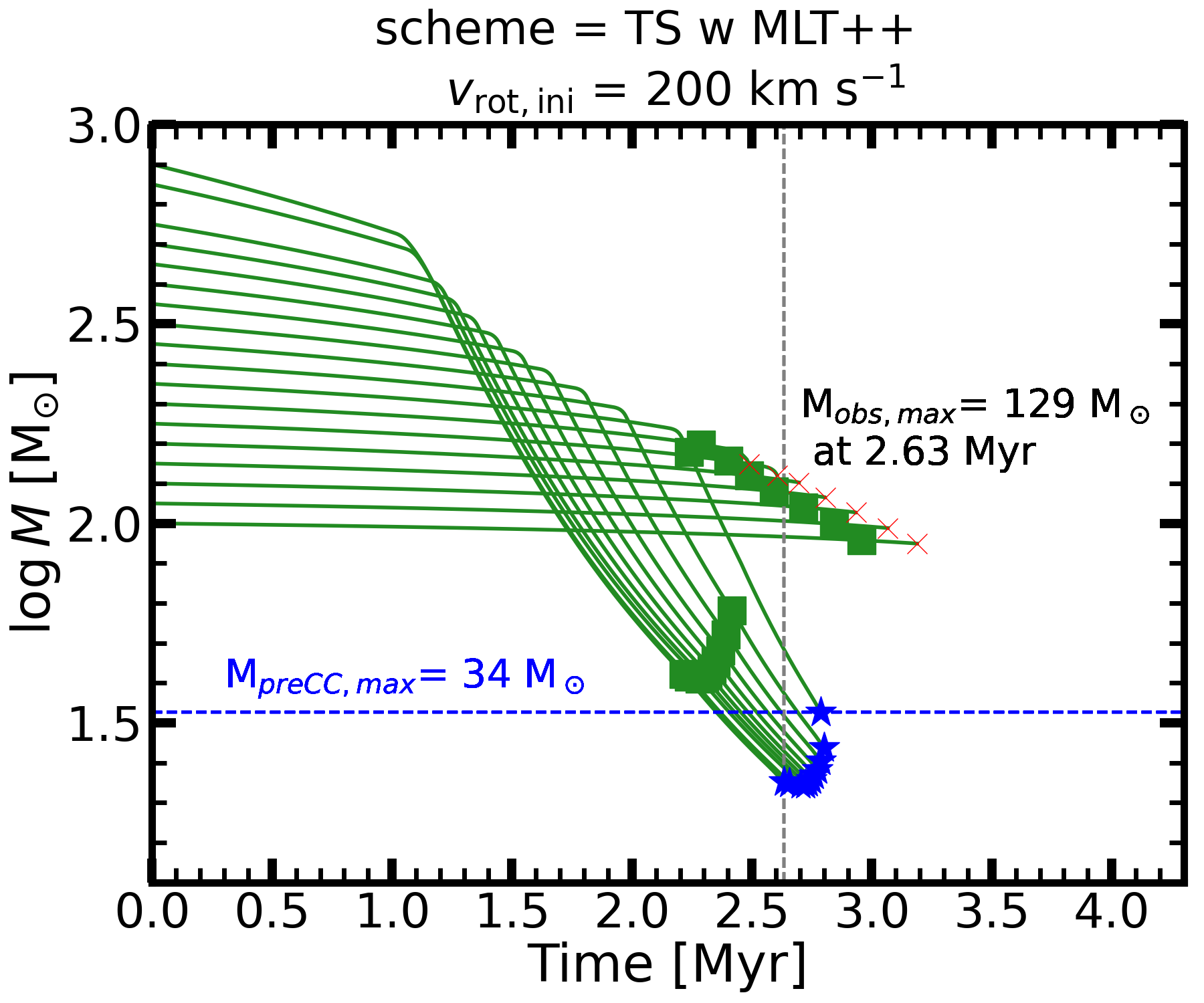}\includegraphics[width=0.3\textwidth]{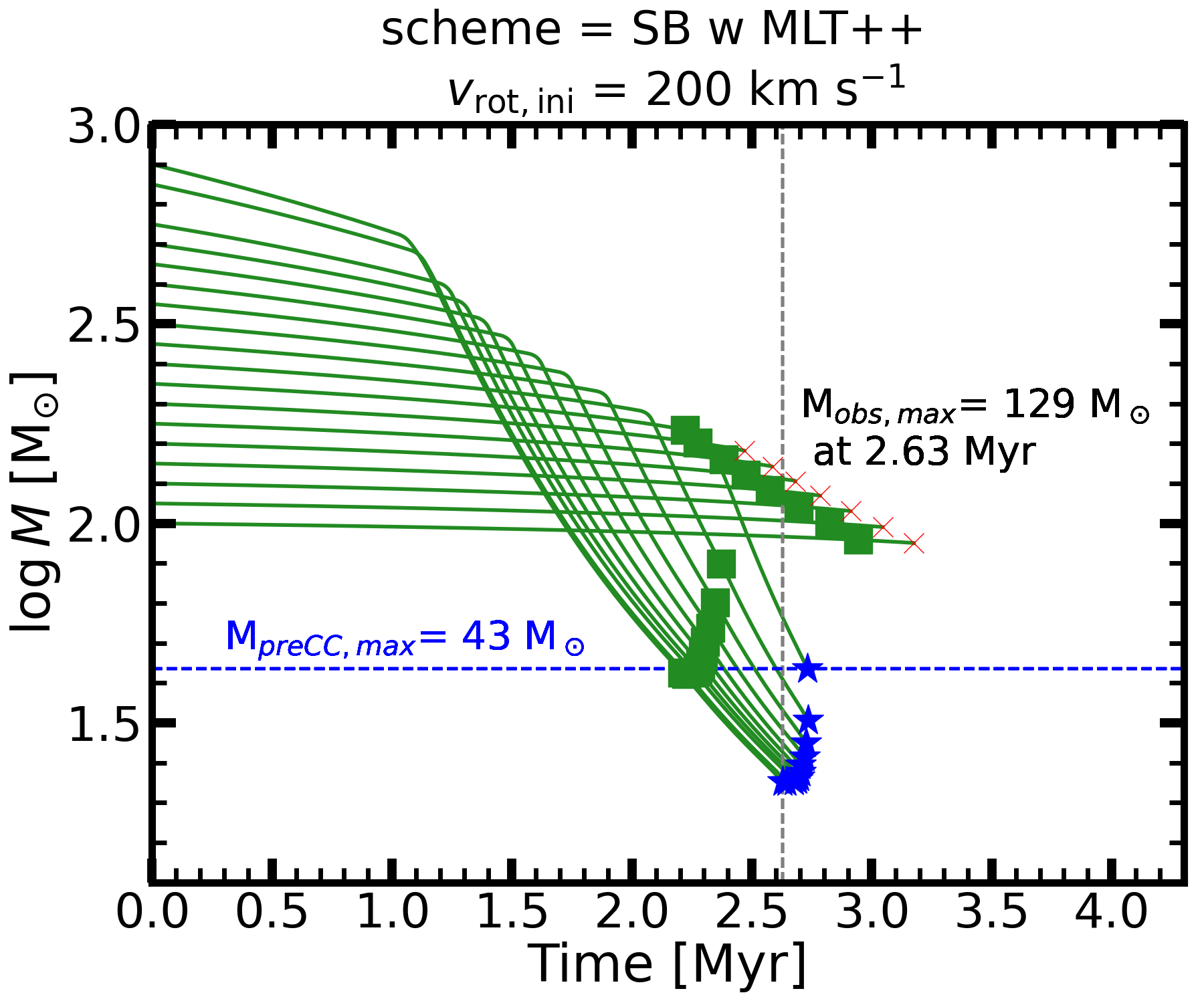}
    \includegraphics[width=0.3\textwidth]{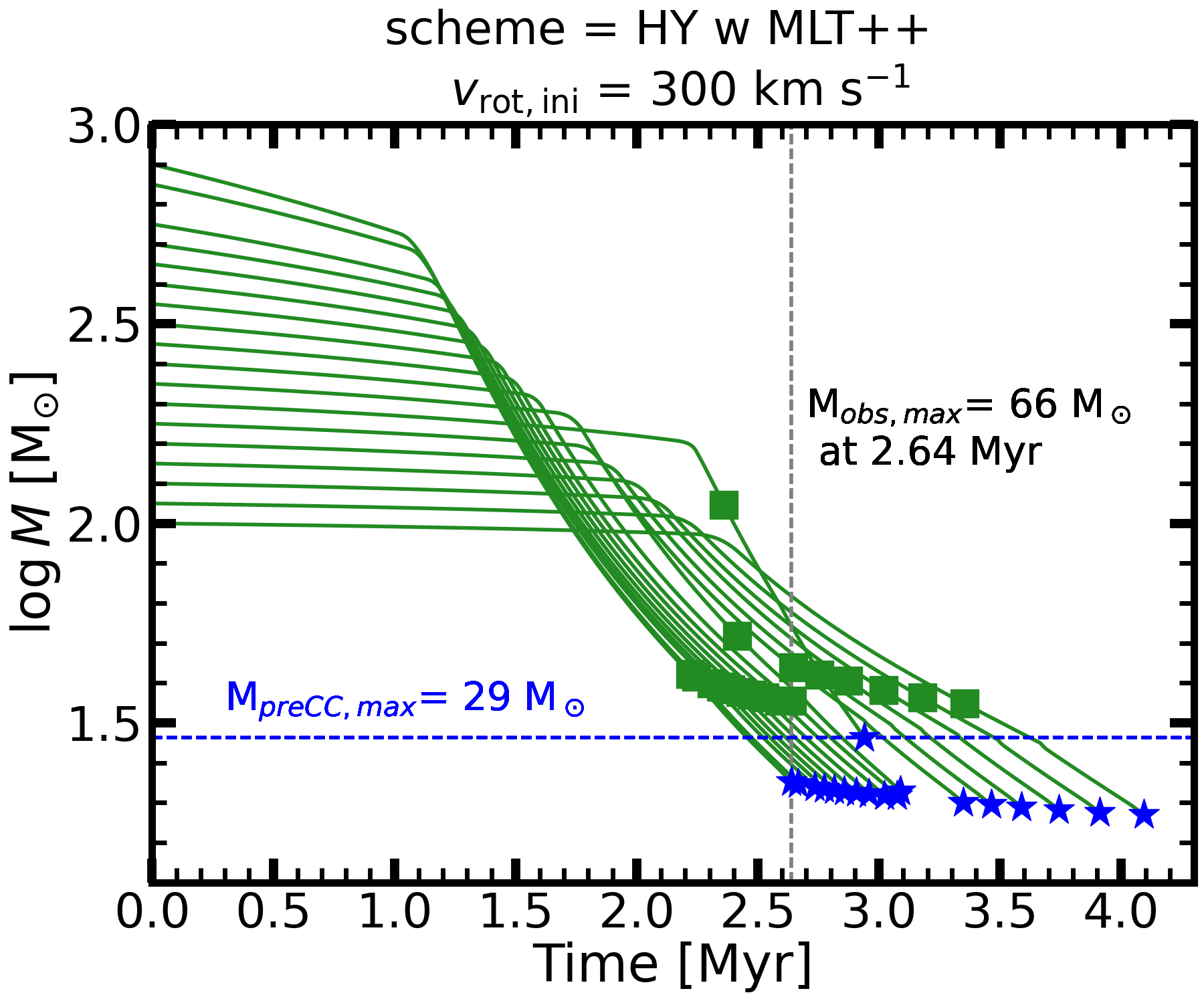}\includegraphics[width=0.3\textwidth]{fig/3_mass/mass_TS_300.png}\includegraphics[width=0.3\textwidth]{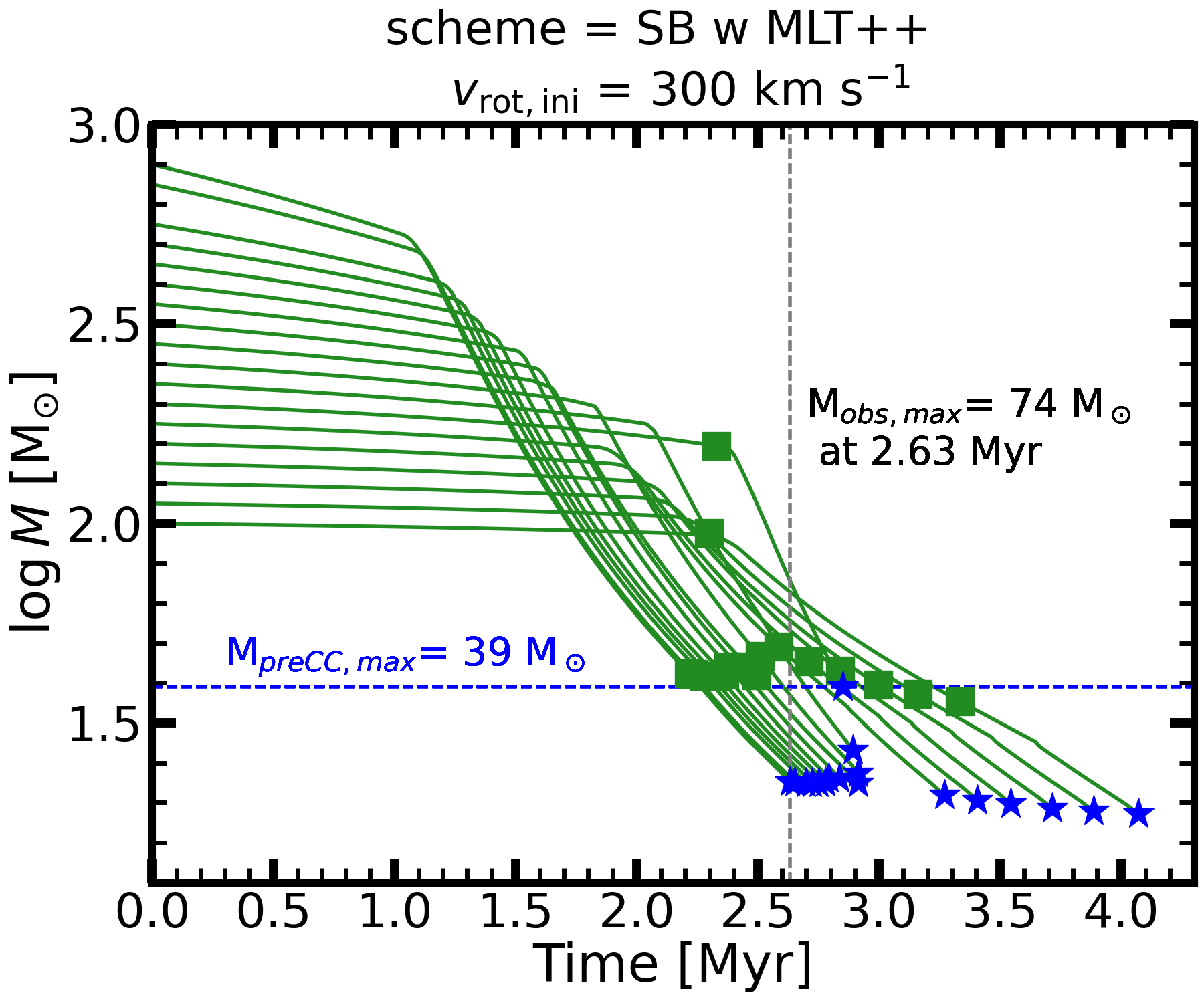}
    \includegraphics[width=0.3\textwidth]{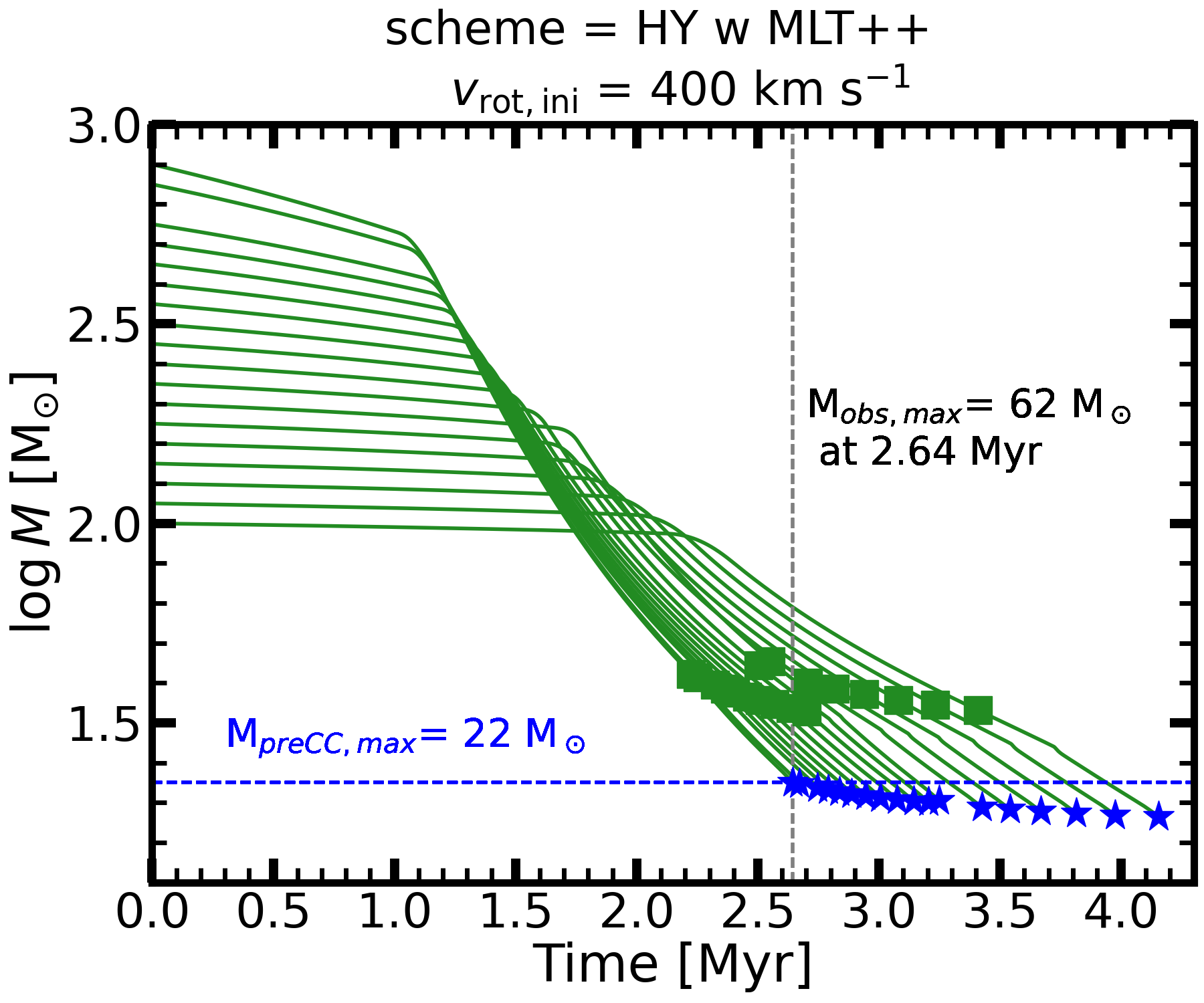}\includegraphics[width=0.3\textwidth]{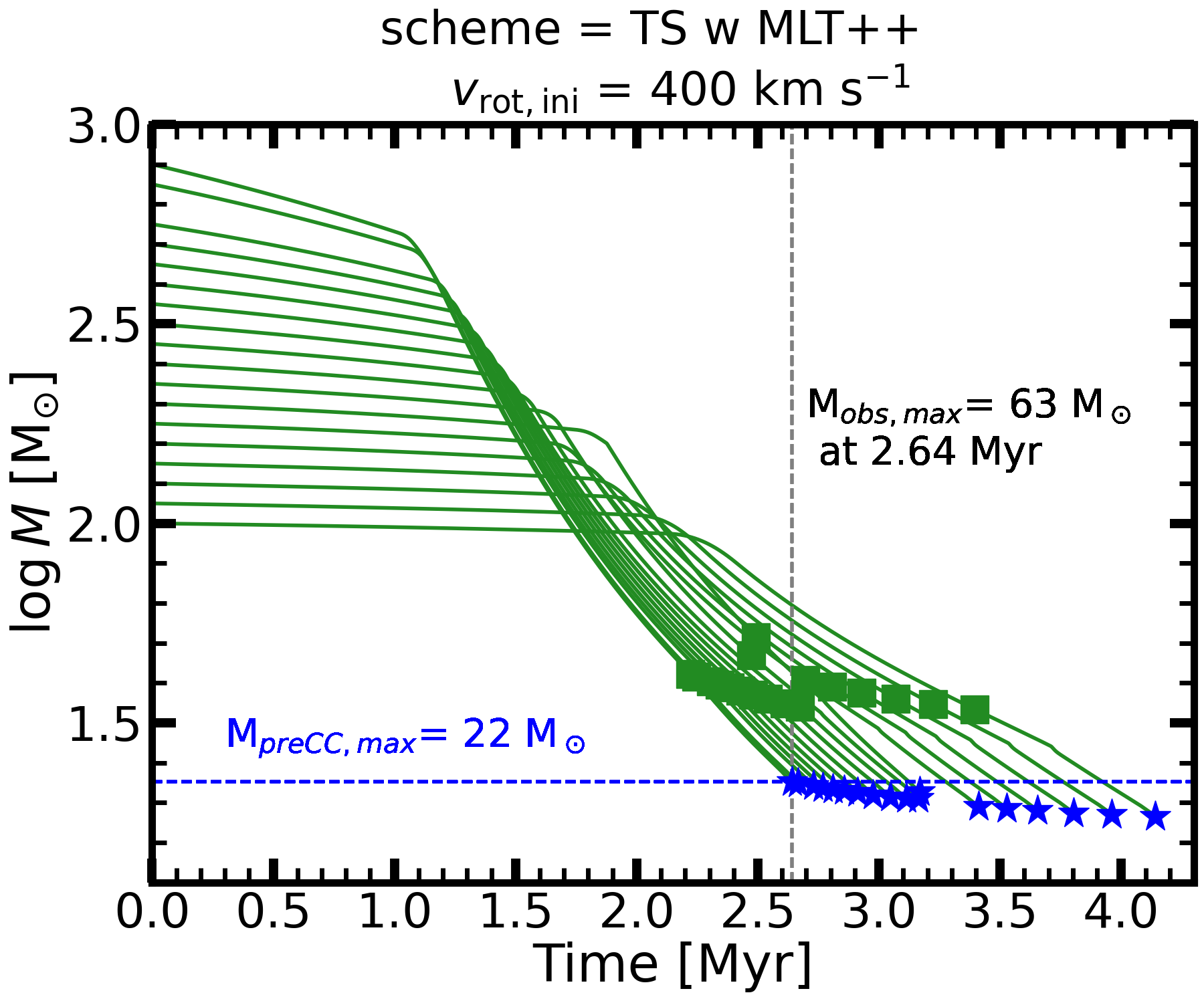}\includegraphics[width=0.3\textwidth]{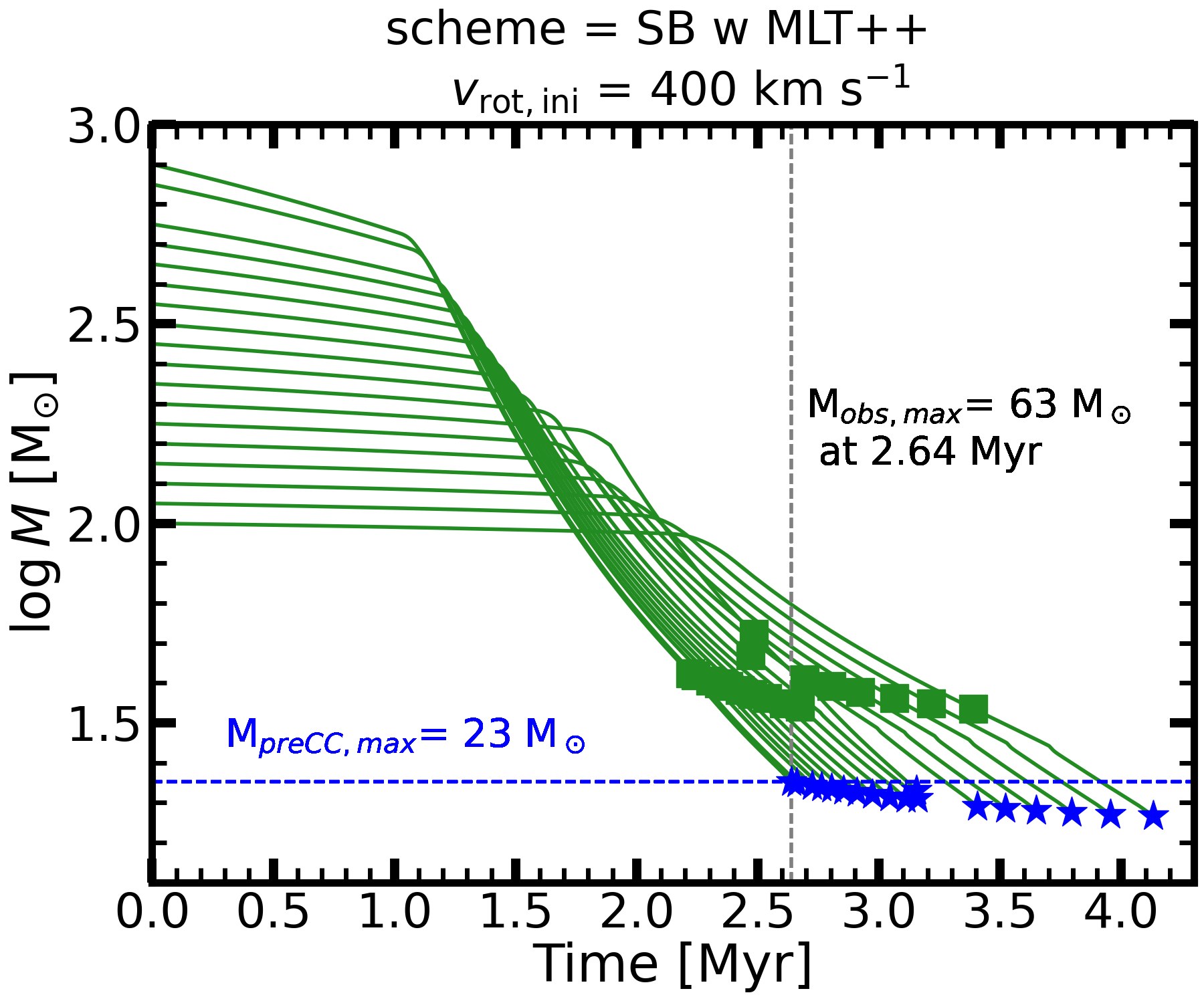}
    \includegraphics[width=0.3\textwidth]{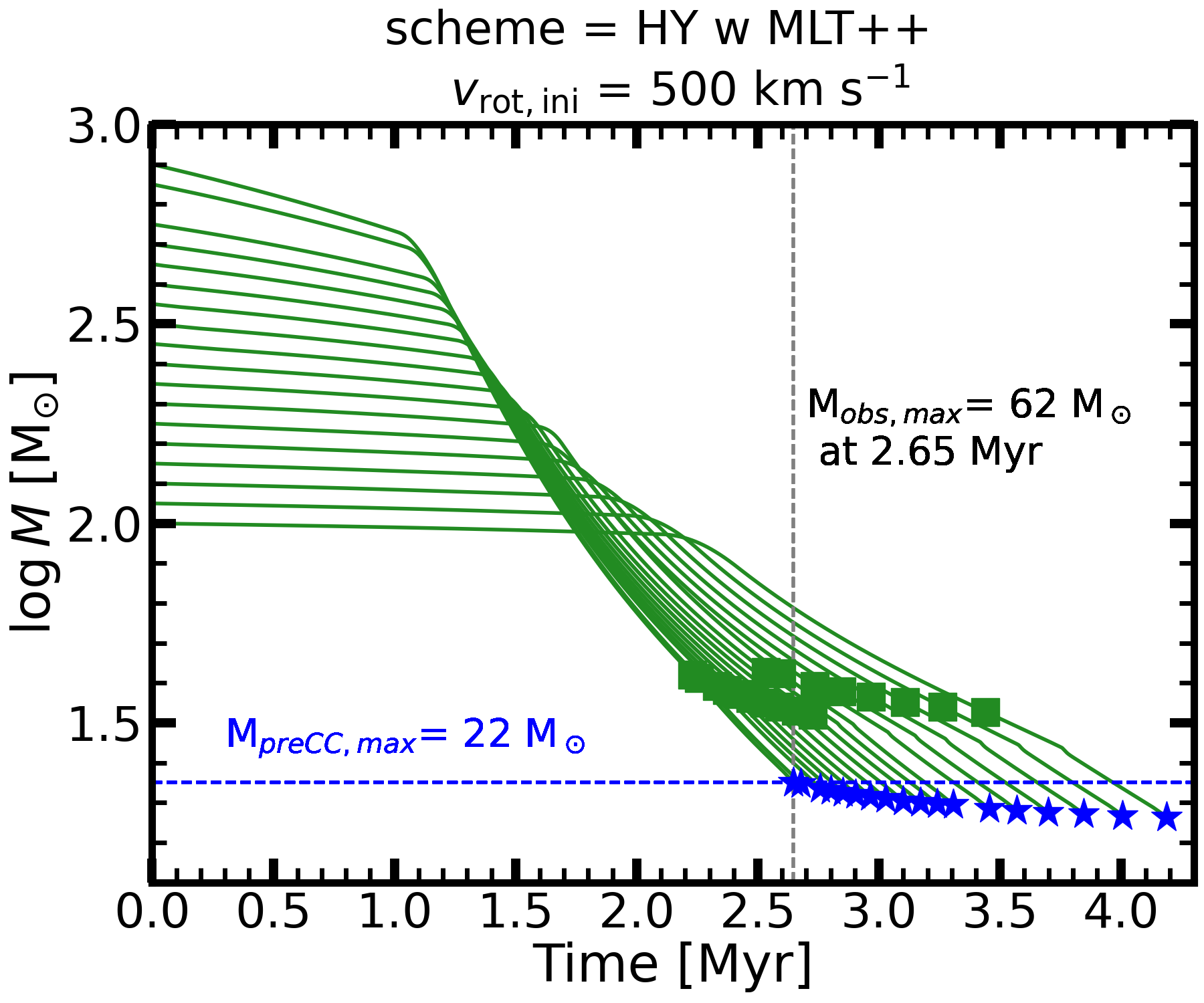}\includegraphics[width=0.3\textwidth]{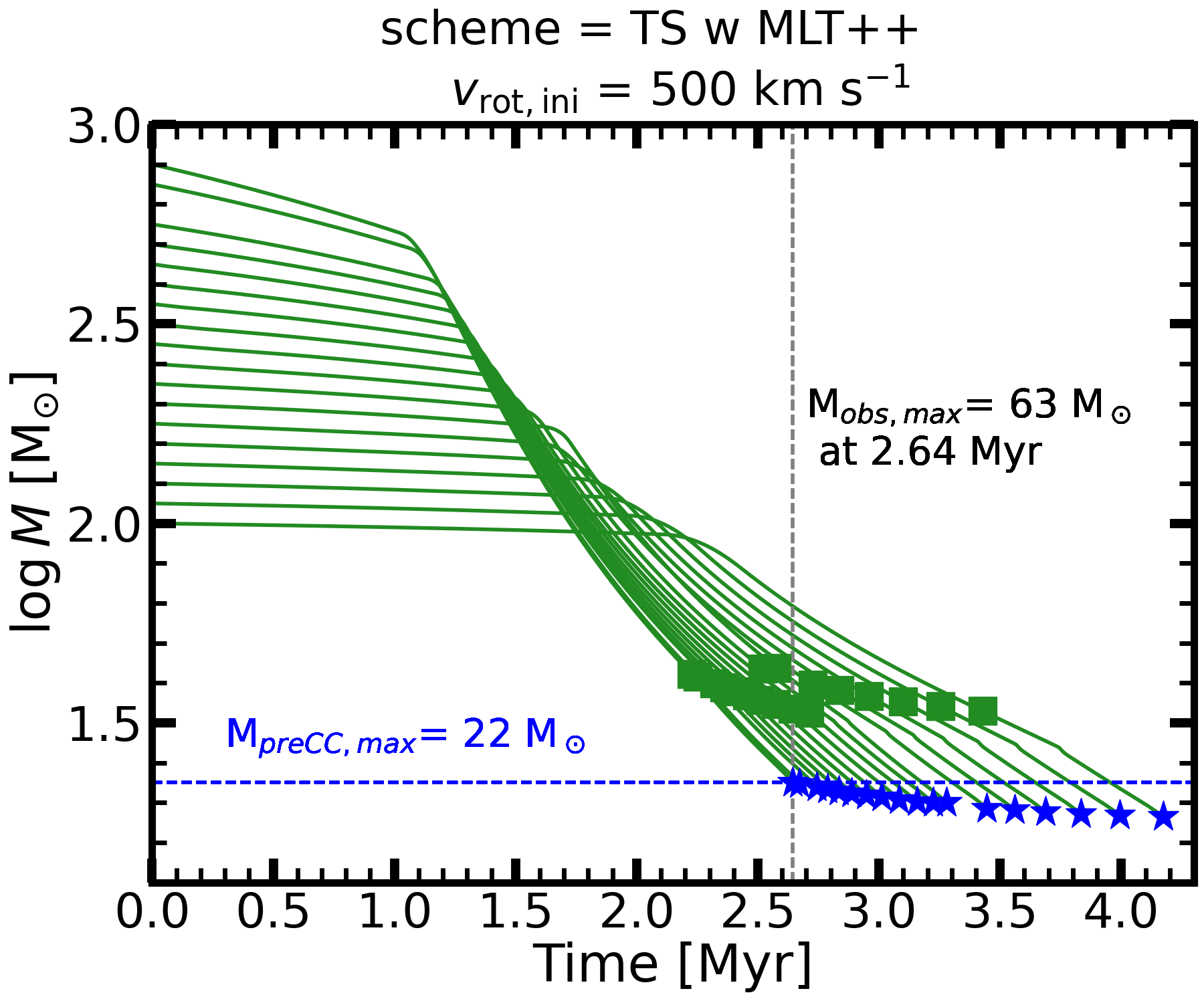}\includegraphics[width=0.3\textwidth]{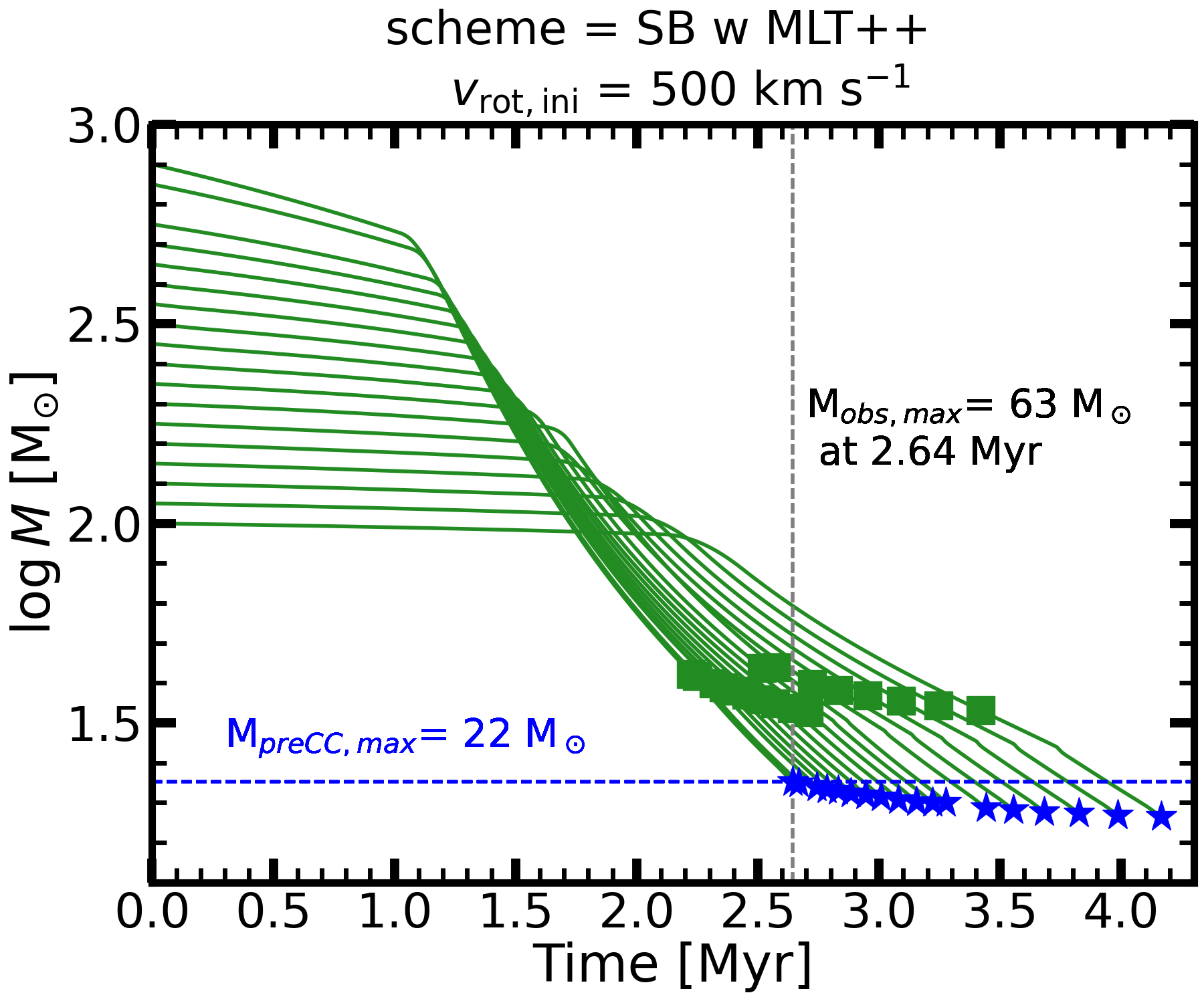}
    \caption{Left to right: HY, TS, and SB angular momentum transport schemes. Top to lower: increasing the initial rotation velocity.}
    \label{fig:massevol_app}
\end{figure*}

In Figure \ref{fig:massevol_app}, we demonstrate the mass evolution for all coupling schemes using MLT++ and all initial rotational velocities in our grid. The maximum pre-supernova mass is higher if the initial rotation velocity is lower or the internal coupling scheme is stronger. For $v_{\rm ini} > 300$~km\,s$^{-1}$, the same trends are true for the maximum observable mass.

%
%
\section{Interpolation framework}\label{sec:app_interp}

Given the computational cost of generating large and dense grids of stellar evolution models, interpolation is commonly applied to facilitate comparison with observational data \citep[e.g.,][]{brott2011b, georgy2014,schneider2014,schneider2017,szecsi2022}.
For this reason, we generate interpolated models and create a denser parameter space in mass and time.

%
%
\subsection{Model selection}\label{sec:modelselection}

We adopt a subset of evolutionary models without MLT++, as models with MLT++ result in effective temperatures that are too high to match observations (see Figure~\ref{fig:ms1} and related discussion). 
In the following, we only discuss models in excess of initially 200~M$_\odot$ because it is clear that models with lower initial masses do not yield sufficiently high luminosities to reproduce the observations of the three WNh stars.
We limit the models to the main sequence phase since in our grid comparison, post-main sequence models cannot match the three WNh stars and also have convergence problems.
In addition, we chose not to generate interpolated grids in the HY scheme due to the larger numerical noise and more prominent convergence problems than in the other two schemes.
In the TS and SB scheme, we omit those models from the interpolation where the TAMS is not resolved properly.

%
%
\subsection{Interpolation method}\label{sec:interpolation}

The interpolation framework employs a two-step linear interpolation method to construct a dense, uniform grid of stellar models:
\begin{enumerate}
    \item \textbf{Time normalisation and interpolation:} For each evolutionary track, we normalise the time axis to the maximum main-sequence evolutionary time of the model. Then, a fixed, high-resolution time grid ($N_t = 20,000$ points) is used to ensure consistent sampling across all models. This is comparable to the time spacing of the evolutionary models. 
    \item \textbf{Interpolation over initial mass:} In the evolutionary model grid, the initial stellar masses are logarithmically distributed, ranging from 225~M$_\odot$ to 800~M$_\odot$. The interpolated dataset is constructed on a dense initial mass grid ($N_m = 1000$ points). This effectively increases the mass resolution by more than two orders of magnitude (since we have 12 different initial masses in the evolutionary model grid in the given parameter space) such that the mass spacing is 1~M$_\odot$ for the lower masses (around 100~M$_\odot$), and 2~M$_\odot$ for the higher masses (around 800~M$_\odot$).
    Linear interpolation is performed across the logarithmic mass grid at each normalised time step. This includes the main stellar parameters of interest: effective temperature, luminosity, surface hydrogen, helium, and nitrogen abundances, mass-loss rate, and terminal velocity.
\end{enumerate}
Since the mapping is onto a $N_m \times N_t$ grid, the number of interpolated points is 20 million. This provides a sufficiently dense coverage. We verified that for a given initial mass, the interpolated data well matches the original evolutionary model. 
The initial rotational velocity rapidly decreases in the models and can follow a non-monotonic trend with initial mass. For this reason, interpolation is only performed over time and initial mass. Likewise, we do not interpolate between the evolutionary tracks that employ different AM schemes. 
This leads to a total of 10 grids of interpolated models for the five initial rotational velocities and two AM schemes in our evolutionary model grid. In the following, we search for solutions in all of the 10 interpolated model grids.

%
%
\section{Markov-Chain Monte Carlo analysis}\label{sec:app_mcmc}

We implement the Bayesian inference framework of Markov-Chain Monte Carlo (MCMC) using the Python package \texttt{emcee}, which employs an affine-invariant ensemble sampler for efficient exploration of high-dimensional parameter spaces \citep{foreman2013}. The MCMC analysis leverages the interpolated models to match observed stellar properties.
To match the observed properties of the WNh stars R136a1, R136a2, and R136a3, we rely on the their luminosity, effective temperature, and surface helium abundance measurements. 

%
%
\subsection{MCMC workflow}

The MCMC calculation consists of the following main steps. 
\begin{enumerate}
    \item Initialize the MCMC walkers at random positions across the parameter grid.
    \item Calculate $\chi^2$ for the selected parameters.
    \item Compute the posterior probability for each walker using the log-likelihood and log-prior functions, discarding those models where the central helium abundance is less than the lower limit of the observed surface helium abundance.
    \item Iteratively update walker positions over a specified number of steps, allowing the ensemble to converge to the posterior distribution.
    \item Discard initial burn-in steps to ensure convergence and extract samples from the posterior distribution.
\end{enumerate}

%
%
\subsection{$\chi^2$ function}

The calculation of the $\chi^2$ function is implemented by first obtaining the individual $\chi^2$ values for each modelled parameter, ensuring logarithmic scale in all instances ($\log L$, $\log T_{\rm eff}$, $\log Y_{\rm surf}$). When the observational uncertainties are not on a logarithmic scale, we calculate them as:
\begin{equation}
\sigma_{\rm param, \, logscale} = 
\frac{\sigma_{\rm param, \, linscale} }{ x_{\rm param}^{\mathrm{obs}} \cdot \ln(10)}
 \, ,
\end{equation}
with $x_{\rm param}^{\mathrm{obs}}$ the observed values of the parameters and $ \sigma_{\rm param, linscale}$ their uncertainties on a linear scale. $\ln$ is the natural (e-based) logarithm. Then, the $\chi^2$ function is:
\begin{equation}
\chi^2 =\frac{\left(\log x_{\rm param}^{\mathrm{obs}} - \log x_{\rm param}^{\mathrm{model}} \right)^2}{\sigma_{\rm param,logscale}^2} \, ,
\end{equation}
where $x_{\rm param}^{\mathrm{model}}$ are the interpolated model predictions.
The individual $\chi^2$ values are then summed up to a total $\chi^2_{\rm total}$ distribution with equal weight, accounting for all considered parameters.
In practice, the individual $\chi^2$ distributions have very low minima (min. $\chi^2_{\rm param} < 10^{-9}$) since the observed parameters are within the range of the parameter values of the dense interpolated model grid. The combined, total $\chi_{\rm total}^2$ distribution requires a compromise between parameters and typical values are between 0.1 and 10.

We explored including additional parameters in the calculations and comparison ($\log \dot{M}$, $\log v_\infty$, $\log X_{\rm nit, surf}$), and decided to avoid a more complex, highly multi-dimensional comparison. On the one hand, the observational uncertainty is more significant on these parameters, which lowers their constraining power. On the other hand, we have found that the effects of mass-loss rates and the surface nitrogen abundances tend to "cancel out" on the $\chi^2_{\rm total}$ distribution. This is because the observed mass-loss rates require a model in its very early evolution, whereas the observed surface nitrogen abundances require a model in its late main sequence phase. These parameters therefore significantly skew the distribution into two different directions, which overall leads to worsening the fit to the other parameters.

%
%
\subsection{Log-probability function}

The logarithmic probability function is constructed from the sum of the logarithmic prior and logarithmic likelihood distributions. 

Uniform priors are assumed for the parameter space, restricting models to physically valid indices within the interpolated dataset. This ensures that the exploration remains consistent with the available evolutionary models.
We assume a Gaussian error distribution, hence the log-likelihood function is described from $\chi^2_{\rm total}$. 
For a given datapoint in the interpolated model grid (representing interpolated parameters at a specific time and mass), the log-likelihood function is defined as:
\begin{equation}
\ln \mathcal{L} = -\frac{1}{2}  \chi^2_{\rm total},
\end{equation}
where $\chi^2_{\rm total} = \chi^2_{\rm logL} + \chi^2_{\rm logT_{eff}} + \chi^2_{\rm log Y_{surf}}$.

%
%
\subsection{Posterior probability density distributions}

We obtain posterior probability density distributions for all parameters included in the interpolated model grid. From these, three parameters (\( \log(L/L_\odot),  \log(T_\mathrm{eff}/\mathrm{K}), \log(Y_{\mathrm{surf}}) \)) are included in the MCMC framework as direct constraints. These provide insights into the most probable values. The joint posterior probability (jpp) density is calculated by normalising the most frequently sampled models in the MCMC run. This effectively synthesises the multi-dimensional parameter space into a coherent statistical representation. While individual posterior distributions are useful for understanding parameter-specific constraints, the joint distribution accounts for the simultaneous fit of all observational constraints. By leveraging the joint probability distribution, the analysis ensures that the inferred stellar properties are robust and consistent across all dimensions of the parameter space. This method provides a statistically rigorous framework for interpreting the selected best models.

%
%
\subsection{Statistical interpretation}
\label{sec:best_model}

The minimum value of the $\chi^2_{\rm total}$ distribution identifies the model with the lowest discrepancy between model predictions and observed values, considering all three parameters ($\log L$, $\log T_{\rm eff}$, $\log Y_{\rm surf}$) simultaneously. In the MCMC analysis, the posterior distribution reflects the probability of each model, given the data and prior assumptions. The peak of the joint posterior probability distribution, also referred to as the Maximum A Posteriori (MAP) estimate, provides a robust statistically interpretation.
In practice, the MAP model does not necessarily have the lowest $\chi^2_{\rm total}$ value; however, it is among the models with the lowest $\chi^2_{\rm total}$ values, as higher posterior probabilities correlate with lower $\chi^2_{\rm total}$.
Generally, the mode of the joint posterior probability distribution, that is, the most frequently sampled model in the MCMC analysis, also recovers the MAP estimate. In some cases (particularly in more complex Gaussian distributions) the mode can represent a secondary peak, which does not align with the highest posterior probability.
In these cases, the mode also tends to have higher $\chi^2_{\rm total}$ values compared to the MAP estimate. 

%
%
In Tables \ref{tab:mcmc_grid_table_TS} and \ref{tab:mcmc_grid_table_SB}, we report our findings from the MCMC analysis, considering the above two statistical interpretations. Here, the uncertainty estimate is obtained from the individual posterior probability distribution functions as a 1~$\sigma$ standard deviation from the mean. In the interpolated grid, each data point can be traced back to an initial mass and as such an evolutionary trajectory can be reconstructed. We make us of this to obtain the initial mass that corresponds to a given model. For example, the model with the lowest $\chi^2_{\rm total}$ has a corresponding initial mass and the model with the highest jpp has a corresponding initial mass. However, unlike the current parameters, the initial mass does not have a posterior probability density distribution because it is a derived quantity. To gauge the uncertainty on the initial mass we construct a reduced $\chi^2_{\rm total}$ distribution against the initial mass as follows. For each initial mass in the interpolated grid, we have $N_t=$20,000 data points along its evolution. These 20,000 models have different current parameters and thus different $\chi^2_{\rm total}$ values. We search for the lowest $\chi^2_{\rm total}$ amongst these 20,000 models and associate it with the initial mass. We repeat this for all $N_m=1,000$ initial masses. This way, we obtain a reduced $\chi^2_{\rm total}$ distribution against all initial masses in the interpolated grid. 
Using this $\chi^2_{\rm total}$ distribution against the initial mass, we identify that R136a1 has well-defined minima in the range of 300--400~M$_\odot$ (see also Figure \ref{fig:R136a1_result}). On the other hand, the $\chi^2_{\rm total}$ distribution becomes almost flat above approximately 500~M$_\odot$ for R136a2 and R136a3. For this reason, we do not describe the initial mass with the $\chi^2_{\rm total}$ or MAP estimate.

\begin{figure*}
    \centering
    \includegraphics[width=0.31\textwidth]{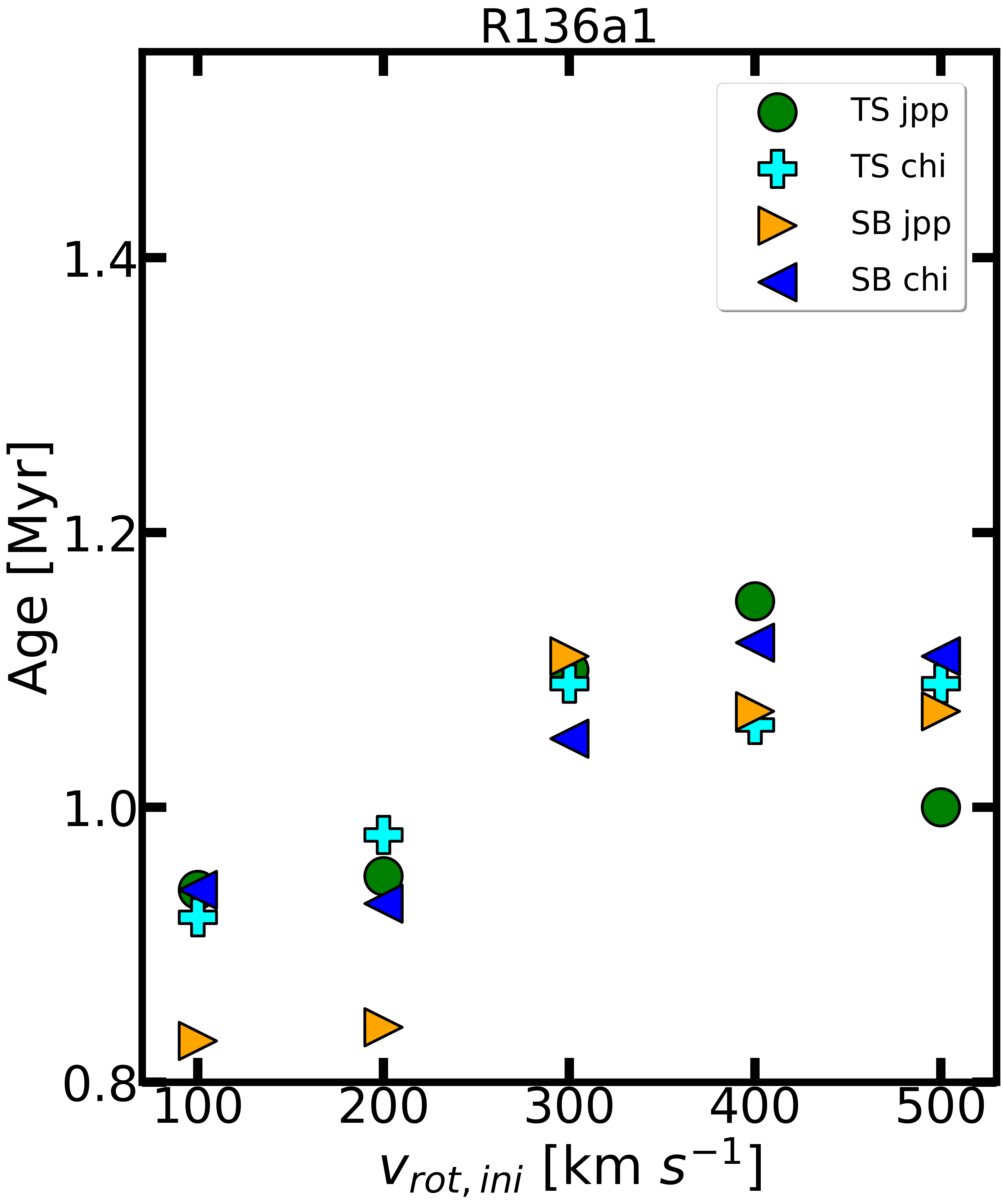}\includegraphics[width=0.31\textwidth]{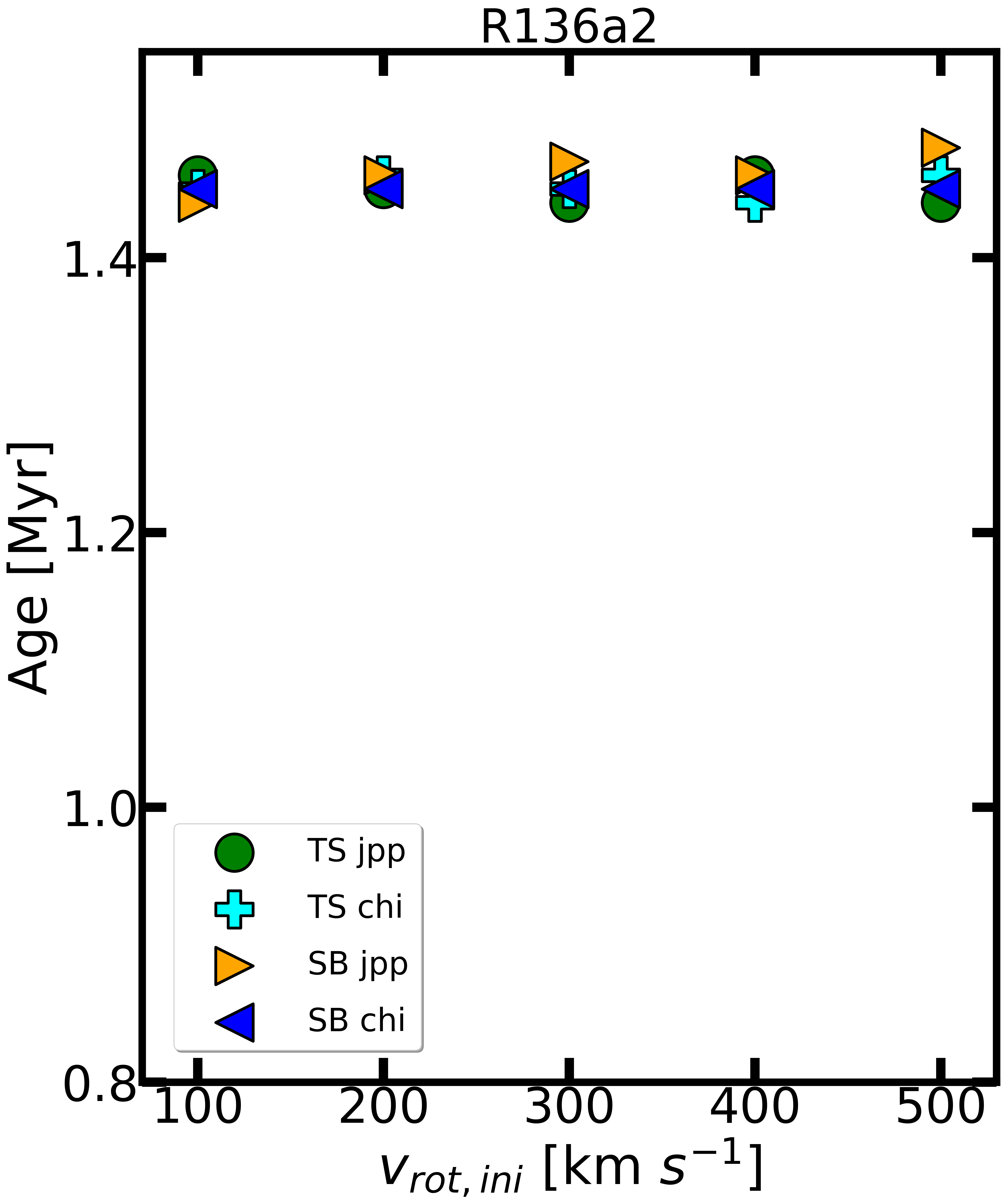}\includegraphics[width=0.31\textwidth]{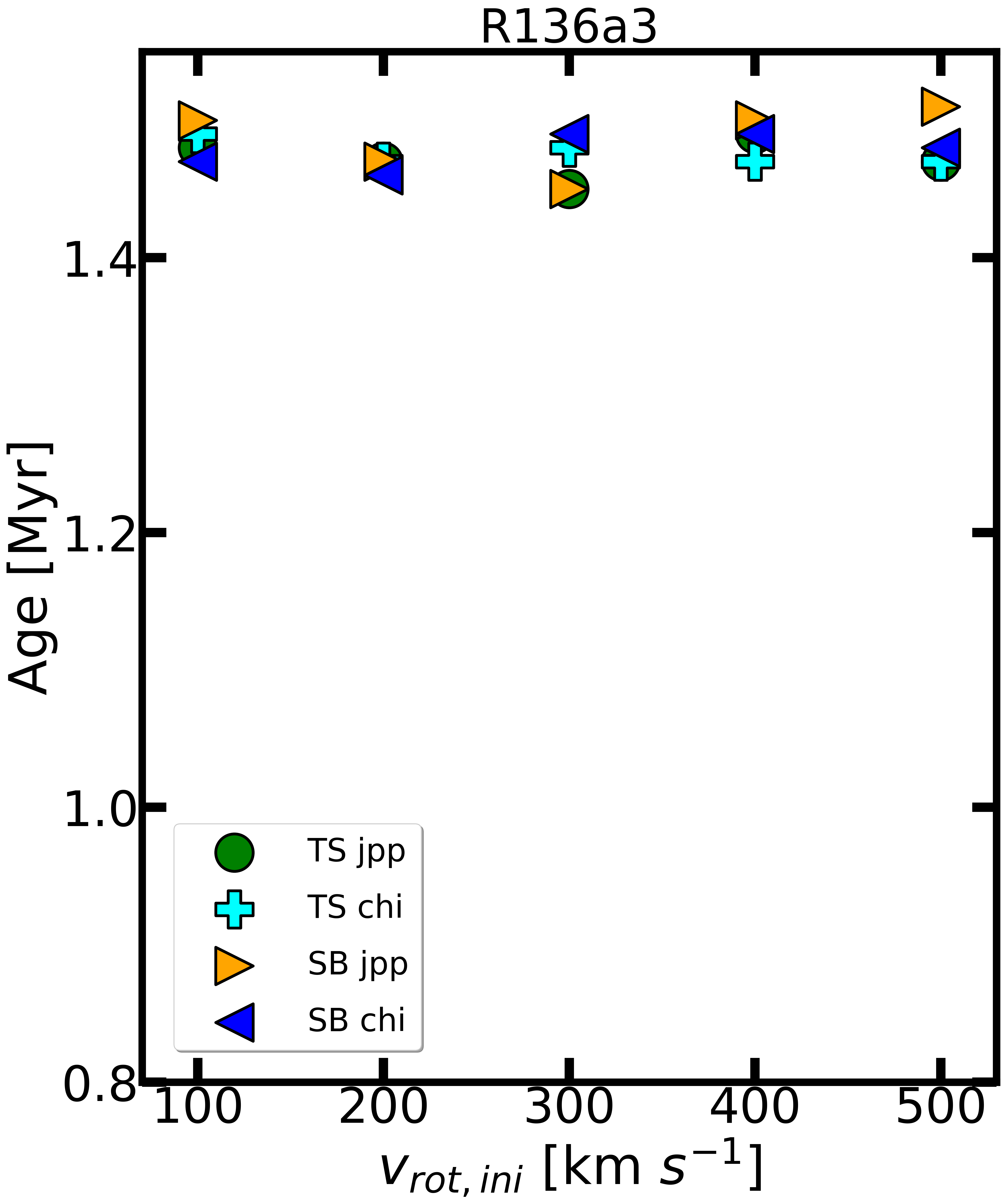}
    \includegraphics[width=0.31\textwidth]{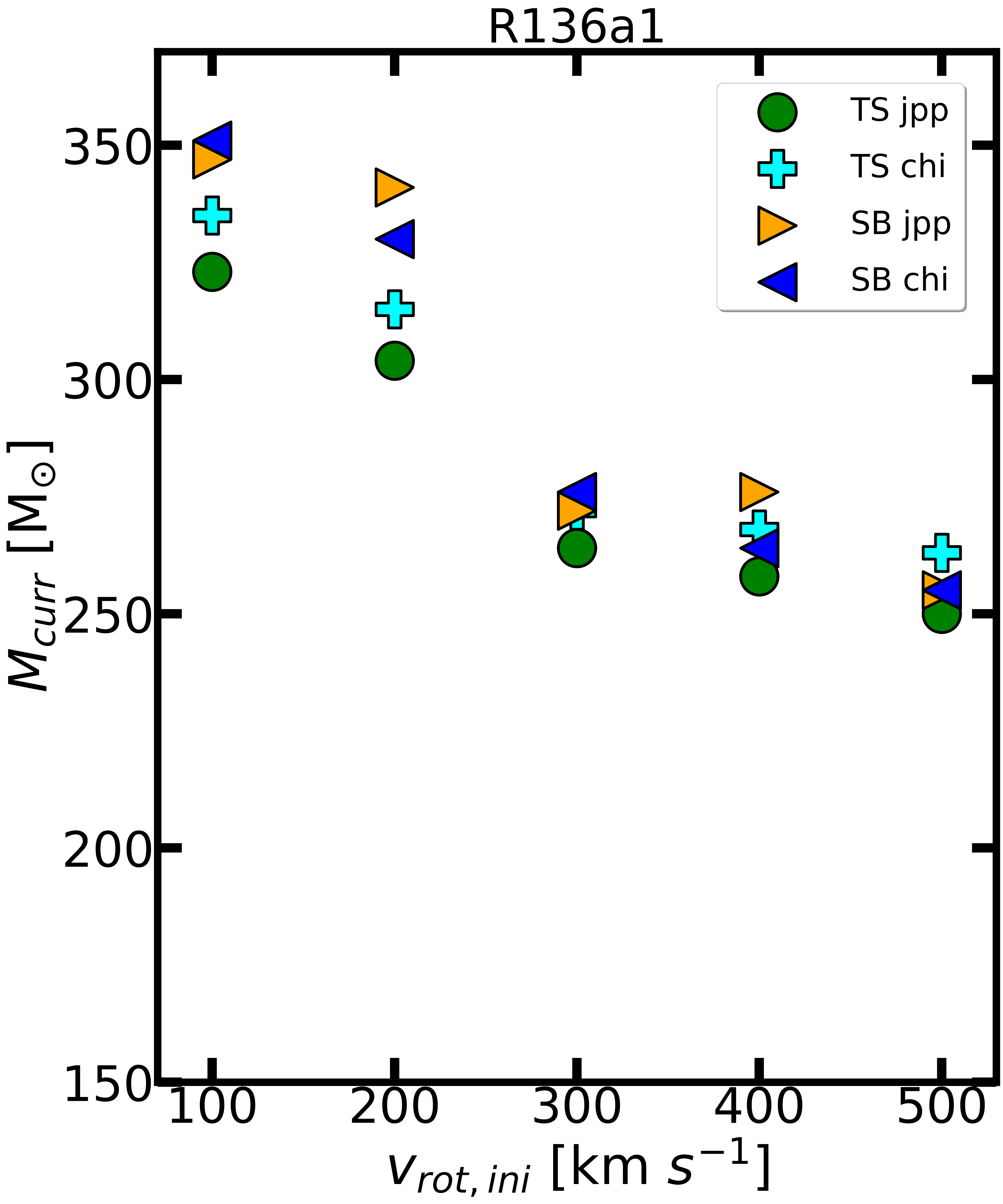}\includegraphics[width=0.31\textwidth]{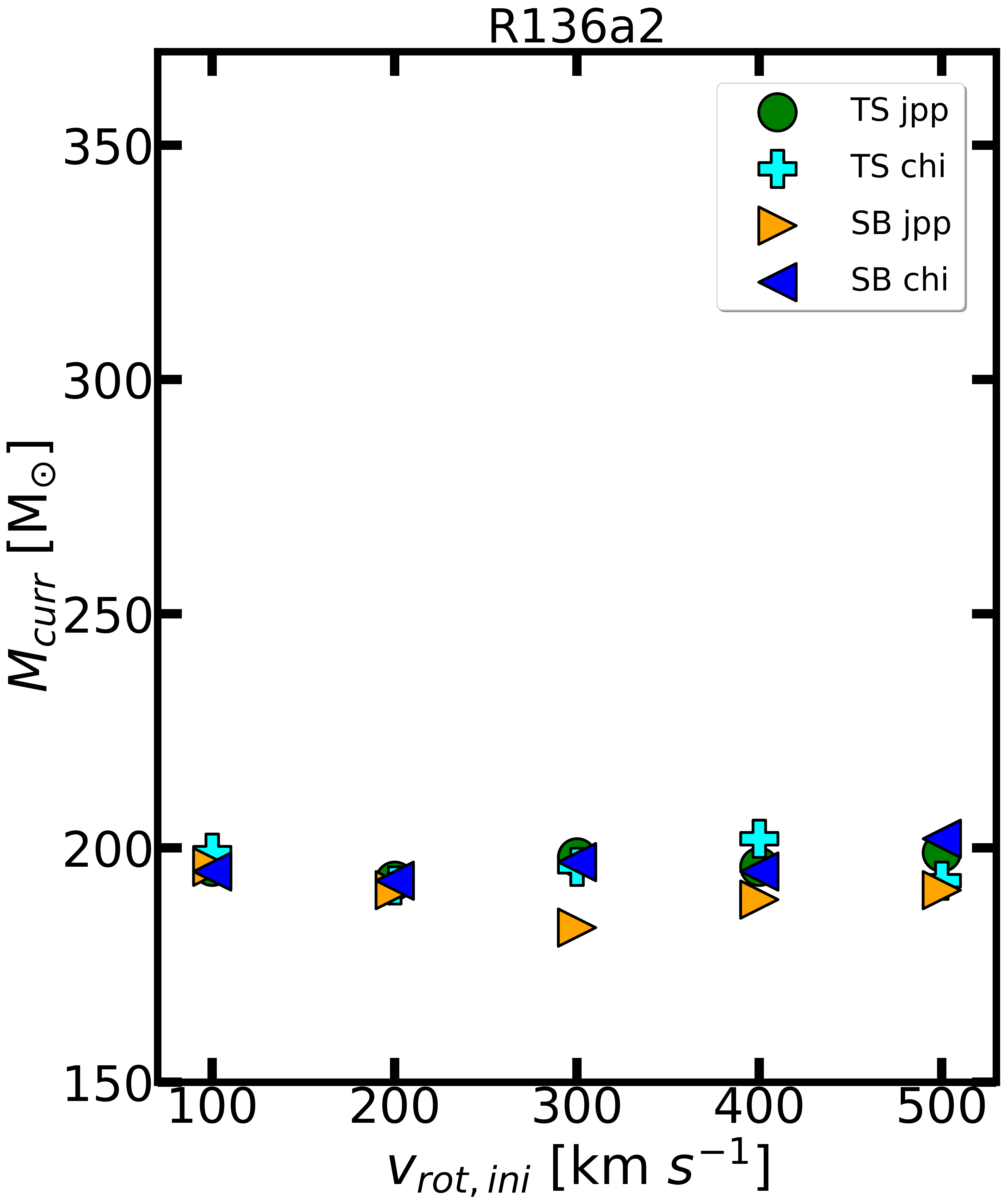}\includegraphics[width=0.31\textwidth]{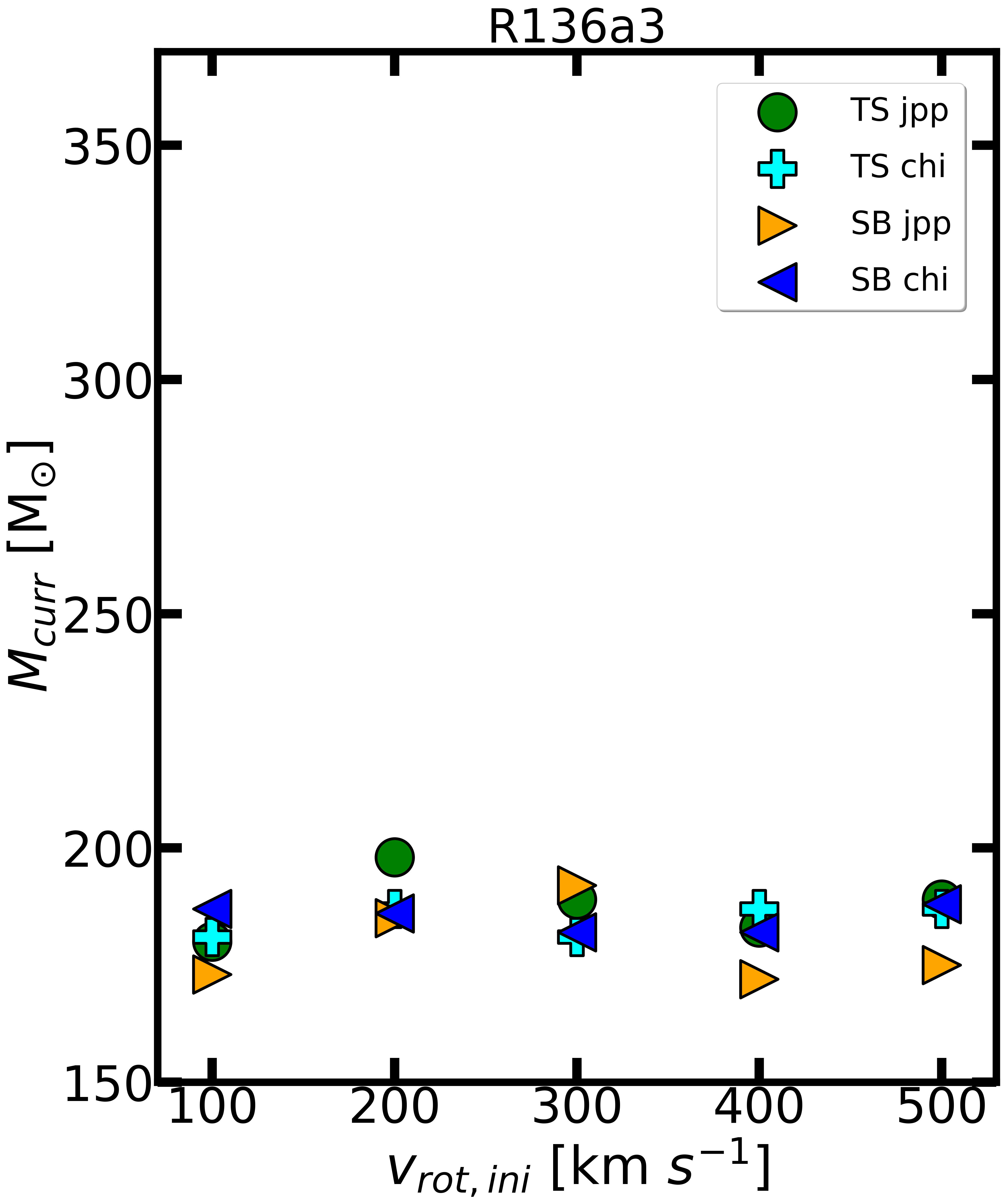}
    \includegraphics[width=0.31\textwidth]{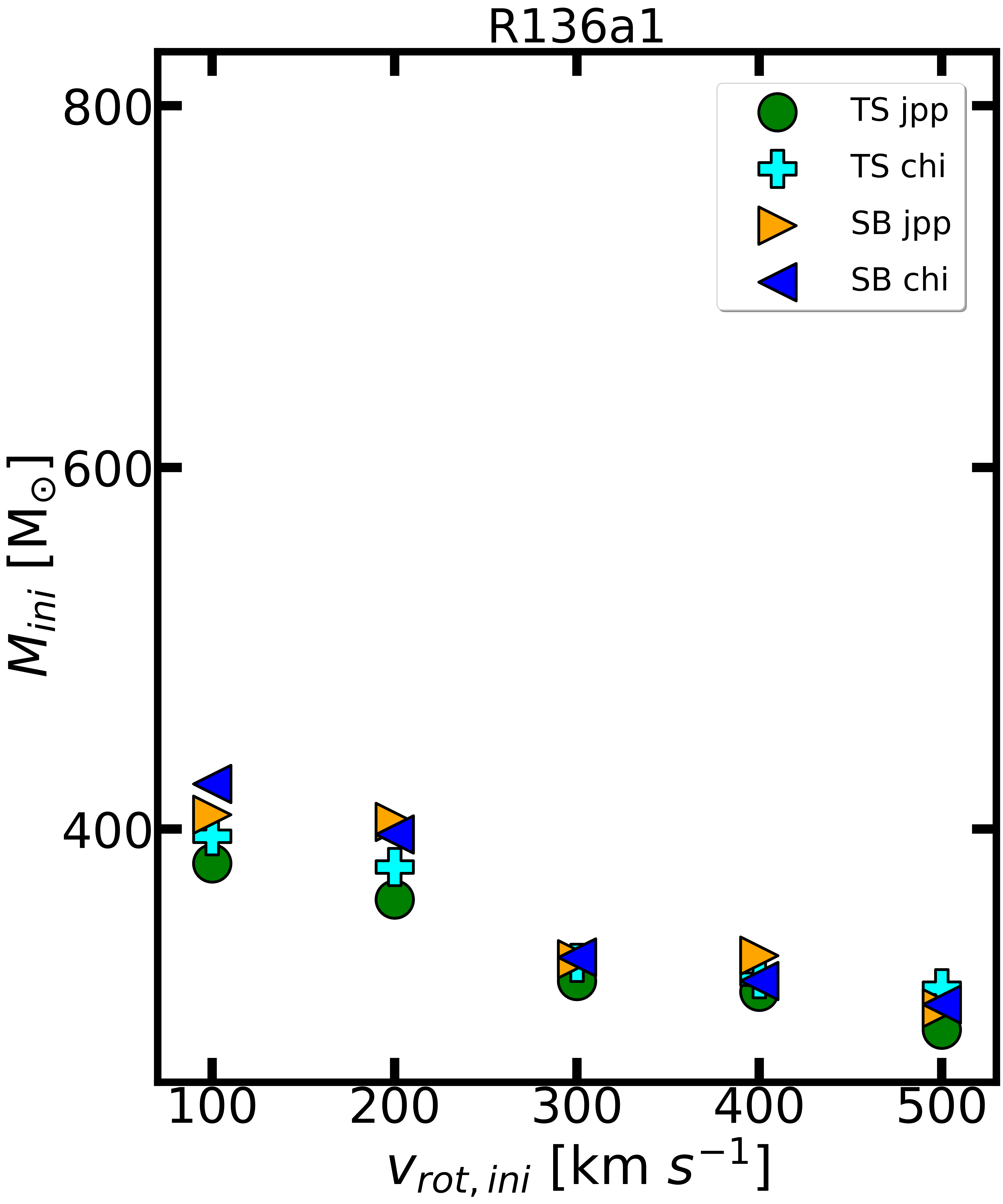}\includegraphics[width=0.31\textwidth]{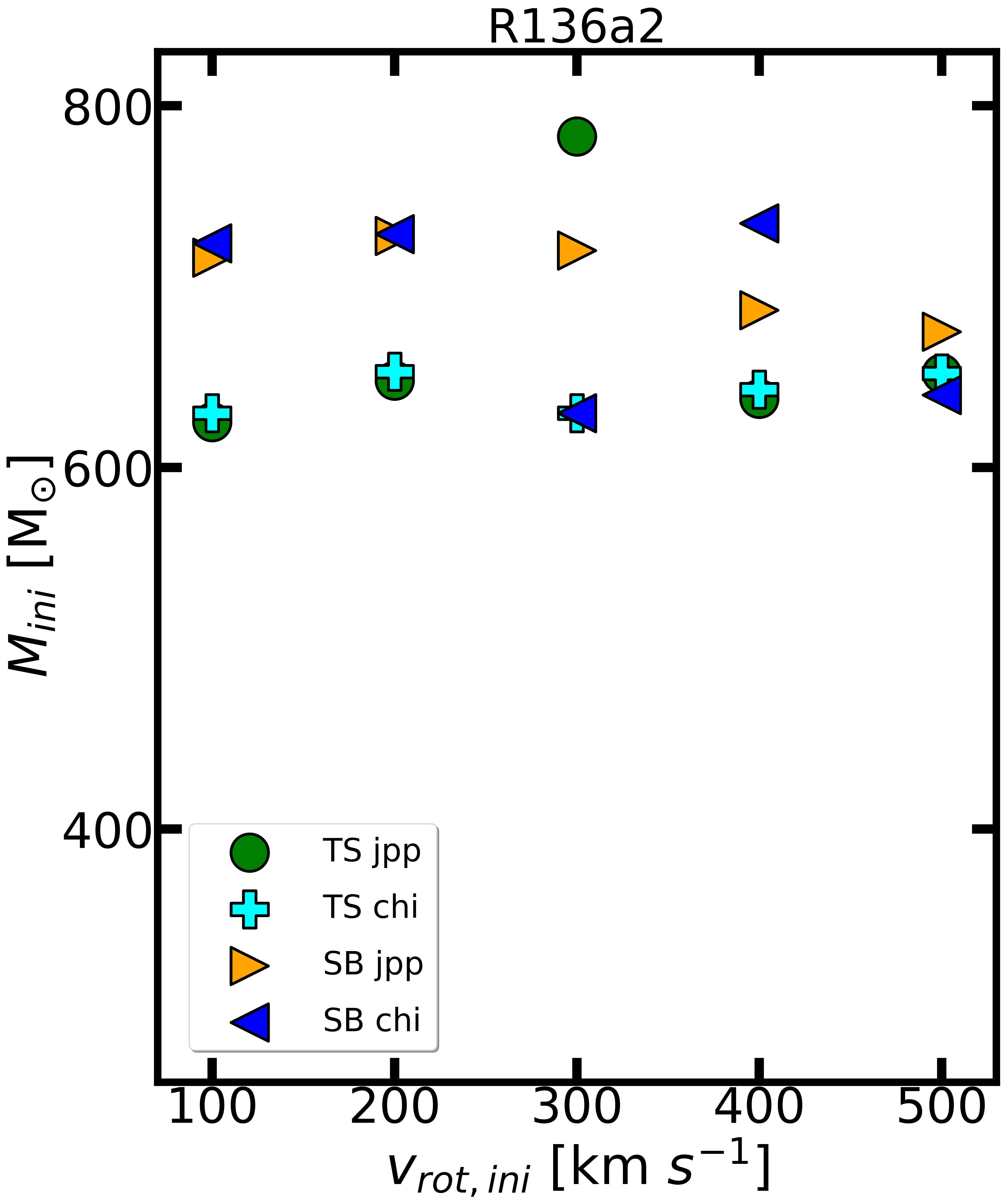}\includegraphics[width=0.31\textwidth]{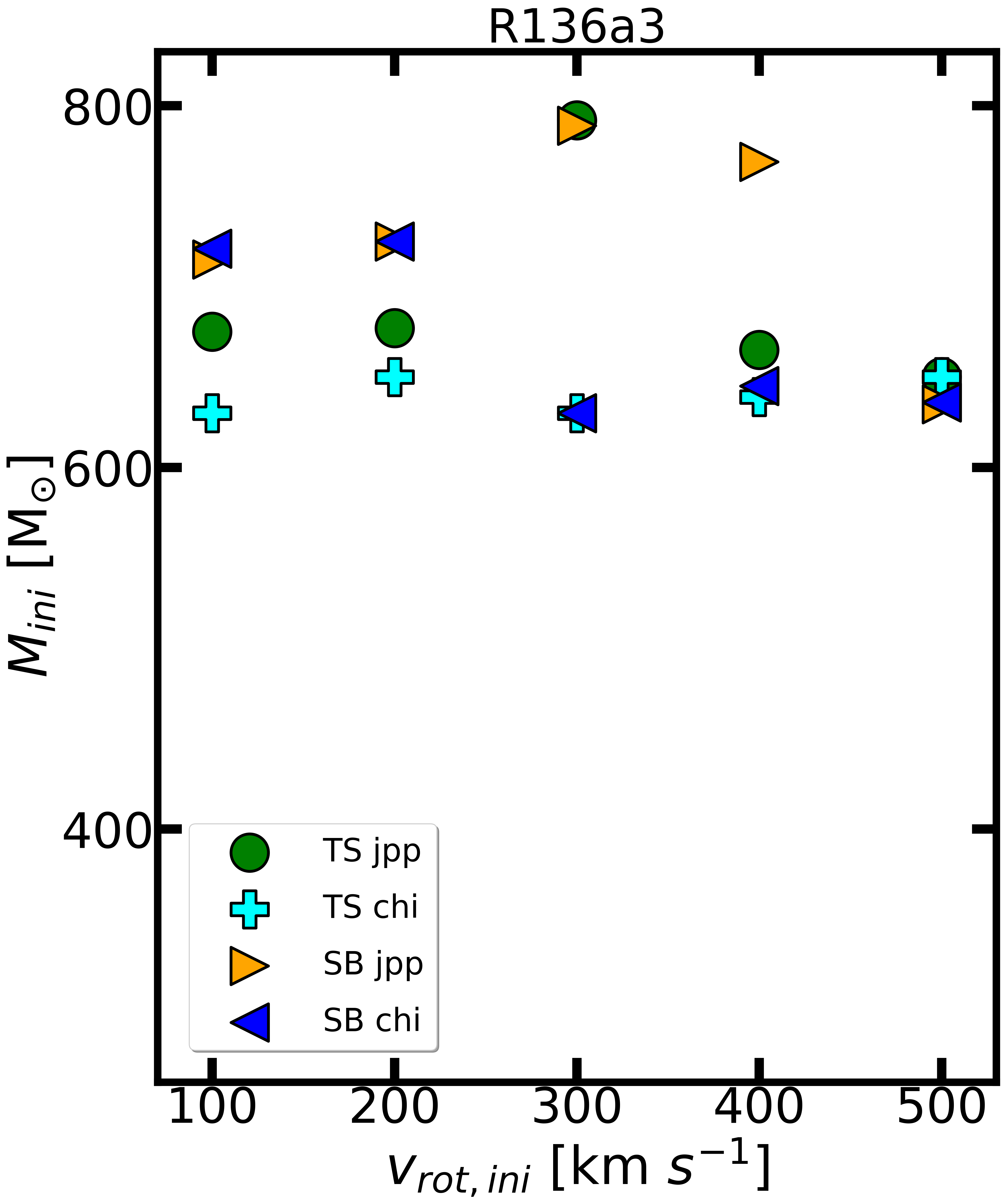}
    \caption{Summary of the results from Tables \ref{tab:mcmc_grid_table_TS} and \ref{tab:mcmc_grid_table_SB}. The three columns correspond to the three WNh stars. The panels from the top to bottom show the estimates on the age, current mass, and initial mass as a function of the initial rotational velocity of the models. The symbol coding shows the AM scheme (Tayler-Spruit and solid body) and statistical interpretation method (minimum $\chi^2$ value and highest joint posterior probability density).}
    \label{fig:sum}
\end{figure*}

%
%
Figure \ref{fig:sum} summarises the results from tables \ref{tab:mcmc_grid_table_TS} and \ref{tab:mcmc_grid_table_SB}, highlighting the scatter between interpolated grids with different schemes, initial rotational velocities, and different statistical interpretations. For R136a2 and R136a3, the results do not show any trends with AM scheme or initial rotational velocity. 
For R136a1, a clear trend is visible as a function of the initial rotational velocity. The age estimate increases, whereas the current and initial mass decreases for higher initial rotational velocities. This can be understood from the well-known degeneracy between initial mass and rotational velocity in typical massive star models evolving redward on the HRD. Namely, a model with higher initial mass but lower rotational velocity has a similar evolutionary trajectory as model with lower initial mass but higher initial rotational velocity. 
This lends further support to describing R136a1 as a redward evolving star, whereas R136a2 and R136a3 are in their down-stream evolution beyond the mass-turnover. 
We stress here that the AM scheme and initial rotational velocity cannot be deduced from our comparison. The scatter in the obtained results simply originates from the different evolution of the three parameters (effective temperature, luminosity, surface helium abundance) that form the basis of the comparison. As demonstrated in Figure \ref{fig:sum} and Tables \ref{tab:mcmc_grid_table_TS}, this scatter is within the $1\sigma$ standard deviation of the posterior probability densities.

%
%
\begin{table*}[htbp]
\centering
\setlength{\tabcolsep}{10.pt} 
\caption{Summary of the results obtained from the MCMC analysis, using the interpolated models in the TS scheme. The first and second values correspond to the model with highest joint posterior probability densities and the lowest $\chi^2_{\rm total}$, respectively. The values in brackets denote the mean value and the $1\sigma$ standard deviation from the individual posterior probability density distributions. For R136a2 and R136a3, *m denotes cases where the $M_{\rm ini}=800$~M$_\odot$ model is omitted from the interpolation, thus the highest initial mass available in the comparison is $M_{\rm ini}=700$~M$_\odot$.}
\begin{tabular}{l cccc }
%
%
\toprule\toprule
\textbf{R136a1} & \multicolumn{4}{c}{TS scheme} \\
\midrule
$v_{rot, ini}$ &  $M_{\rm ini}$ & $M_{\rm curr}$ &  $t$ & $\chi^2_{\rm total}$  \\ \midrule\midrule
100  & 381, 396 & 323, 335 (337$^{+51}_{-41}$) & 0.94. 0.92 (0.95$\pm$0.15) & 5.34, 5.10  \\
200  & 361, 379 & 304, 315 (324$^{+49}_{-40}$) & 0.95, 0.98 (0.96$\pm$0.16) & 4.24, 3.90  \\
300  & 316, 326 & 264, 272 (301$^{+49}_{-41}$) & 1.10, 1.09 (0.99$\pm$0.16) & 2.61, 2.56  \\
400  & 310, 317 & 258, 268 (284$^{+53}_{-37}$) & 1.15, 1.06 (1.00$\pm$0.17) & 1.91, 1.60   \\
500  & 289, 312 & 250, 263 (281$^{+52}_{-36}$) & 1.00, 1.09 (1.03$\pm$0.17) & 2.53, 1.50 \\ 
\bottomrule
%
%
\toprule
\textbf{R136a2} \\\midrule\midrule
100  & >500 (625, 630)*m &  196, 199 (192$^{+38}_{-31}$) & 1.46, 1.45 (1.50$\pm$0.11) & 4.37, 4.35    \\
200  & >500 (648, 653) &  193, 192 (195$^{+38}_{-32}$) & 1.45, 1.46 (1.50$\pm$0.11) & 2.99, 2.88  \\
300  & >500 (783, 630) &  198, 196 (193$^{+38}_{-32}$) & 1.44, 1.45 (1.52$\pm$0.11) & 4.60, 4.27  \\
400  & >500 (638, 643)*m &  196, 202 (195$^{+37}_{-32}$) & 1.46, 1.44 (1.50$\pm$0.12) & 3.77, 3.62    \\
500  & >500 (652, 652)*m &  199, 193 (195$^{+37}_{-32}$) & 1.44, 1.46 (1.50$\pm$0.12) & 2.87, 2.83   \\ 
\bottomrule
%
%
\toprule
\textbf{R136a3} \\\midrule\midrule
100 & >500 (675, 630)*m & 180, 181 (183$^{+43}_{-33}$) & 1.48, 1.49 (1.54$\pm$0.13)  & 4.14, 4.08   \\
200 & >500 (677, 650) &  198, 187 (187$^{+46}_{-34}$) & 1.47, 1.47 (1.54$\pm$0.14) & 4.29 (3.15) \\
300 & >500 (792, 630) & 189, 181 (187$^{+46}_{-37}$) & 1.45, 1.48 (1.55$\pm$0.13) & 4.49, 3.98  \\
400 & >500 (665, 639)*m & 183, 187 (192$^{+45}_{-38}$) & 1.49, 1.47 (1.55$\pm$0.16) & 4.04, 3.75  \\
500 &  >500 (650, 650)*m & 189, 187 (193$^{+44}_{-39}$) & 1.47, 1.47 (1.55$\pm$0.16) & 3.11, 3.10 \\ 
\bottomrule\bottomrule
\end{tabular}
\label{tab:mcmc_grid_table_TS}
\end{table*}

%
%
\begin{table*}[htbp]
\centering
\setlength{\tabcolsep}{10.pt} 
\caption{Same as Table \ref{tab:mcmc_grid_table_TS} but using the SB models. }
\begin{tabular}{l cccc }
%
%
\toprule\toprule
\textbf{R136a1} & \multicolumn{4}{c}{SB scheme} \\
\midrule
$v_{rot, ini}$ & $M_{\rm ini}$ &  $M_{\rm curr}$ &  $t$ & $\chi^2_{\rm total}$ \\ \midrule\midrule
100 & 408, 425 & 347, 351 (350$^{+53}_{-44}$) & 0.83, 0.94 (0.92$\pm$0.16) & 7.80, 7.01  \\
200 & 404, 397 & 341, 330 (330$^{+52}_{-42}$) & 0.84, 0.93 (0.93$\pm$0.15) & 5.94, 5.37  \\
300 & 328, 329 & 272, 276 (299$^{+53}_{-39}$) & 1.11, 1.05 (0.97$\pm$0.16) & 3.77, 3.47  \\
400 & 330, 316 & 276, 264 (282$^{+53}_{-34}$) & 1.07, 1.12 (1.00$\pm$0.16) & 2.40, 2.27  \\
500 & 301, 303 & 255, 255 (280$^{+51}_{-33}$) & 1.07, 1.11 (1.00$\pm$0.16) & 2.04, 1.95 \\ 
\bottomrule
%
%
\toprule
\textbf{R136a2} \\\midrule\midrule
100  & >550 (716, 724) & 196, 195 (195$^{+38}_{-33}$) & 1.44, 1.45 (1.50$\pm$0.10) & 3.85, 3.58  \\
200  & >570 (728, 729) & 191, 193 (194$^{+38}_{-32}$) & 1.46, 1.45 (1.50$\pm$0.10) & 3.31, 3.30  \\
300  & >520 (720, 630) & 183, 197 (192$^{+38}_{-32}$) & 1.47, 1.46 (1.50$\pm$0.10) & 4.65, 4.47  \\
400  & >510 (687, 735) & 189, 195 (193$^{+38}_{-32}$) & 1.46, 1.45 (1.50$\pm$0.11) & 4.54, 4.39  \\
500  & >500 (675, 640)*m & 191, 202 (193$^{+38}_{-32}$) & 1.48, 1.45 (1.50$\pm$0.11) & 4.49, 4.02  \\ 
\bottomrule
%
%
\toprule
\textbf{R136a3} \\\midrule\midrule
100 & >510 (715, 721) & 173, 187 (184$^{+53}_{-23}$) & 1.50, 1.47 (1.52$\pm$0.11) & 4.00, 3.81  \\
200 & >510 (725, 725) & 185, 186 (185$^{+41}_{-34}$) & 1.47, 1.46 (1.53$\pm$0.13) & 3.58, 3.57 \\
300 & >530 (789, 630) & 192, 182 (183$^{+46}_{-34}$) & 1.45, 1.49 (1.52$\pm$0.12) & 4.58, 4.25  \\
400 & >510 (769, 645) & 172, 182 (188$^{+35}_{-36}$) & 1.50, 1.49 (1.52$\pm$0.13) & 4.51, 4.25  \\
500 & >530 (635, 636)*m & 175, 188 (191$^{+47}_{-38}$)  & 1.51, 1.48 (1.54$\pm$0.14) & 4.24, 4.12 \\ 
\bottomrule\bottomrule
\end{tabular}
\label{tab:mcmc_grid_table_SB}
\end{table*}

%
%

%
%
\subsection{Comparison with the TS-300 grid}
\label{sec:ts-300}

\begin{figure*}
    \centering
    \includegraphics[width=0.33\textwidth]{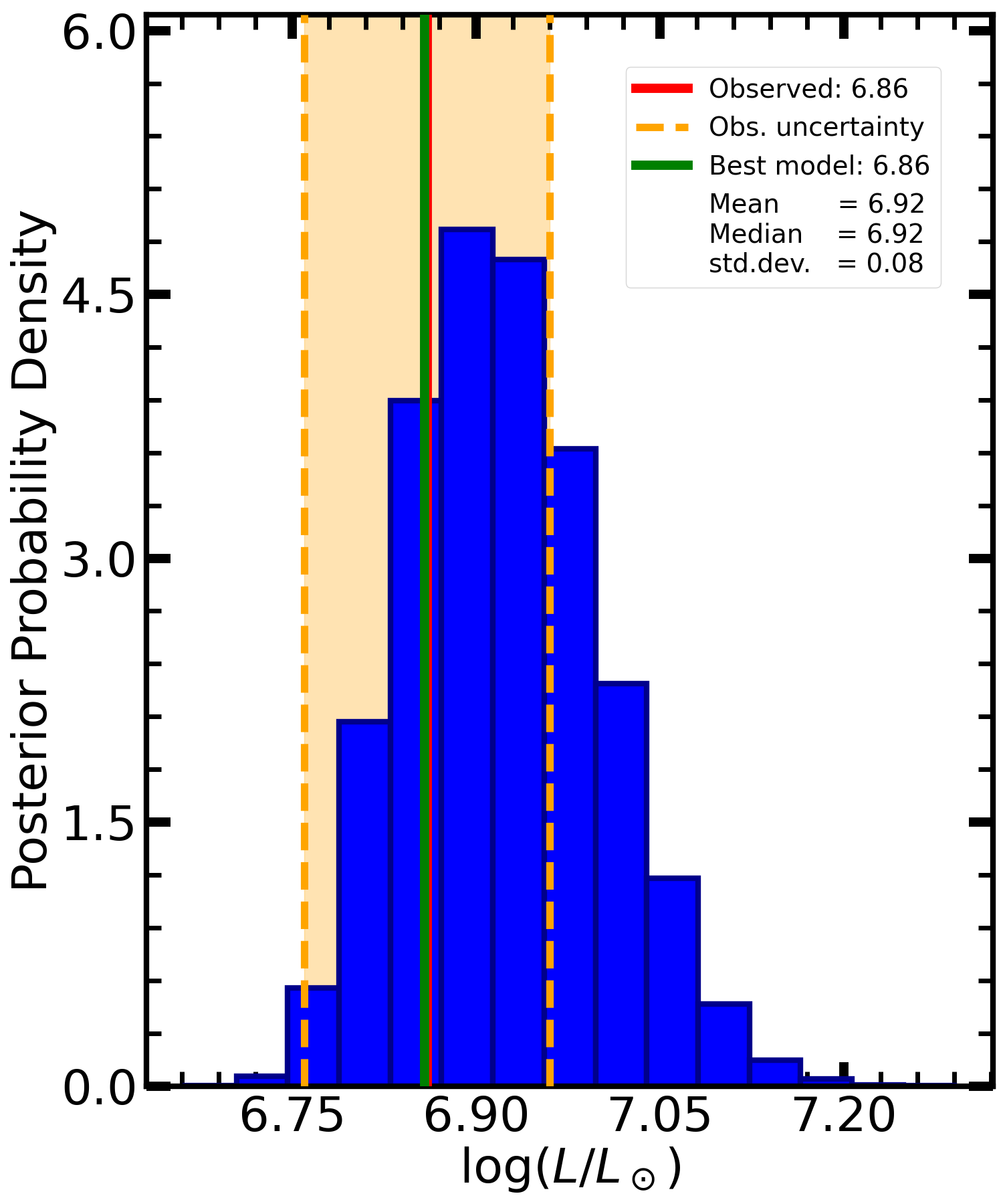}\includegraphics[width=0.33\textwidth]{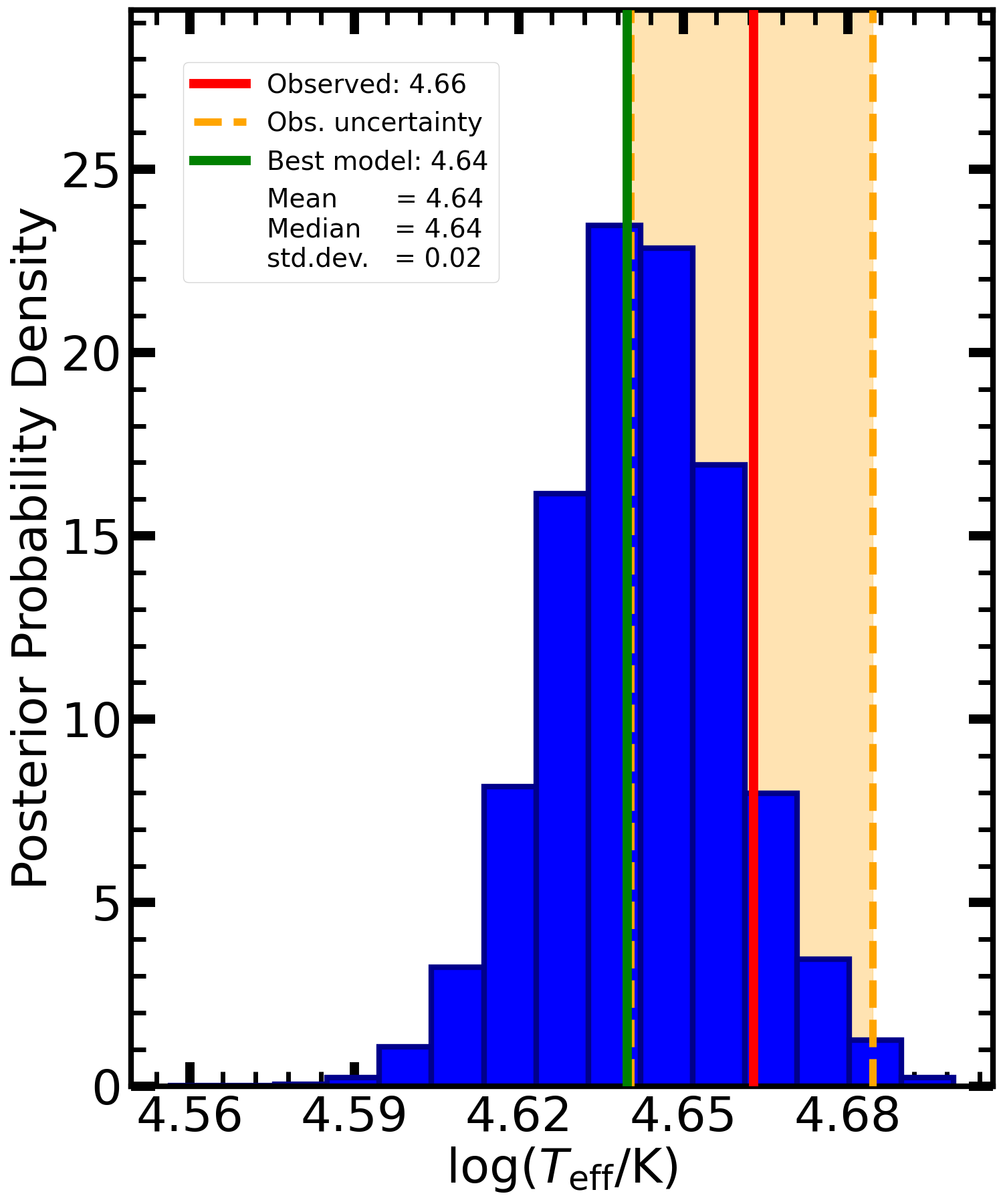}\includegraphics[width=0.33\textwidth]{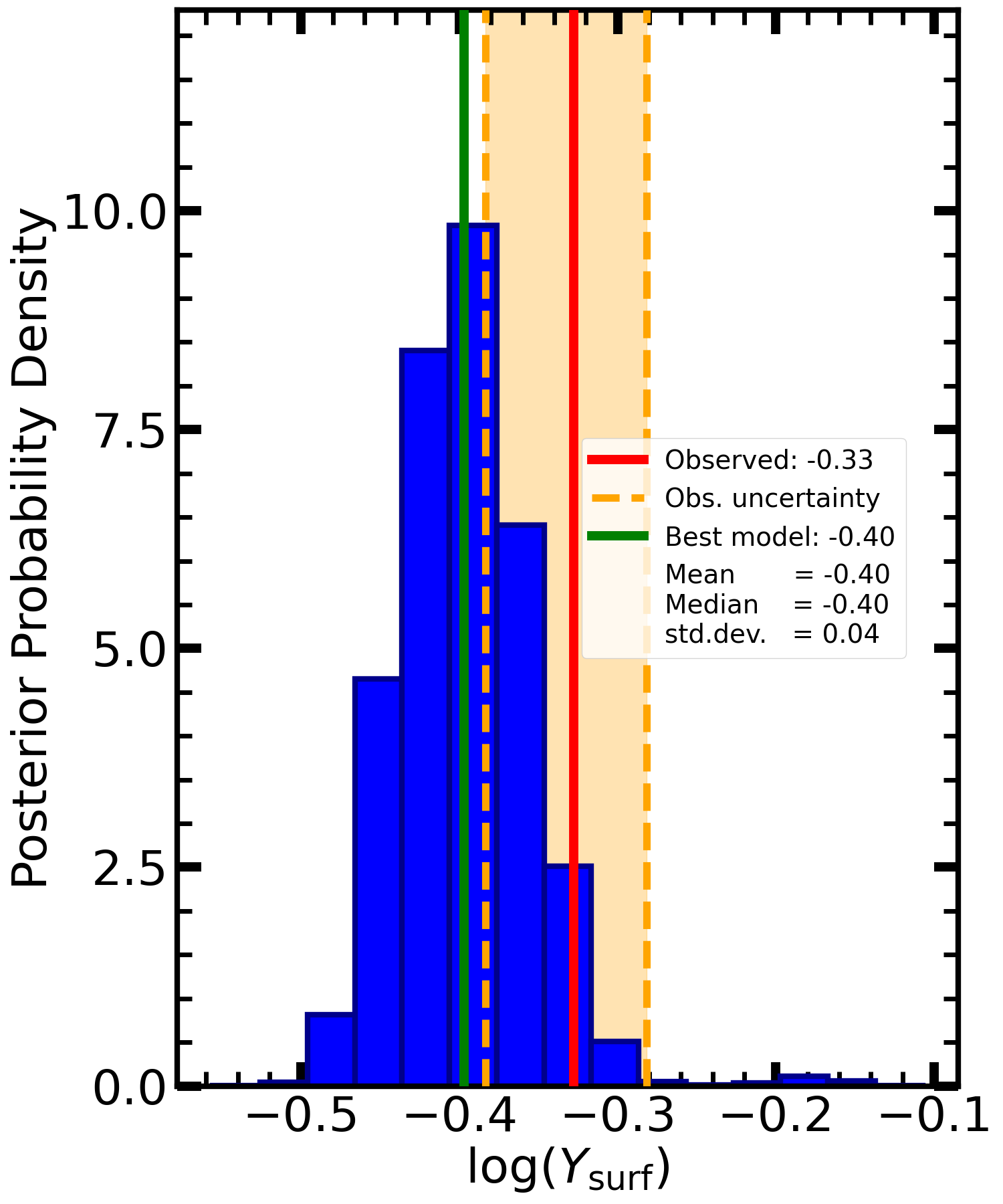}
    \includegraphics[width=0.33\textwidth]{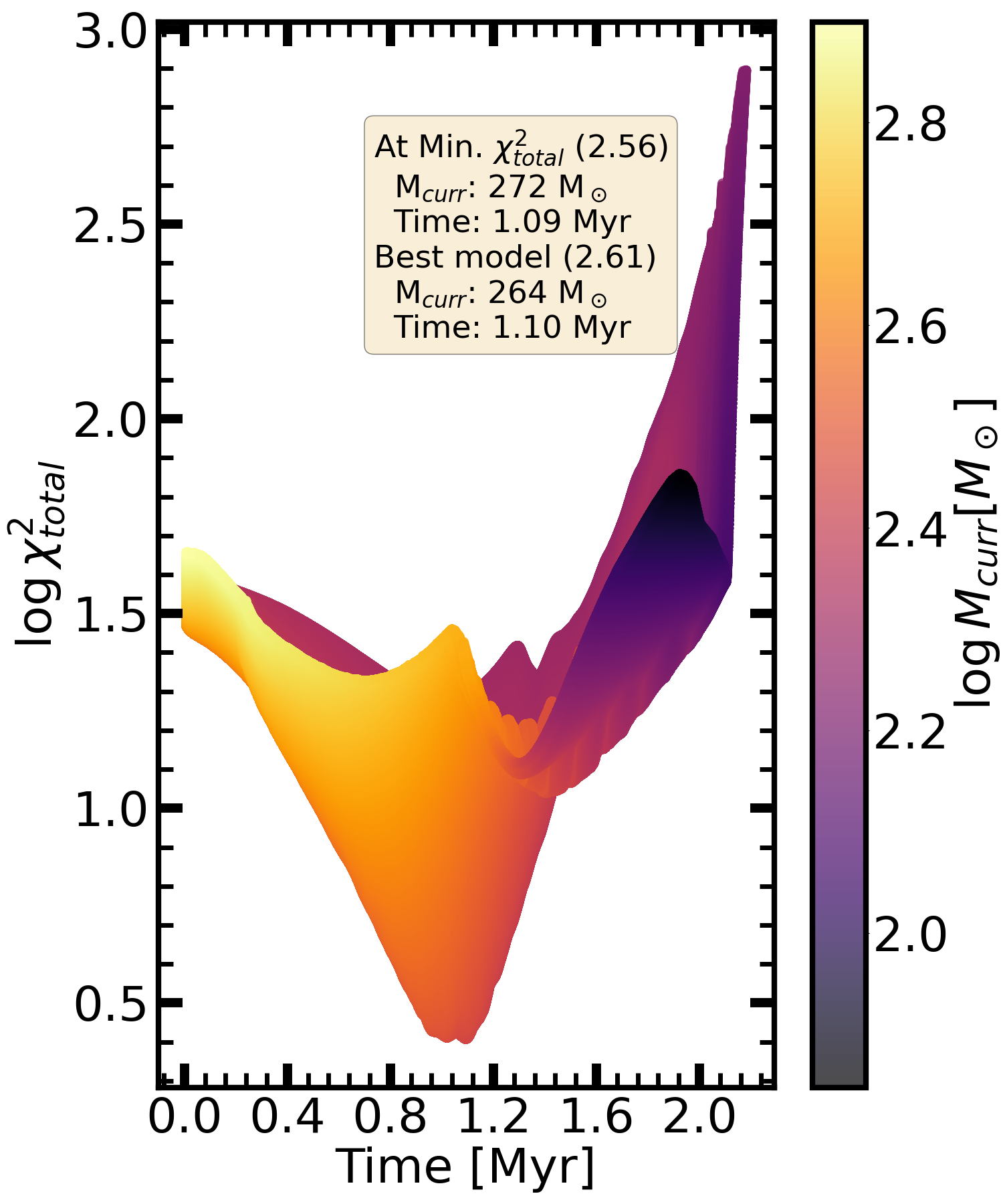}\includegraphics[width=0.33\textwidth]{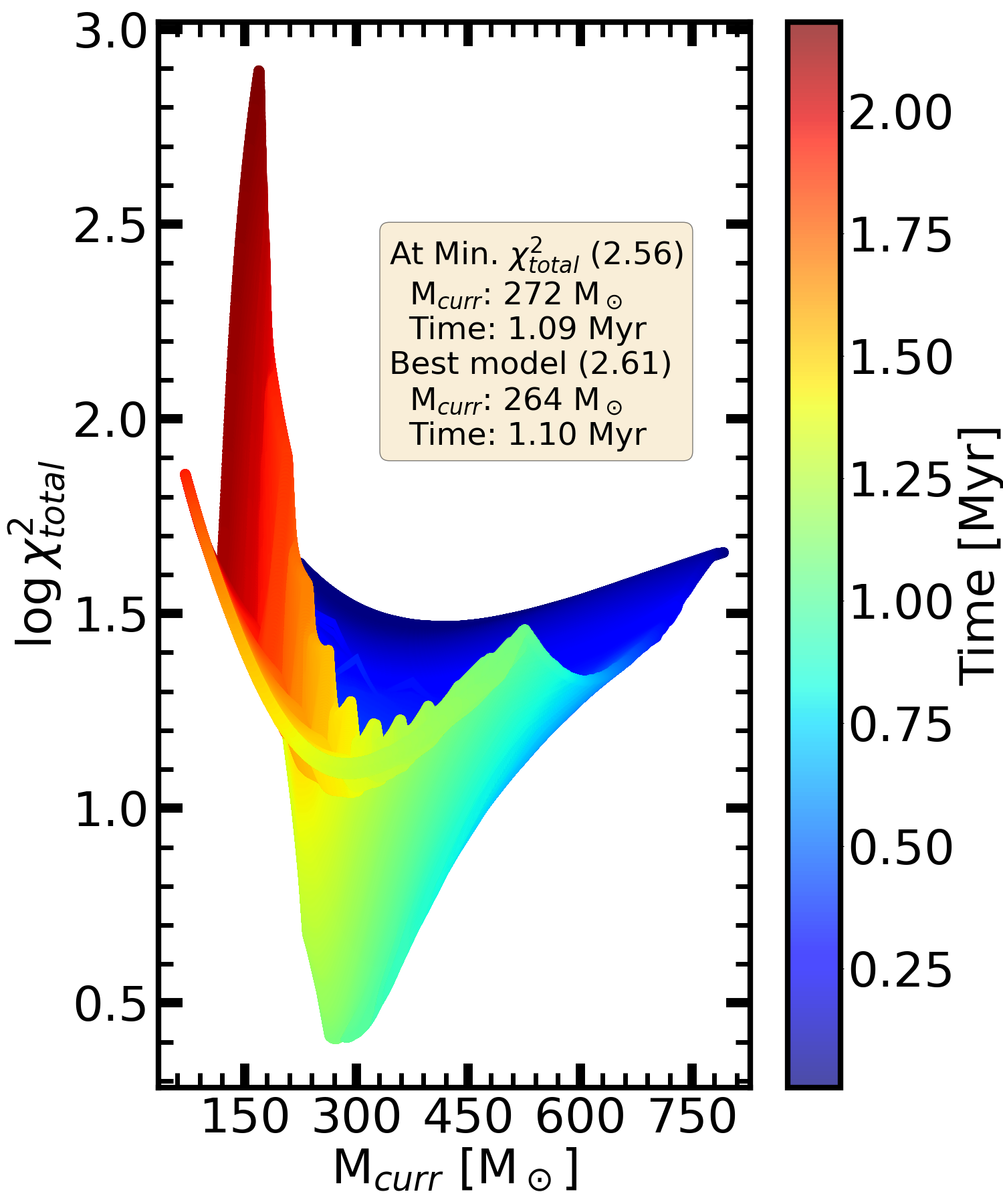}\includegraphics[width=0.33\textwidth]{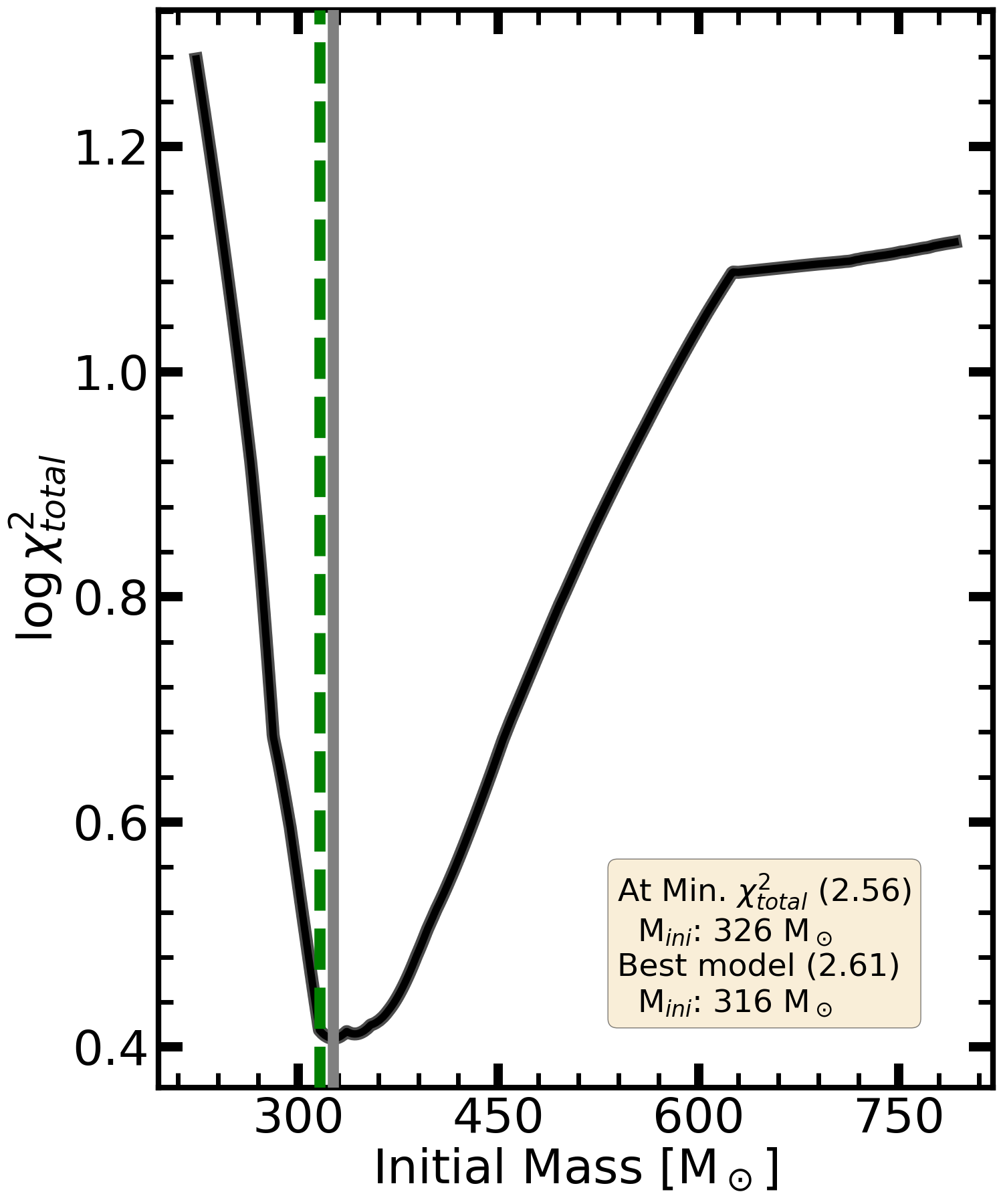}
    \caption{MCMC analysis results for R136a1, considering the TS-300 interpolated grid as an example. The top panels show the individual posterior probability density distributions for the three parameters considered in the analysis. The lower left and middle panels show $\chi^2_{\rm total}$ as a function of time and current mass (with colour-coding of current mass and time, respectively). The lower right panel shows a $\chi^2_{\rm total}$ distribution projected for the initial masses. The grey line corresponds to the lowest $\chi^2_{\rm total}$ value, the green line shows the MAP estimate. }
    \label{fig:R136a1_result}
\end{figure*}

\begin{figure*}
    \centering
    \includegraphics[width=0.33\textwidth]{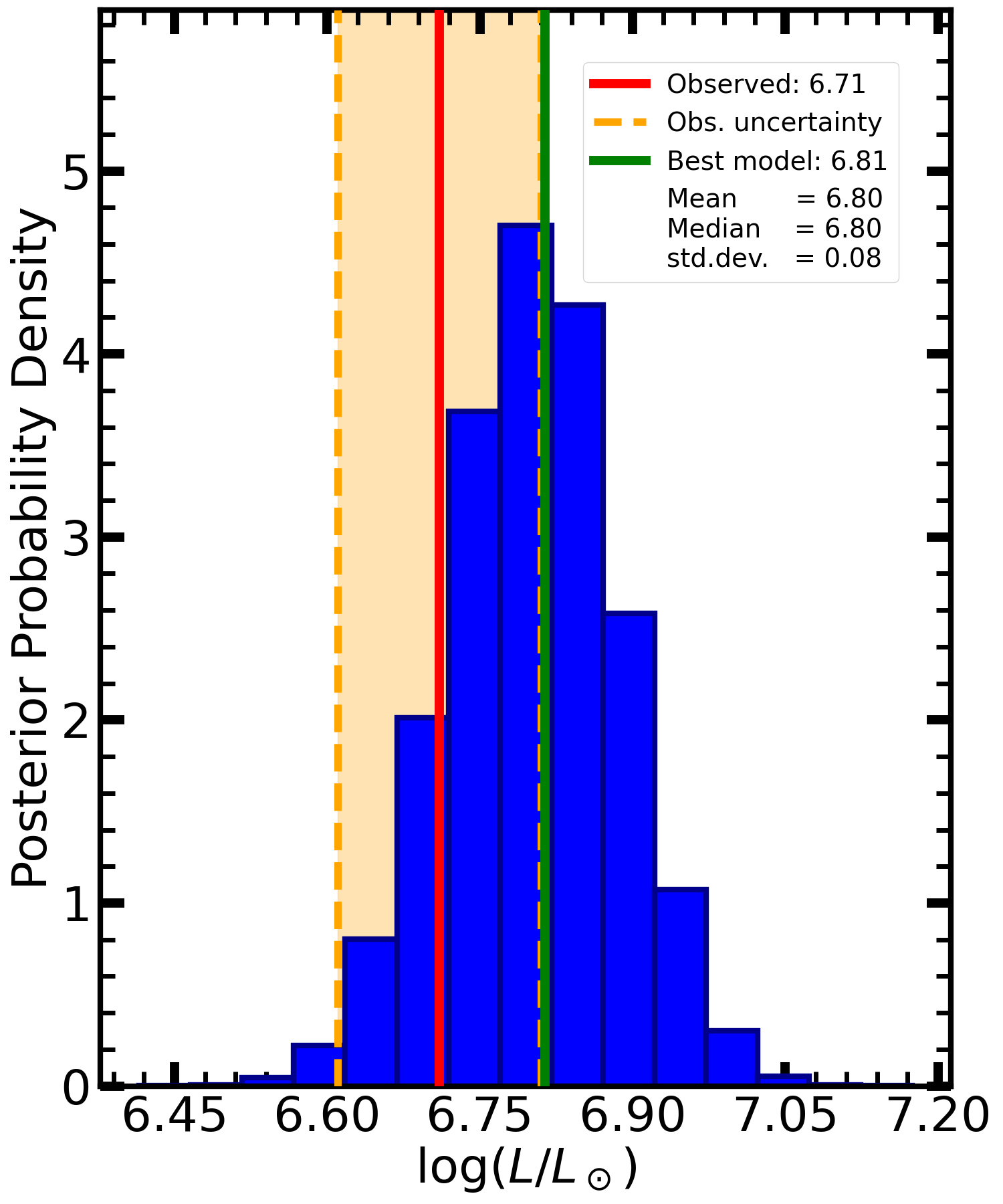}\includegraphics[width=0.33\textwidth]{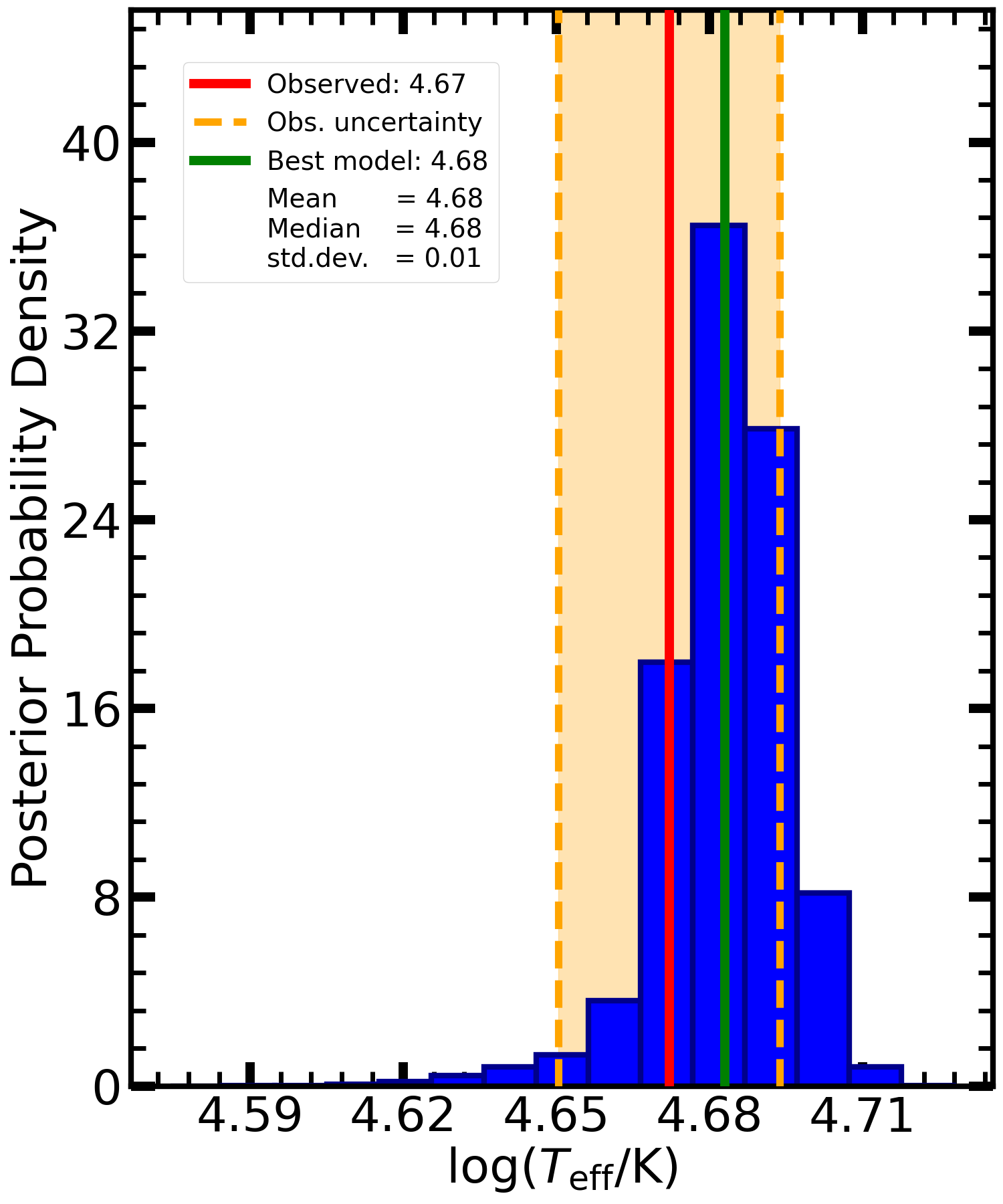}\includegraphics[width=0.33\textwidth]{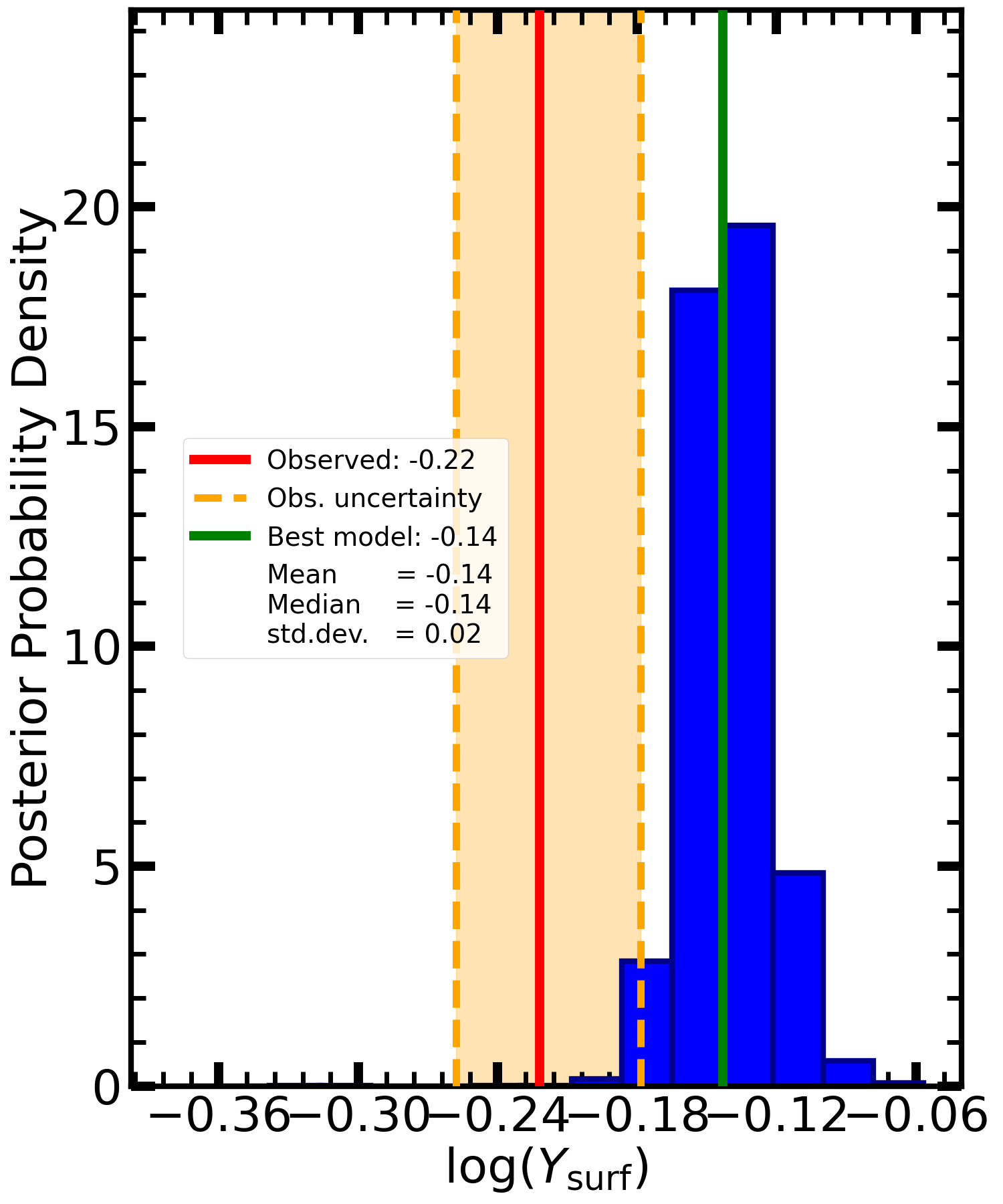}
    \includegraphics[width=0.33\textwidth]{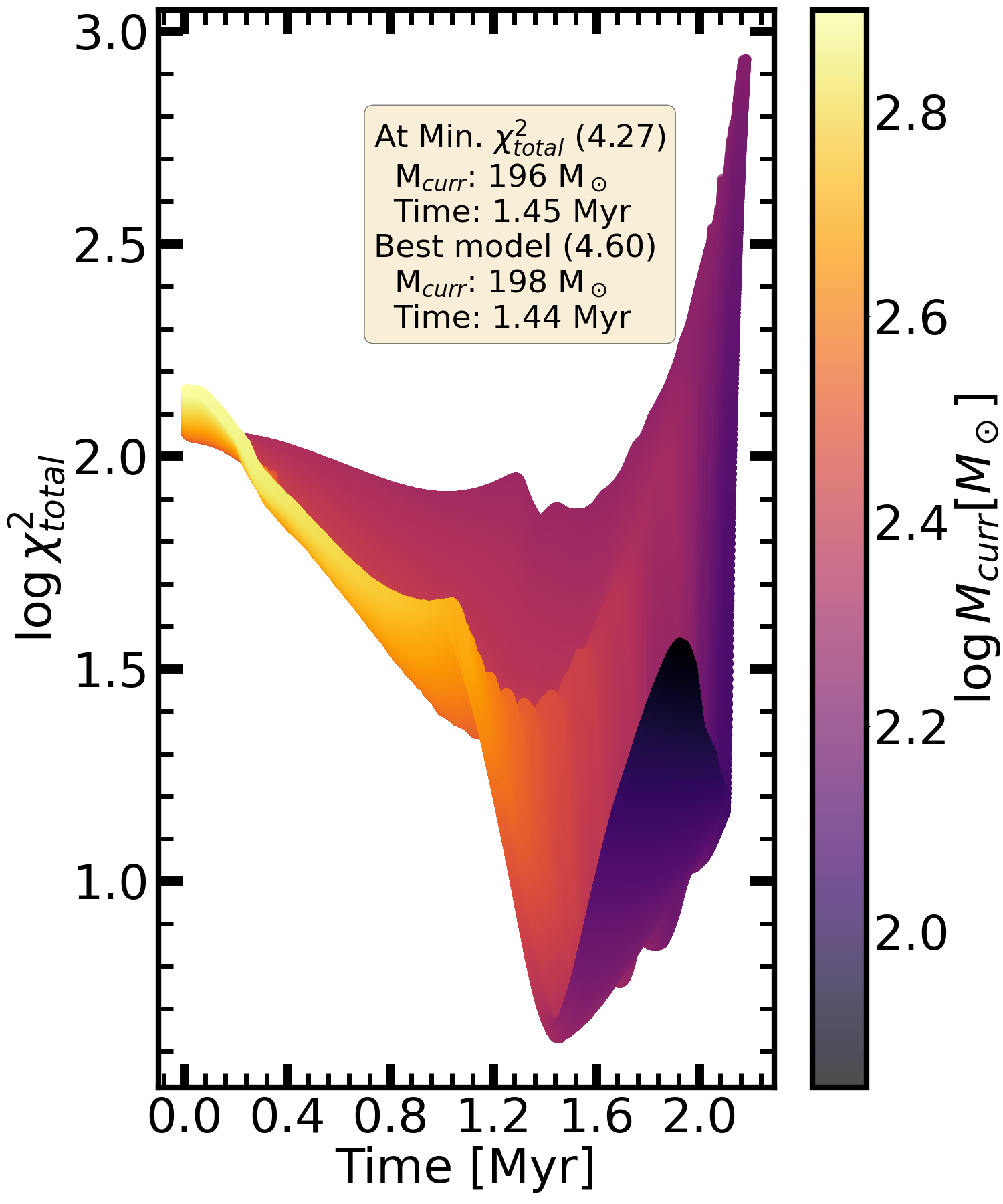}\includegraphics[width=0.33\textwidth]{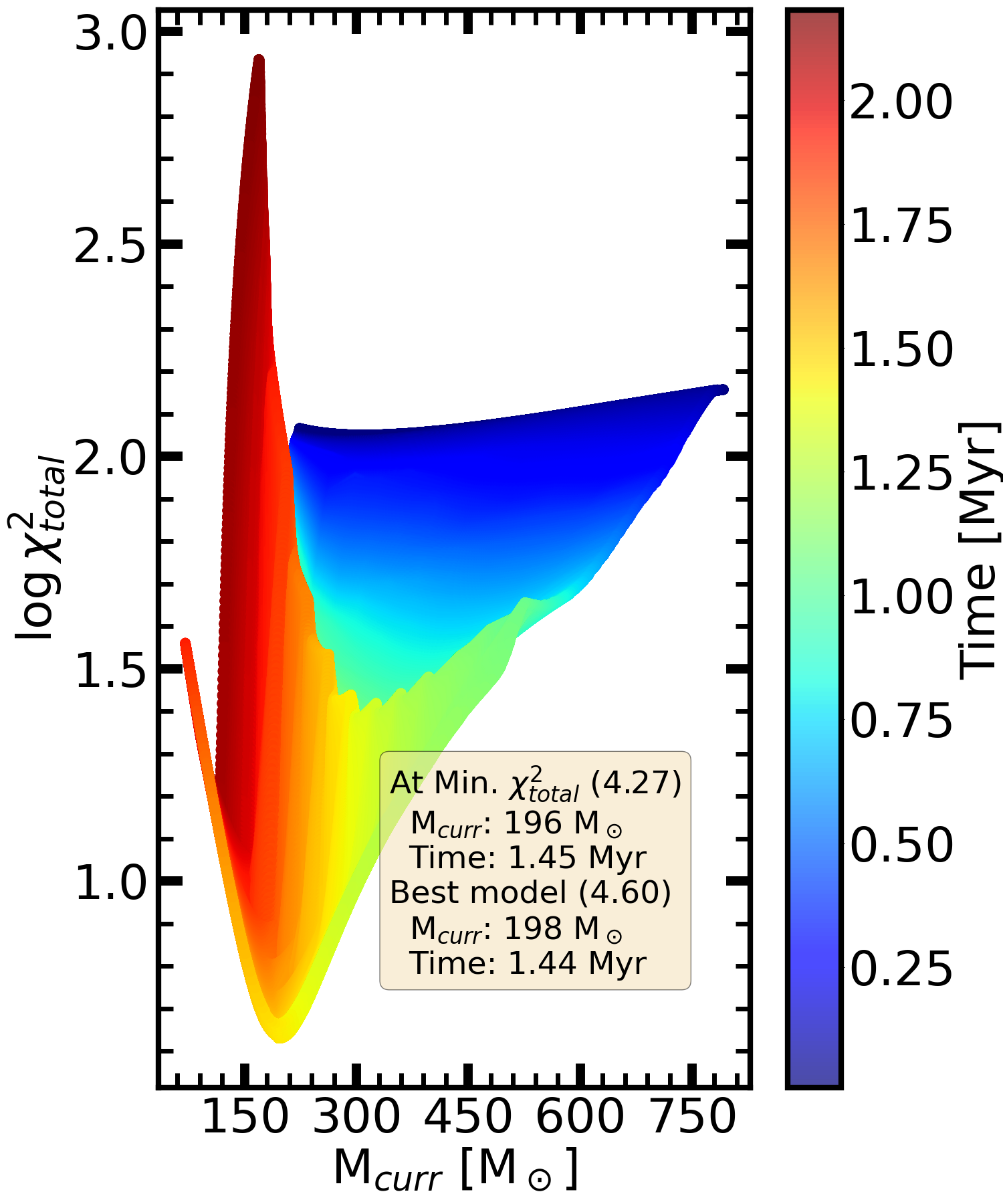}\includegraphics[width=0.33\textwidth]{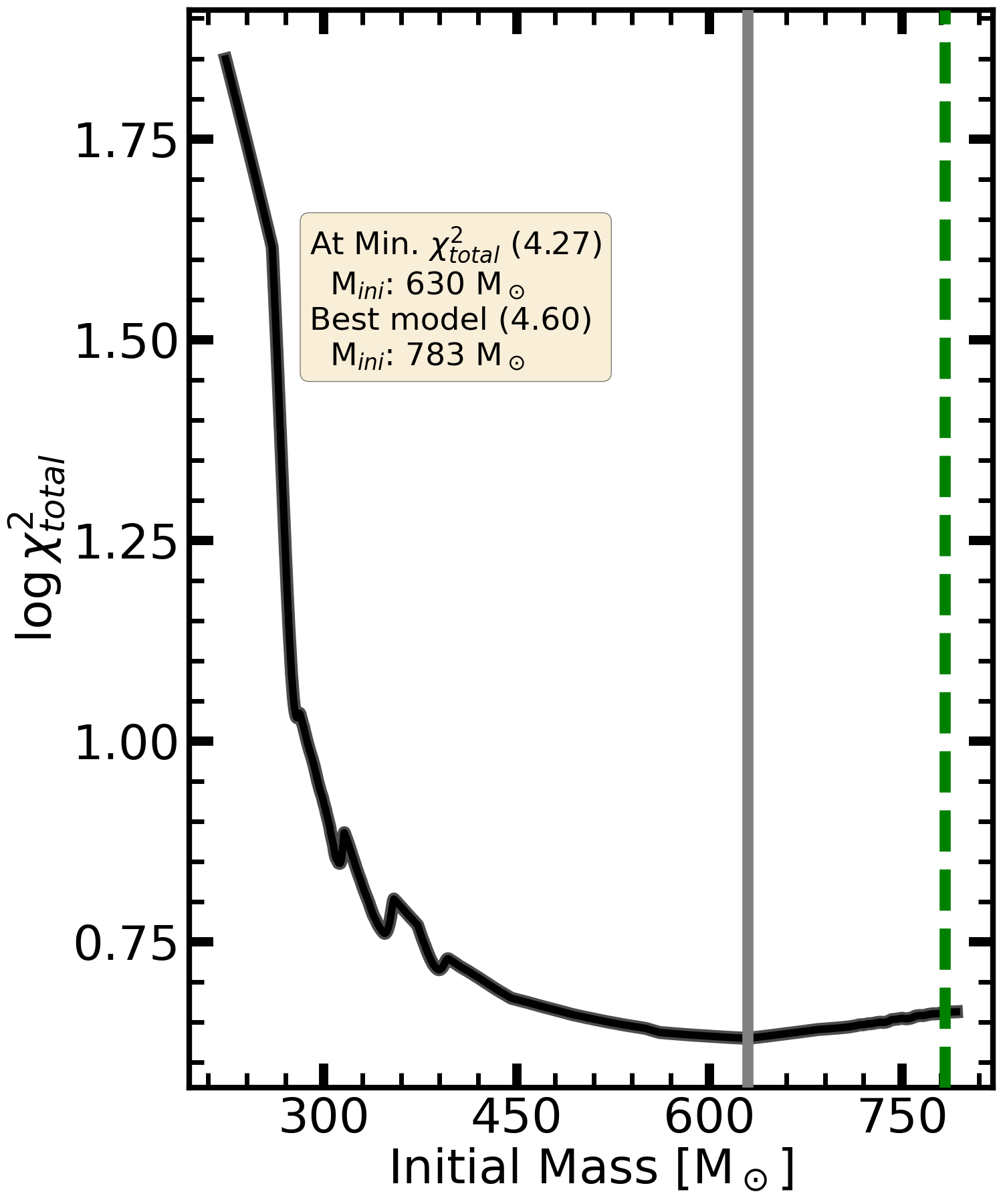}
    \caption{Same as Figure \ref{fig:R136a1_result} but for R136a2. }
    \label{fig:R136a2_result}
\end{figure*}

\begin{figure*}
    \centering
    \includegraphics[width=0.33\textwidth]{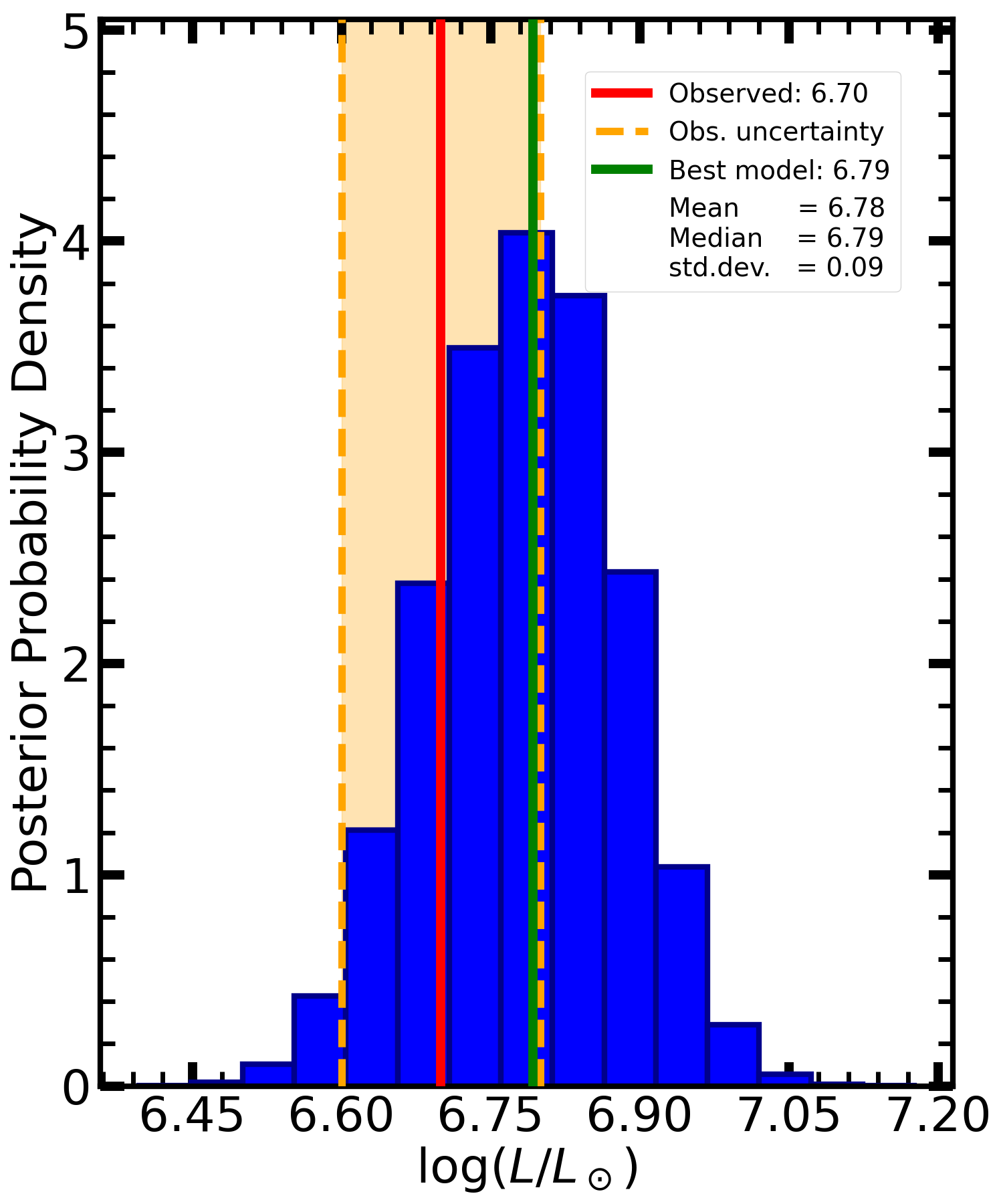}\includegraphics[width=0.33\textwidth]{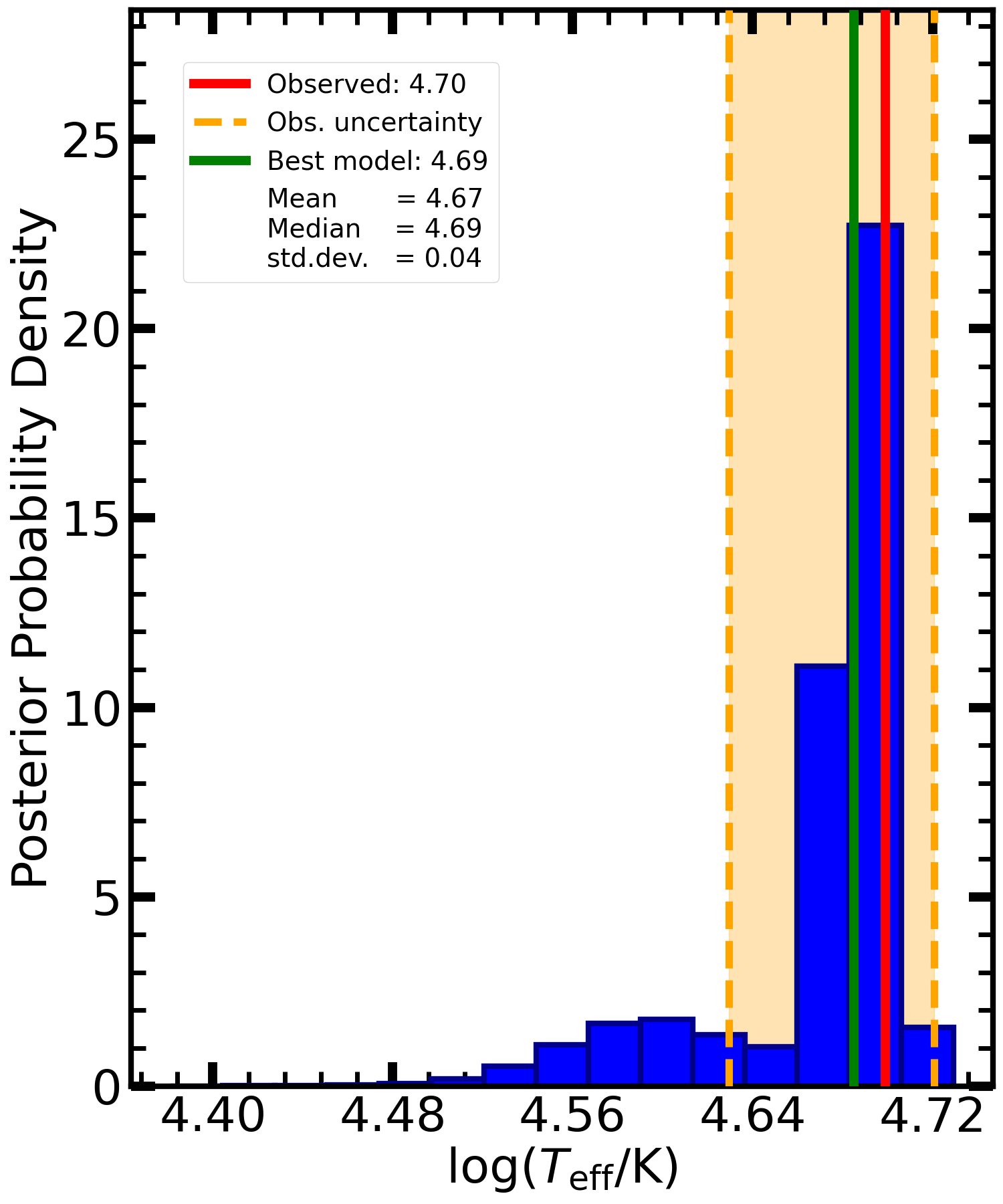}\includegraphics[width=0.33\textwidth]{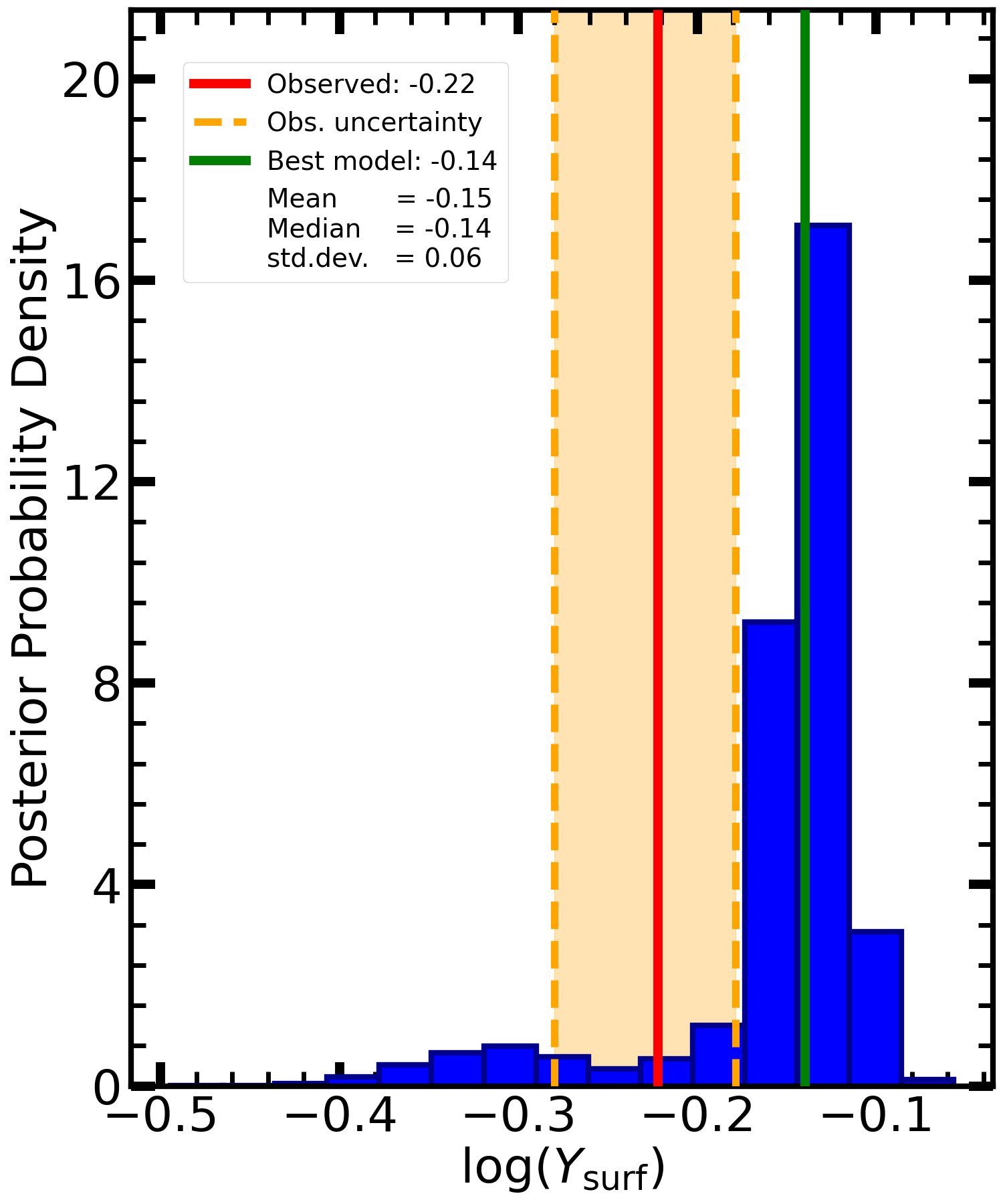}
    \includegraphics[width=0.33\textwidth]{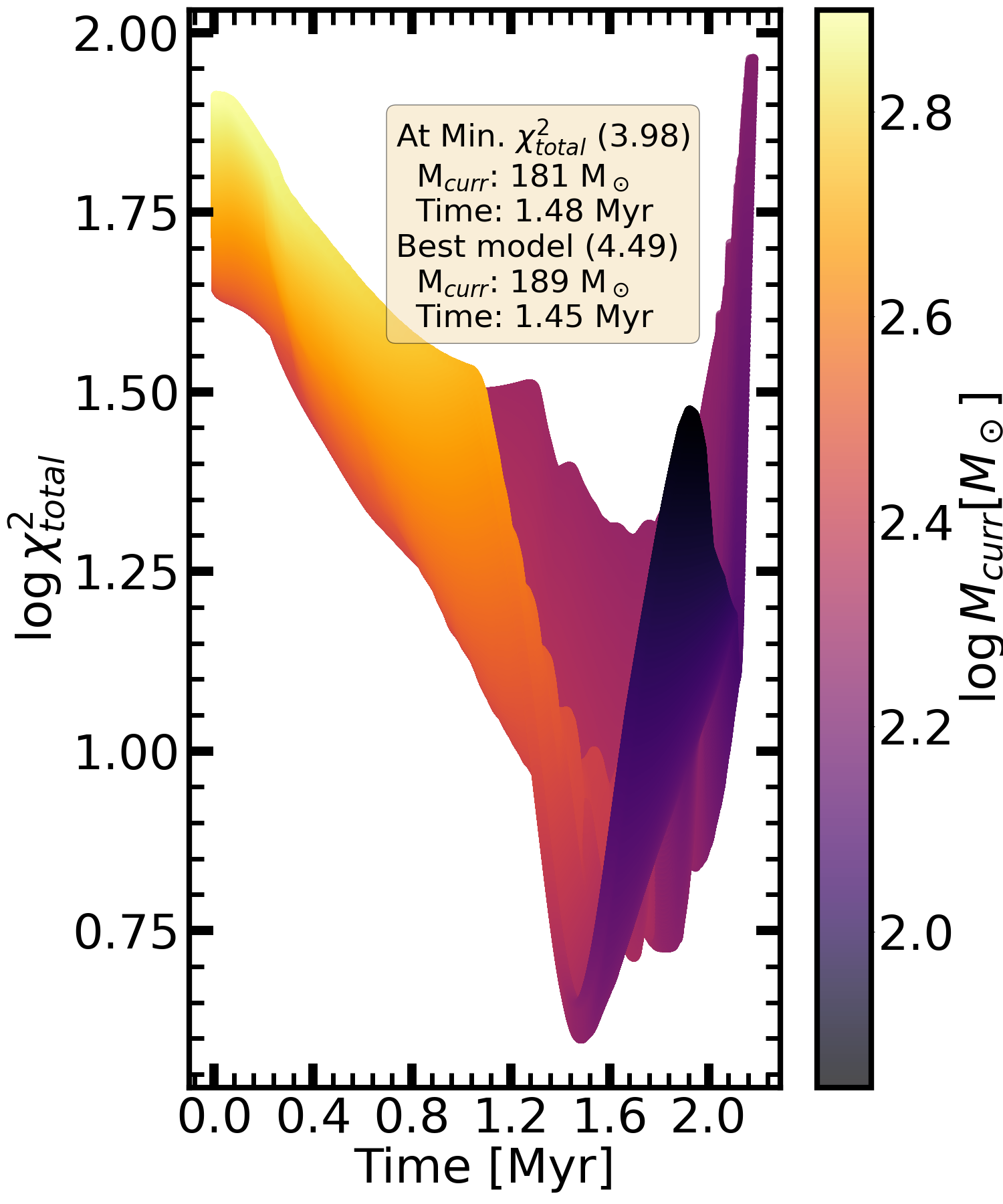}\includegraphics[width=0.33\textwidth]{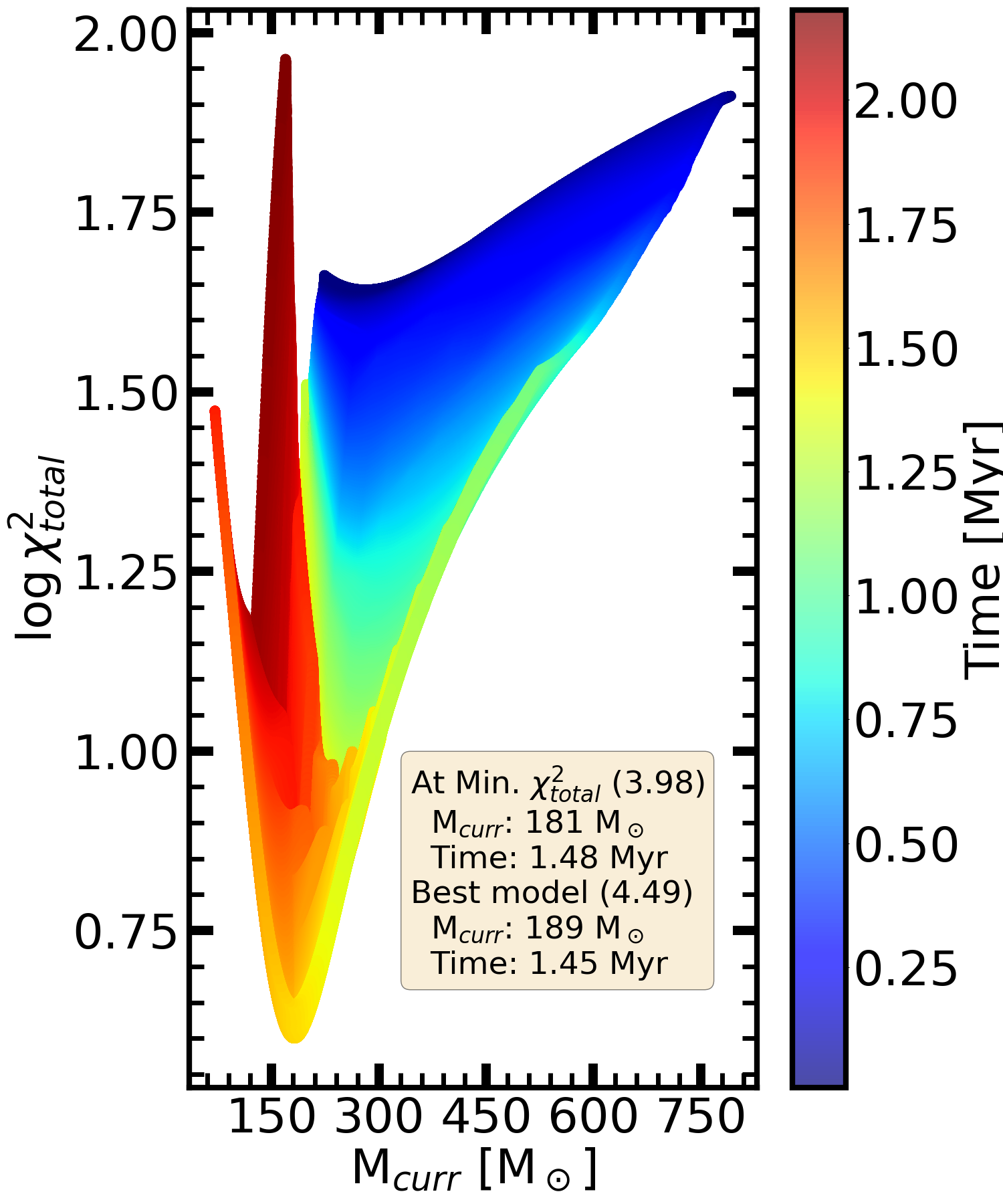}\includegraphics[width=0.33\textwidth]{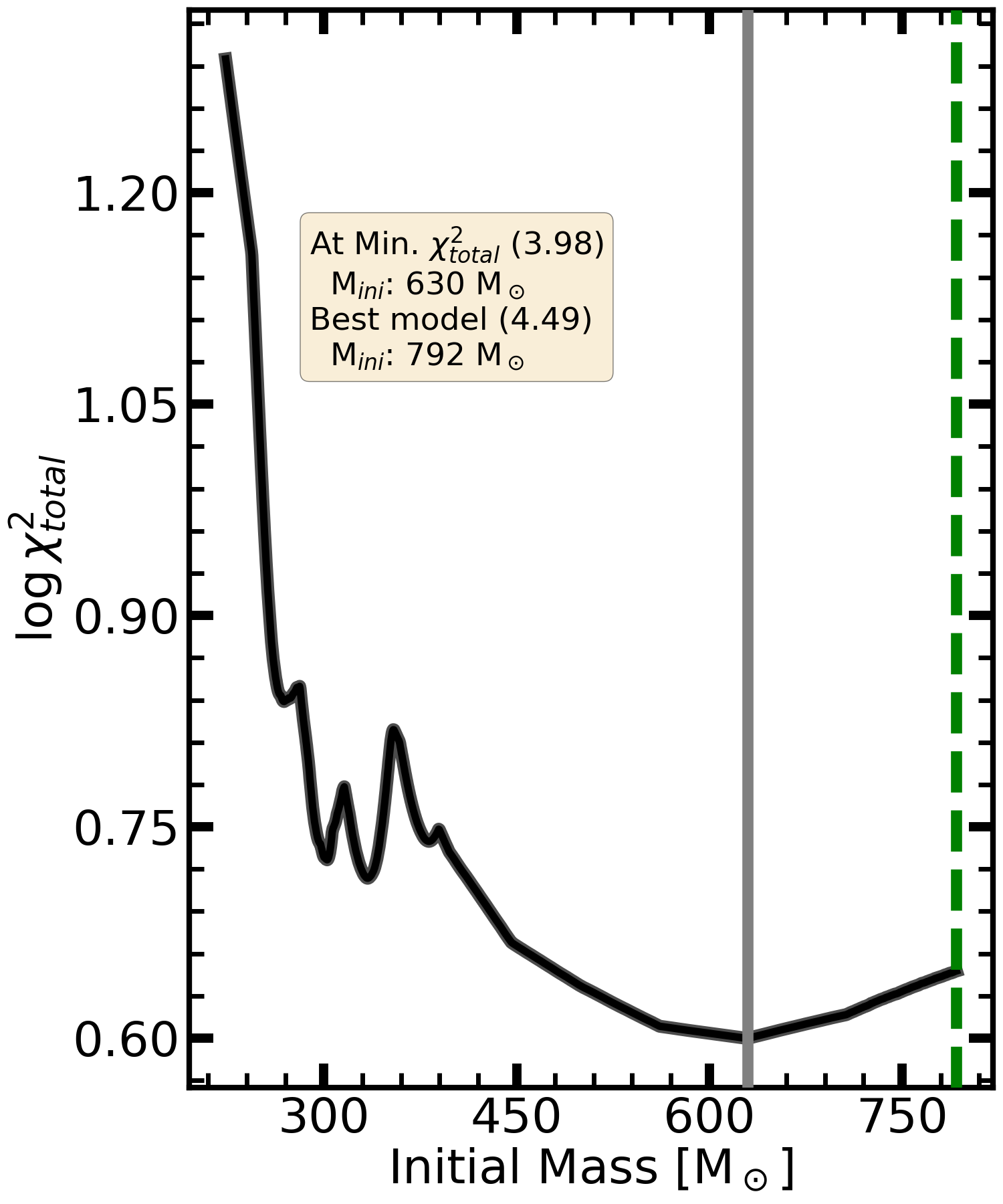}
    \caption{Same as Figure \ref{fig:R136a1_result} but for R136a3.}
    \label{fig:R136a3_result}
\end{figure*}

Figures \ref{fig:R136a1_result}, \ref{fig:R136a2_result}, and \ref{fig:R136a3_result} show the MCMC results based on the TS scheme and an initial rotational velocity of 300~km\,s$^{-1}$ as a demonstration. As highlighted above, the SB scheme and other initial rotational velocities produce broadly consistent results.

%
%
%
%

%
%
%
%

%
%
%
%

\section{\textsc{mesa} microphysics}\label{sec:app_mesamicro}
The \textsc{mesa} equation of state (EOS) is a blend of the OPAL \citep{Rogers2002}, SCVH
\citep{Saumon1995}, FreeEOS \citep{Irwin2004}, HELM \citep{Timmes2000},
PC \citep{Potekhin2010}, and Skye \citep{Jermyn2021} EOSes.

Radiative opacities are primarily from OPAL \citep{Iglesias1993,
Iglesias1996}, with low-temperature data from \citet{Ferguson2005}
and the high-temperature, Compton-scattering dominated regime by
\citet{Poutanen2017}.  Electron conduction opacities are from
\citet{Cassisi2007}.

Nuclear reaction rates are from JINA REACLIB \citep{Cyburt2010}, NACRE \citep{Angulo1999} and
additional tabulated weak reaction rates \citep{Fuller1985, Oda1994,
Langanke2000}. Plasma screening is included via the prescription of \citet{Chugunov2007}.
Thermal neutrino loss rates are from \citet{Itoh1996}.

\end{appendix}  

\end{document}